\RequirePackage{fix-cm}
\documentclass[twocolumn,epjc3]{svjour3}  
\smartqed  
\RequirePackage{graphicx}
\usepackage{graphicx}	
\usepackage{amsmath}	
\usepackage{subfigure}
\usepackage{ifthen}
\usepackage{verbatim}
\usepackage{multicol}
\usepackage{pdflscape}
\usepackage{tabularx}
\usepackage{booktabs} 
\usepackage{rotating} 
\usepackage{adjustbox} 
\usepackage{lipsum} 
\usepackage{float} 
\usepackage{enumitem} 
\usepackage{makecell} 
\usepackage{url}
\usepackage{xcolor}
\usepackage{tikz}
\newcommand{\wignerThree}[6]{
  \begin{pmatrix}
    #1 & #2 & #3 \\
    #4 & #5 & #6
  \end{pmatrix}
}

\journalname{Eur. Phys. J. C}
\begin{document}

\title{Delensing for Precision Cosmology: Optimizing Future CMB B-mode Surveys to Constrain $r$
}


\author{Wen-Zheng Chen\thanksref{addr1,addr2}
        \and
        Yang Liu\thanksref{e1,addr1} 
        \and
        Yi-Ming Wang\thanksref{addr1,addr2}
        \and
        Hong Li\thanksref{e2,addr1,addr2}
}

\thankstext{e1}{e-mail: liuy92@ihep.ac.cn}
\thankstext{e2}{e-mail: hongli@ihep.ac.cn}

\institute{Key Laboratory of Particle Astrophysics, Institute of High Energy Physics, Chinese Academy of Sciences, Beijing 100049, People's Republic of China \label{addr1}
           \and
           University of Chinese Academy of Sciences, Beijing 100049, People's Republic of China \label{addr2}
}

\date{Received: date / Accepted: date}

\maketitle

\begin{abstract}
The detection of primordial B-modes, a key probe of cosmic inflation, is increasingly challenged by contamination from weak gravitational lensing B-modes induced by large-scale structure (LSS). We present a delensing pipeline designed to enhance the sensitivity to the inflationary parameter $r$, minimizing reliance on foreground mitigation during lensing reconstruction. Using simulations of Simons Observatory-like CMB observations and \emph{Euclid}-like LSS surveys in the Northern hemisphere, we demonstrate that excluding low-$\ell$ modes ($\ell<200$) effectively mitigates foreground biases, enabling robust lensing potential reconstruction using observed CMB polarization maps.
We reconstruct the lensing potential with a minimum-variance (MV) quadratic estimator (QE) applied to CMB polarization data and combine this with external LSS tracers to improve delensing efficiency. Two complementary methods—the Gradient-order template and the Inverse-lensing approach—are used to generate lensing B-mode templates, which are cross-correlated with observed B-modes. This achieves a 40\% reduction in the uncertainty of r with CMB-only reconstruction, improving to 60\% when incorporating external LSS tracers.
We validate our results using both the Hamimeche \& Lewis likelihood and a Gaussian approximation, finding consistent constraints on $r$. Our work establishes a streamlined framework for ground-based CMB experiments, demonstrating that synergies with LSS surveys significantly enhance sensitivity to primordial gravitational waves.

\end{abstract}

\section{Introduction}
\label{intro}
The Cosmic Microwave Background (CMB) radiation, a faint glow remaining from the early universe, serves as a crucial window into the cosmos' infancy, offering a snapshot of the primordial universe frozen in time. A particularly exciting aspect of CMB research is the quest to detect primordial gravitational waves (PGWs). These ripples in the fabric of spacetime are believed to have been generated during the inflationary epoch, a period of rapid expansion that occurred a fraction of a second after the Big Bang. Detecting these waves would provide direct evidence for inflation, with profound implications for our understanding of fundamental physics and the origin of the universe.

Primordial gravitational waves leave a unique imprint on the CMB, manifesting as a distinct pattern of polarization known as B-modes. Unlike E-mode polarization, which can be produced by both density fluctuations and gravitational waves, primordial B-modes are the distinctive signature of gravitational waves. 
However, the measurement of primordial CMB B-modes is heavily contaminated due to the long journey from the last scattering surface to the Earth, during which the inherently weak signal is further affected.

One of the most significant challenges in achieving accurate measurements of CMB polarization is contamination from foreground emissions, particularly Galactic polarized foregrounds, which are highly complex and difficult to mitigate.
Depending on the sky region, their amplitudes correspond roughly to $r$=0.01–0.1 at their minimum frequency range (70–90 GHz)\cite{krachmalnicoff2016characterization}. 
Galactic polarized foregrounds are primarily composed of two components: synchrotron radiation, which dominates at lower frequencies (below 100 GHz), and thermal dust emission, which becomes significant at higher frequencies (above 100 GHz). Synchrotron radiation is produced by relativistic electrons spiraling in the Galactic magnetic field, while thermal dust emission arises from dust grains aligned with the same magnetic field.
Fortunately, several component separation methods have been developed to disentangle foregrounds and the CMB based on their different frequency dependence, such as \texttt{Commander} \cite{eriksen2008joint,adam2016planck}, \texttt{NILC} \cite{basak2012needlet}, \texttt{SEVEM} \cite{fernandez2012multiresolution}, and \texttt{SMICA} \cite{cardoso2008component}, among others.  In this paper, we use the NILC method to examine the foreground components in the large aperture telescope (LAT) observations.

With the improvements in detector sensitivity, both current and future CMB experiments have enabled to measure the primordial fluctuations in the CMB B-mode polarization. The sensitivity is now such that contamination from lensing B-modes, rather than instrumental noise, is becoming the primary source of limitation.
During the travel of CMB photons from the last scattering surface to us, they are gravitationally deflected by large-scale structures (LSS) in the universe. This gravitational lensing effect distorts the observed pattern of CMB anisotropies, with a typical deflection angle of about 2 arcminutes \cite{lewis2006weak}. A portion of primordial E-modes is converted into B-modes during weak gravitational lensing, and the resulting B-modes contaminate measurements of primordial gravitational waves, making lensing B-mode a source of noise for primordial B-mode detection \cite{manzotti2017cmb}. Therefore, reducing the contamination introduced by lensing becomes increasingly crucial when instrumental noise becomes subdominant compared to the lensed B-mode, which occurs at around 5 $\mu$K-arcmin.

The removal of lensing contamination from CMB anisotropy observation maps has garnered significant attention and has been widely studied in the literature \cite{smith2012delensing,carron2017maximum,sherwin2015delensing,simard2015prospects}. Prior knowledge of the mass distribution  (i.e., lensing potential $\phi$), whether obtained using a quadratic estimator from the two CMB fields (internal) \cite{carron2017maximum,Okamoto:2003zw} or from a properly filtered external datasets, such as the Cosmic Infrared Background \cite{sherwin2015delensing} or high-redshift galaxies \cite{smith2012delensing}, is the prerequisite for measuring the effect of lensing and reduce its contamination to primordial B-mode observation. The former has been extensively studied in the literature  \cite{namikawa2012full,namikawa2014lensing,story2015measurement,Han:2023gvr,Liu:2022beb}. Using a minimum variance estimator, the Planck 2018 temperature and polarization maps improve the measurement of the lensing potential to a confidence level of 40 $\sigma$ \cite{aghanim2020planck}.

Additionally, the presence of foregrounds can also interfere with the delensing process. Two intuitive approaches have been proposed: one skips foreground cleaning before delensing, using frequency bands with weaker foreground polarization instead, as in \cite{manzotti2017cmb,han2021atacama,ade2021demonstration}; the other performs component separation first, followed by delensing, as in \cite{belkner2024cmb}. The former saves the computational cost of foreground cleaning without significantly sacrificing performance, while the latter appears essential for observations with high sensitivities, aimed at more precise detection of $r$.
In 2016, the BICEP/Keck and SPTpol collaborations jointly constrained $r$ by incorporating a lensing template into their likelihood analysis \cite{ade2016improved}. By 2021, they achieved a $10\%$ reduction in the uncertainty on $r$ by adding a lensing B-mode template \cite{ade2021demonstration}.
For next-generation ground-based CMB experiments like CMB-S4, literature \cite{belkner2024cmb} outlines a strategy where a harmonic internal linear combination (ILC) is applied to the data prior to delensing, followed by iterative internal delensing. This approach has proven highly effective in reducing lensing contamination and tightening constraints on $r$. Achieving the target sensitivity of $\sigma(r)\approx5 \cdot 10^{-4}$ will require an efficient delensing procedure, highlighting its critical role in enabling high-precision measurements.

Likelihood functions are fundamental to CMB analysis, as they encapsulate the statistical relationship between theoretical models and observed data. Traditionally, a Gaussian approximation to the likelihood has been widely used due to its simplicity and computational efficiency, assuming that the underlying data (e.g., power spectra) follow a multivariate normal distribution. However, for partial sky coverage and large-scale observations involving polarization, the non-Gaussian nature of the data must be properly accounted for. Hamimeche and Lewis (HL) \cite{hamimeche2008likelihood} proposed a fast and accurate approximation that effectively captures the non-Gaussian shape of the likelihood. In this work, we employ both the Gaussian and HL approximations to analyze and compare their impact on parameter estimation.

Under the current conditions of CMB experiments, the internal lensing reconstruction typically results in a low signal-to-noise ratio (S/N) on small scales, due to the presence of the reconstruction noise (primarily $N^0$). Therefore, it is promising to combine the external LSS tracers, such as CIB and galaxy number density, to improve the S/N of the lensing reconstruction, especially on small scales. Even though the primary contribution to large-scale lensed B-modes originates from lensing on intermediate scales ($L \sim \mathcal{O}(100)$, see Fig.2 of \cite{simard2015prospects}), incorporating external data has been shown to indeed enhance delensing efficiency, as demonstrated in \cite{yu2017multitracer,namikawa2024litebird,manzotti2018future,baleato2022delensing,hertig2024simons}.

In this paper, we present a comprehensive approach aimed at refining the constraints on $r$ for future ground-based CMB polarization observations, leveraging simulated data as the foundation. The structure of the paper is as follows: Section \ref{sec:method} introduces the two delensing methods utilized in our work, followed by a comprehensive analysis of all delensing biases, with a particular focus on the contributions from foregrounds and lensing reconstruction noise. Section \ref{sec:sims} provides details of the simulation setup for both the CMB maps, the lensing potential maps, the foreground maps and the LSS tracers. Our implementation and main results of NILC, lensing reconstruction, and Lensing B-mode Template (LT) construction are presented in Section \ref{sec:result}. We also clarify the feasibility of the pipeline, where no prior foreground cleaning is performed. We then constrain the parameter $r$ by adding the lensing template as an additional pseudo-channel. Section \ref{sec:conclusion} concludes the paper. In Appendix \ref{sec:nilc}, we describe the theory of NILC in the context of observations contaminated by diffuse Galactic and Extragalactic emissions. In Appendix \ref{sec:phirec}, we introduce our lensing reconstruction pipeline following Planck 2018 baseline \cite{aghanim2020planck}. 
Some potentially useful algorithms for delensing operations are summarized in Appendix \ref{sec:algorithms}. Appendix \ref{app:z_binning} provides a brief analysis of the improvement achieved through redshift binning when combining LSS tracers that include redshift information. We conduct a toy test for our generalized model used in parameter constraints in Appendix \ref{app:model_toy}.

\section{CMB lensing and delensing}\label{sec:method}
The lensing effect of large-scale structures (LSS) on the primordial CMB can be described as a remapping of its temperature and polarization anisotropies. This effect causes primordial CMB photons to reach us from directions deviating from their original line of sight, which can be mathematically expressed as:
\begin{equation}\label{EQ:lensedT}
    \begin{aligned}
        \tilde{\Theta}(\hat{\mathbf{n}}) &= \Theta(\hat{\mathbf{n}} + \mathbf{d}(\hat{\mathbf{n}})) \\
        &= \Theta(\hat{\mathbf{n}}) + \nabla_i\Theta(\hat{\mathbf{n}})\mathbf{d}^i(\hat{\mathbf{n}}) + \mathcal{O}(\phi^2),
    \end{aligned}
\end{equation}
\begin{equation}\label{EQ:lensedQU}
    \begin{aligned}
        \relax[\tilde Q\pm i\tilde U](\hat{\mathbf{n}}) &= [Q\pm iU](\hat{\mathbf{n}} + \mathbf{d}(\hat{\mathbf{n}})) \\
            &= [Q\pm iU](\hat{\mathbf{n}}) + \nabla_i[Q\pm iU](\hat{\mathbf{n}})\mathbf{d}^i(\hat{\mathbf{n}}) + \mathcal{O}(\phi^2),
    \end{aligned}
\end{equation}
where tildes indicate lensed quantities and \( \mathbf{d} \) is the deflection angle, given by the gradient of the lensing potential \( \mathbf{d}(\hat{\mathbf{n}}) = \nabla \phi(\hat{\mathbf{n}}) \) (here we ignore the curl modes) \cite{namikawa2014lensing}, which correlates to the lensing convergence as:
\[
\kappa = -\frac{1}{2} \nabla \cdot \mathbf{d}.
\]

One interesting aspect is that in harmonic space, the remapping will convert part of the E-modes into B-modes, known as lensing B-modes. At linear order in $\phi$, these are given by (neglecting the tensor B-modes) \cite{namikawa2022simons}:
\begin{equation}\label{EQ:lensed_B}
	\begin{aligned}
		{B}_{\ell m}^\text{lens} &= i \sum_{\ell' m'}\sum_{LM} \wignerThree{\ell}{\ell'}{L}{m}{m'}{M} p^{-}F^{(2)}_{\ell L \ell'} E^*_{\ell'm'}\kappa^*_{LM},
	\end{aligned}
\end{equation}
where $p^{+} (p^{-})$ is unity if $\ell+L+\ell'$ is even (odd) and zero otherwise. Lensing induced mode coupling for spin-$s$ fields can be expressed as:
\begin{equation}\label{EQ:mode_coupling}
	\begin{aligned}
	F^{(s)}_{\ell Ll'} = \frac{2}{L(L+1)} \left[ \ell'(\ell' + 1) + L(L + 1) - \ell(\ell + 1) \right] \\
    \times \sqrt{\frac{(2\ell + 1)(2\ell' + 1)(2L + 1)}{16\pi}} \wignerThree{\ell}{\ell'}{0}{-s}{s'}{0}
	\end{aligned}
\end{equation}

As instrumental sensitivity improves, lensing-induced B-modes become an increasingly significant source of contamination in the observation of primordial gravitational waves (PGWs). In Section \ref{sec: method_intro}, we introduce two fundamental delensing methods, with a more detailed discussion available in \cite{chen2025enhancing}. Furthermore, a comprehensive analysis of biases in the delensing procedure is presented in Section \ref{sec:bias}.

\subsection{A brief introduction to the methods}\label{sec: method_intro}
CMB B-mode delensing can be implemented using two approaches: one constructs a template based on a gradient-order approximation, while the other directly remaps the observed field to its original orientation. Below, we provide a brief overview of both methods.

\paragraph{\textbf{Gradient-order Template Method}} {The linear-order approximation (Eq. (\ref{EQ:lensed_B})) is known to be a good approximation for the lensing B-modes on large scales \cite{challinor2005lensed,baleato2021limitations}. Therefore, it is feasible to construct a lensing B-mode template on large scales using only gradient-order terms. This method is particularly convenient when working in the map domain, as shown by Eq. (\ref{EQ:lensedQU}). In this case, the gradient-order template can be expressed as the product of the deflection angle and the gradients of the polarization fields , namely: 
$\nabla_i[Q \pm iU](\hat{\mathbf{n}}) \nabla^i \phi(\hat{\mathbf{n}})$.
In practice, the gradient of a field can be easily obtained using the ladder operators \cite{Okamoto:2003zw}:
\begin{equation}
	\begin{gathered}
		D_i [{_s}f(\hat{\mathbf{n}})] = -\frac{1}{\sqrt{2}} \left\{ \sharp {_s}f(\hat{\mathbf{n}}) \bar m + \flat {_s}f(\hat{\mathbf{n}}) m \right\},
	\end{gathered}
\end{equation}
where $\sharp$ and $\flat$ are the ladder operators, and the complex-conjugated vectors $\bar{m}$ and $m$ serve as the basis. The key point to note is that the Stokes parameters $Q$ and $U$ used to construct the template should exclude the B-modes, as including them would result in their re-lensing, introducing redundant terms. The template can then be subtracted from the observation (map-level delensing) or treated as a pseudo-channel in the likelihood when constraining parameters (cross-spectral method delensing), as discussed in Section \ref{sec: baseline_fit}.}

\paragraph{\textbf{Inverse-lensing Method}} By reversing the lensing effect through remapping the observed photons back to their original positions, we can achieve a more accurate and optimal delensing approach.
The inverse deflection angle is well-defined because the points are remapped onto themselves after being deflected back and forth, as discussed in \cite{Diego-Palazuelos:2020lme}:
\begin{equation}\label{hat_n}
	\hat{\mathbf{n}} + \mathbf{d^{inv}}(\hat{\mathbf{n}}) + \mathbf{d}(\hat{\mathbf{n}} + \mathbf{d^{inv}}(\hat{\mathbf{n}})) = \hat{\mathbf{n}},
\end{equation}
where the primary CMB fields can be recovered from the lensed field by remapping the latter using the inverse deflection angle $\mathbf{d^{inv}}(\hat{\mathbf{n}})$:
\begin{equation}
	\tilde{X}(\hat{\mathbf{n}}) = X(\hat{\mathbf{n}} + \mathbf{d}(\hat{\mathbf{n}})) \Leftrightarrow X(\hat{\mathbf{n}}) = \tilde{X}(\hat{\mathbf{n}} + \mathbf{d^{inv}}(\hat{\mathbf{n}})).
\end{equation}
In practice, the inverse deflection angle is approximately the negative of the deflection at low-order approximation. A more accurate estimate can be obtained by solving Eq. (\ref{hat_n}) using a Newton-Raphson scheme, where the inverse deflection angle $\mathbf{d^{inv}}(\hat{\mathbf{n}})$ is iteratively computed as described in \cite{Carron:2017vfg}. A lensing B-mode template can be obtained by subtracting the inverse-lensed B-mode map from the observed data.

In practice, lensed E-modes are typically used in the gradient template method, which has been shown to perform better than using the unlensed E-modes due to the cancellation of high-order terms \cite{baleato2021limitations}. As discussed in \cite{chen2025enhancing}, inverse-lensing is believed to be more optimal than the gradient template method, as it incorporates more high-order lensing terms. However, in practice, we find that inverse-lensing does not significantly outperform the gradient-order approximation when performing large-scale B-mode delensing. In this regime, the gradient-order approximation is quite accurate, and higher-order terms tend to introduce more noise. On small scales, where the gradient-order approximation becomes less accurate, the inverse-lensing method would show its optimality.

\subsection{Bias Analysis}\label{sec:bias}
Although essential, the delensing procedure may introduce biases in the measurement of $r$. In this work, we systematically analyze these biases and develop debiasing techniques to correct them, thereby enhancing the accuracy of $r$ measurements.

In the following derivation, we comprehensively analyze contributions from all potential sources to facilitate effective debiasing. Building on our previous work \cite{chen2025enhancing}, we extend the analysis to include biases introduced by both the foreground and lensing reconstruction processes. As highlighted in the literature \cite{lizancos2021impact,teng2011cosmic}, the latter—arising from the correlation between the B-modes utilized in lensing reconstruction and those targeted for delensing—plays a significant role in shaping the delensed power spectrum.
Notably, the primary B-mode (if present) is not explicitly incorporated in this analysis, as we have previously demonstrated that its impact on our results is negligible \cite{chen2025enhancing}.

The observed E-modes, B-modes and estimated lensing potential with Wiener filters (for a detailed description, see Section \ref{subsec:mapproc}), are as follows:
\begin{equation}
	\begin{aligned}
		&E = \mathcal{W}^E(E^\text{lens}+E^\text{noise}+E^\text{fg}), \\
		&B = \mathcal{W}^E(B^\text{lens}+B^\text{noise}+B^\text{fg}), \\
		&\hat \phi = \mathcal{W}^{\phi}(\phi + \phi^\text{noise}),
    \end{aligned}
\end{equation}
where $\mathcal{W}^E$ and $\mathcal{W}^{\phi}$ are the Wiener filters applied to CMB polarization observation and reconstructed lensing potential, respectively.

\subsubsection{Bias Analysis of the Gradient-Order Template Method}\label{sec:bias1}
Starting from the expression above, the lensing B-mode template can be written as:
\begin{equation}\label{EQ:template}
	\begin{aligned}
    		B^\text{temp} &= \mathcal{B}^{(1)}[\mathcal{W}^E(E^\text{lens}+E^\text{noise}+E^\text{fg}) \ast \mathcal{W}^{\phi}(\phi + \phi^\text{noise})] \\
		  &= \begin{aligned}[t] &\left\{\mathcal{B}^{(1)}[\mathcal{W}^E E^\text{lens} \ast \mathcal{W}^{\phi}\phi]\right\} + \mathcal{B}^{(1)}[\mathcal{W}^E E^\text{noise} \ast \mathcal{W}^{\phi}\phi] \\
		  	&+ \mathcal{B}^{(1)}[\mathcal{W}^E (E^\text{lens}+E^\text{noise}) \ast \mathcal{W}^{\phi}\phi^\text{noise}] \\
            &+ \mathcal{B}^{(1)}[\mathcal{W}^E E^\text{fg} \ast \mathcal{W}^{\phi}(\phi + \phi^\text{noise})] 
		  \end{aligned} \\
		  &= B^\text{temp}_S + B^\text{temp}_N.
	\end{aligned}
\end{equation}

\begin{figure*}
	\includegraphics[width=\textwidth]{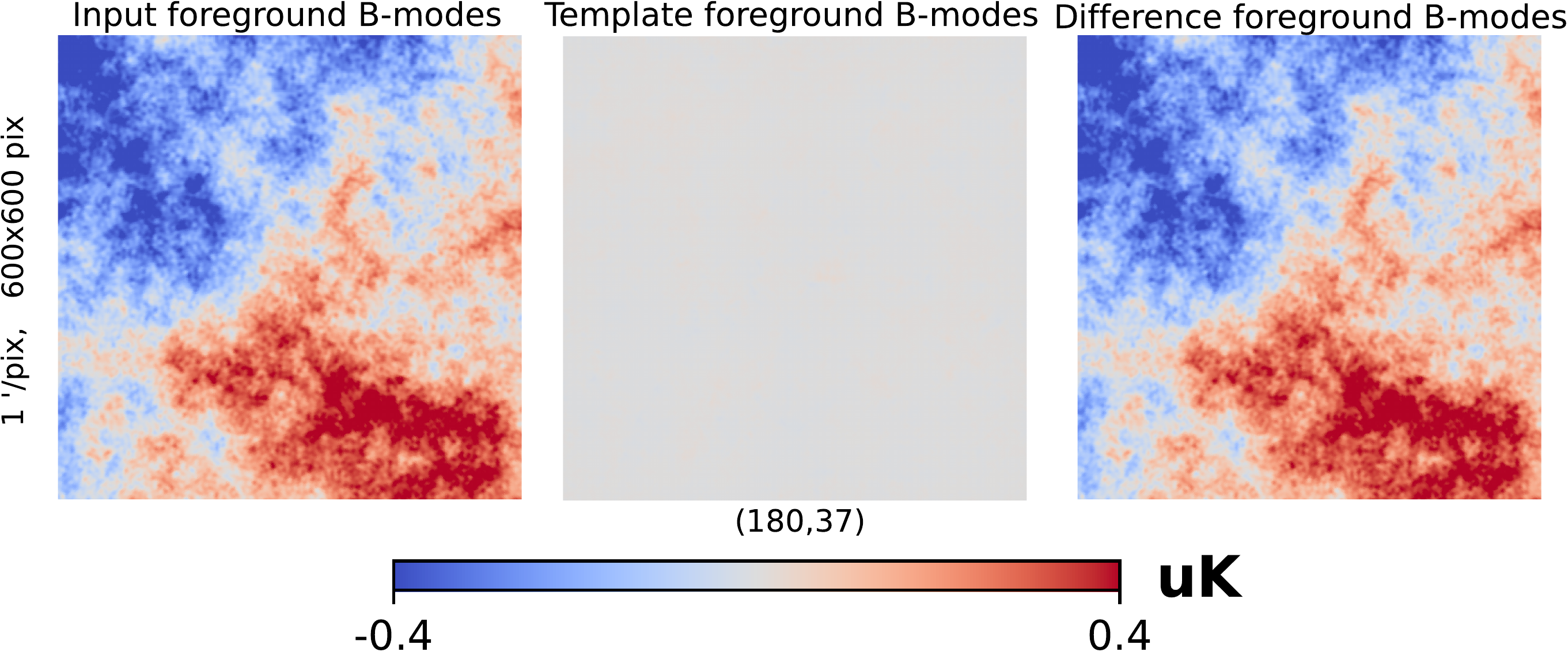}
	\caption{The B-mode maps of the template (middle panel), constructed from the foreground E-mode and the input foreground at 93 GHz (left panel), are presented. We observe that only a negligible portion of the E-modes is converted to B-modes, and the difference map (right panel)—referred to as the delensed map—shows almost no change compared to the input foreground. The maximum multipole is set to be  $\ell_\text{max}=3000$.}
	\label{fig:temp_fg}
\end{figure*}

\begin{figure}
	\includegraphics[width=\columnwidth]{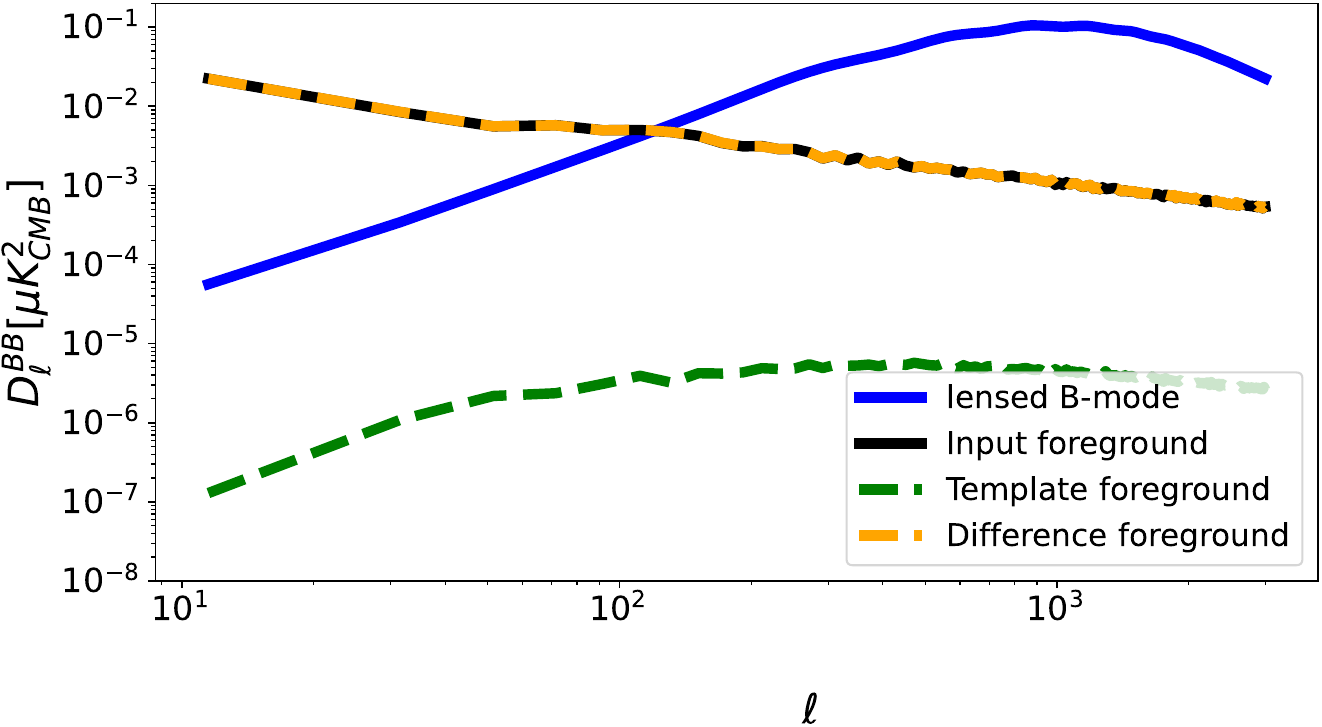}
	\caption{The B-mode power spectra of the template constructed from the foreground E-mode, input foreground, and their difference at 93 GHz is shown. We observe that, at the scales presented, the power of the template constructed from the foreground E-mode is very small, indicating negligible lensing effects (as demonstrated by the perfect overlap of the orange dashed line with the black solid line). It is important to note that the power of the foreground difference is related to the intensity of the input foreground.}  
	\label{fig:temp_fg_cl}
\end{figure}

Here, $\mathcal{B}^{(1)}[E \ast \phi]$ represents the operation of constructing the gradient-order template using $E$ and $\phi$. We artificially define the signal part ($B^{\text{temp}}_S$), which consists of the signal (the term in the braces), and the noise part ($B^{\text{temp}}_N$) introduced during the delensing procedure. 
In practice, we find that the last term from the foreground has a negligible contribution.
As shown in Fig.~\ref{fig:temp_fg}, only a negligible fraction of the foreground E-mode is converted into B-mode. The power spectrum of the foreground lensing B-mode template, represented by the green dashed line in Fig.~\ref{fig:temp_fg_cl}, is substantially smaller than that of the CMB lensed B-mode. This result aligns with expectations, as both the lensing potential and the lensing efficiency within our galaxy is exceptionally weak, leading to minimal interaction with foreground emissions.
Besides, the power of the polarized foreground is close to scale-invariant, characterized by \( \ell (\ell+1) C_{\ell} \approx \text{const} \). Lensing has a minimal effect when the background structures are scale-invariant. In contrast, for backgrounds exhibiting characteristic features such as acoustic oscillations and small-scale damping in the CMB, lensing tends to introduce important distortions.
For a detailed discussion, see \cite{lewis2006weak}.

The delensed B-modes then become,
\begin{equation}\label{EQ:template de}
	\begin{aligned}
		B^\text{del} &= B^\text{obs} - B^\text{temp} \\
				&= (B^\text{lens}+B^\text{noise}+B^\text{fg}) - (B^\text{temp}_S + B^\text{temp}_N) \\
				&= (B^\text{lens}-B^\text{temp}_S) + (B^\text{noise}-B^\text{temp}_N) + B^\text{fg} , \\
    \end{aligned}
\end{equation}
where the first term is defined as the delensing residual due to the method itself, $B^{\text{res}} = B^{\text{lens}} - B^{\text{temp}}_S$, and the second term is related to the presence of noise. 
Due to the presence of these non-primordial B-modes, bias terms are introduced.

These effects can be analyzed by calculating the auto- and cross-power spectra of the lensing template and the observations. The auto-power spectrum of the lensing template is expressed as:
\begin{equation}\label{EQ:template_auto}
	\begin{aligned}
        C_S^\text{temp} = C^\text{temp} - C_N^\text{temp} - 2\langle B^\text{temp}_SB^\text{temp}_N \rangle,
	\end{aligned}
\end{equation}
where $C^\text{temp} = \langle B^\text{temp}B^\text{temp}\rangle$ and $C^\text{temp}_N = \langle B^\text{temp}_NB^\text{temp}_N\rangle$ represent the power spectra of the noisy lensing B-mode template and the noise component of the lensing B-mode template, respectively. 
With \( B^{\text{temp}} = \mathcal{T}^{-1} B^{\text{lens}} + B_N^{\text{temp}} \) and \( B^{\text{obs}} = B^{\text{lens}} + B^{\text{noise}} + B^{\text{fg}} \), the cross-power spectrum between them can be expressed as:
\begin{equation}\label{EQ:template_cross}
	\begin{aligned}
        C^\text{cross} &= \mathcal{T}^{-1} C^\text{lens} + \langle B^\text{lens}B_N^\text{temp} \rangle + \langle B^\text{noise}B^\text{temp} \rangle \\
         &\quad + \langle B^\text{fg}B^\text{temp} \rangle.
	\end{aligned}
\end{equation}


For Eq. (\ref{EQ:template_auto}), we find that the cross-term between $B^{\text{temp}}_S$ and $B^{\text{temp}}_N$ becomes significant, as expected due to the correlation between lensing reconstruction noise $\phi^\text{noise}$ and the lensed B-modes. This phenomenon has been widely discussed in the literature, including \cite{lizancos2021impact,teng2011cosmic}.
For Eq. (\ref{EQ:template_cross}), we observe that the second term can be significant, primarily due to the similar underlying cause discussed above. In contrast, the contributions from the last two terms remain relatively minor.

To estimate the dominant bias terms, we can either compute them directly through simulations or extract them by averaging the difference between the baseline result and the results obtained from signal-only simulations, where the noisy observation and reconstructed \( \hat{\phi} \) in the baseline simulations are replaced with signal-only CMB polarization maps and signal \( \phi \).

We have verified that both methods are effective in debiasing, with the first method is adopted in this work.

\subsubsection{Bias analysis of inverse-lensing method}\label{sec:bias2}
For the inverse-lensing method, we analyze its bias by starting with the remapped B-modes,
\begin{equation}\label{EQ:remap}
	\begin{aligned}
    		B^\text{del} &= \mathcal{B} [\mathcal{W}^E (B^\text{lens} + B^\text{noise} +B^\text{fg}) \star \mathbf{d^{inv}} ] \\
				& \approx  \mathcal{B} [\mathcal{W}^E (B^\text{lens} + B^\text{noise} +B^\text{fg}) \star (\mathbf{d^{inv}_\phi} + \mathbf{d^{inv}_\text{noise}}) ] \\
				& \approx \left\{ \mathcal{B} [\mathcal{W}^E B^\text{lens} \star \mathbf{d^{inv}_\phi} ] \right\} +  \left\{ \mathcal{B} [\mathcal{W}^E B^\text{fg} \star \mathbf{d^{inv}} ] \right\}\\
				&\quad + \left\{ \mathcal{B} [\mathcal{W}^E B^\text{lens} \star \mathbf{d^{inv}_\text{noise}} ] + \mathcal{B} [\mathcal{W}^E  B^\text{noise} \star \mathbf{d^{inv}} ]  - \mathcal{W}^E B^\text{lens} \right\} \\
				&= B^\text{del}_S + B^\text{fg,re} + B^\text{del}_{N}, 
	\end{aligned}
\end{equation}
where $\mathbf{d^{inv}_\phi}$ and $\mathbf{d^{inv}_\text{noise}}$ are the inverse deflection angle obtained from the lensing potential and its associated noise.

\begin{figure*}
	\includegraphics[width=\textwidth]{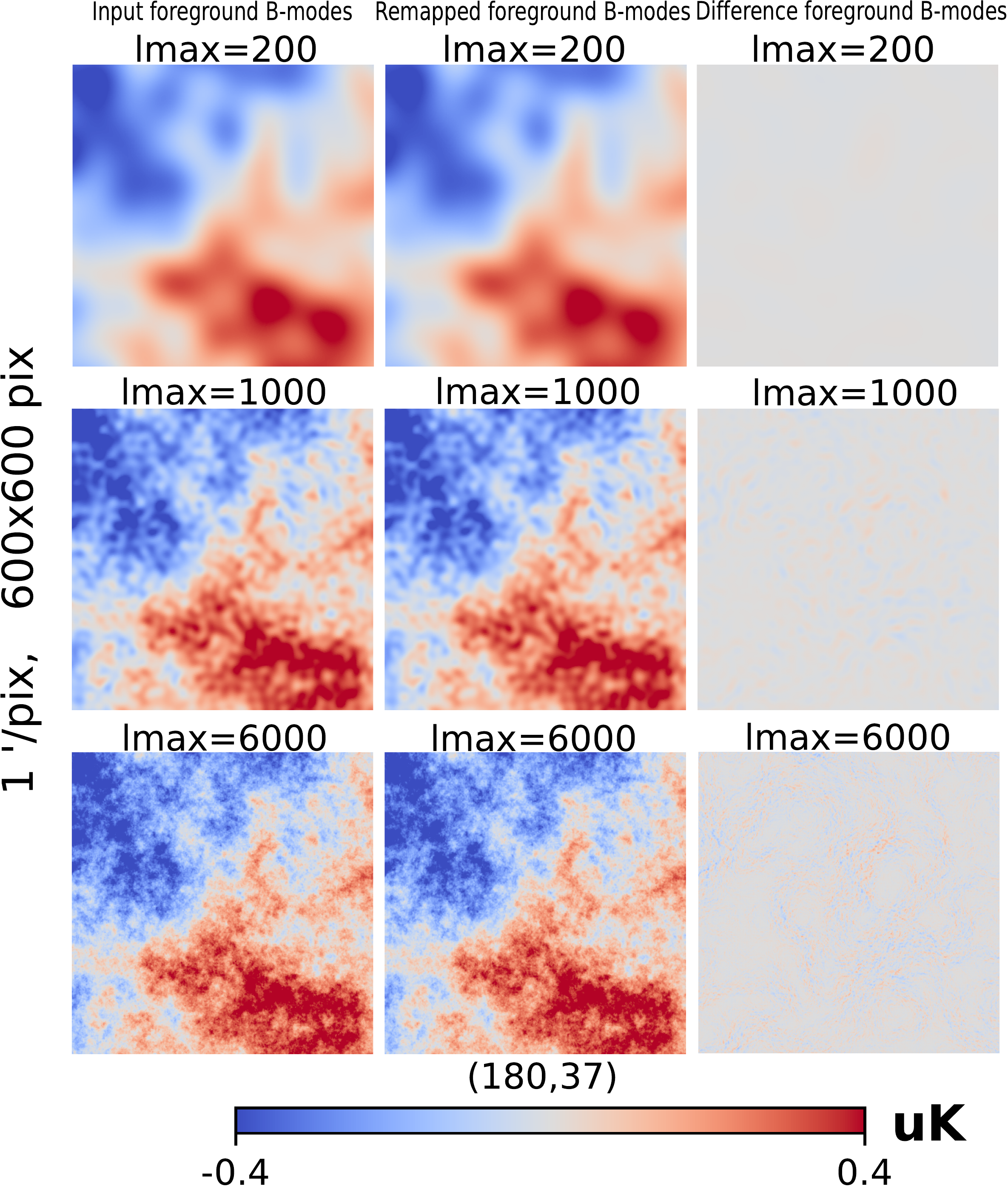}
	\caption{B-mode map comparisons of remapped foreground (middle panel) and input foreground at 93 GHz(left panel) at different multipole ranges. The first column shows the input foreground B-mode maps, the second column displays the remapped foreground B-mode maps, and the third column illustrates their difference. Results are presented for three multipole ranges: $\ell_{\text{max}} = 200$, $1000$, and $6000$. On large scales ($\ell_{\text{max}} = 200$), the differences are negligible, indicating accurate template construction. However, as the multipole range increases, the differences grow, reflecting more significant deviations on smaller angular scales due to lensing effect.}
	\label{fig:remap_fg}
\end{figure*}

The approximations in the second and third lines are detailed in the \ref{sec:algorithms}. Here, $\mathcal{B}[B \star \phi]$ represents the inverse remapping operation of lensed B mode with $\phi$ to obtain the delensed B-modes. 
In this context, $B^\text{del}_S$ represents the delensed B-mode signal (the first brace), which is the desired outcome and arises from the method itself, thus existing even in the absence of noise. On the other hand, $B^\text{del}_N$ captures the noise component arising from the intrinsic noise introduced by the delensing process (the third brace). This term broadly encompasses the noise B-mode, which is not explicitly written for simplicity. 
The term $B^\text{fg,re}$ represents the foreground that has undergone "delensing" via the inverse remapping operation. From Fig.~\ref{fig:remap_fg} it is evident that this "delensing" process only slightly modifies the input foreground on small scales and has negligible impact on large scales.

\begin{figure}
	\includegraphics[width=\columnwidth]{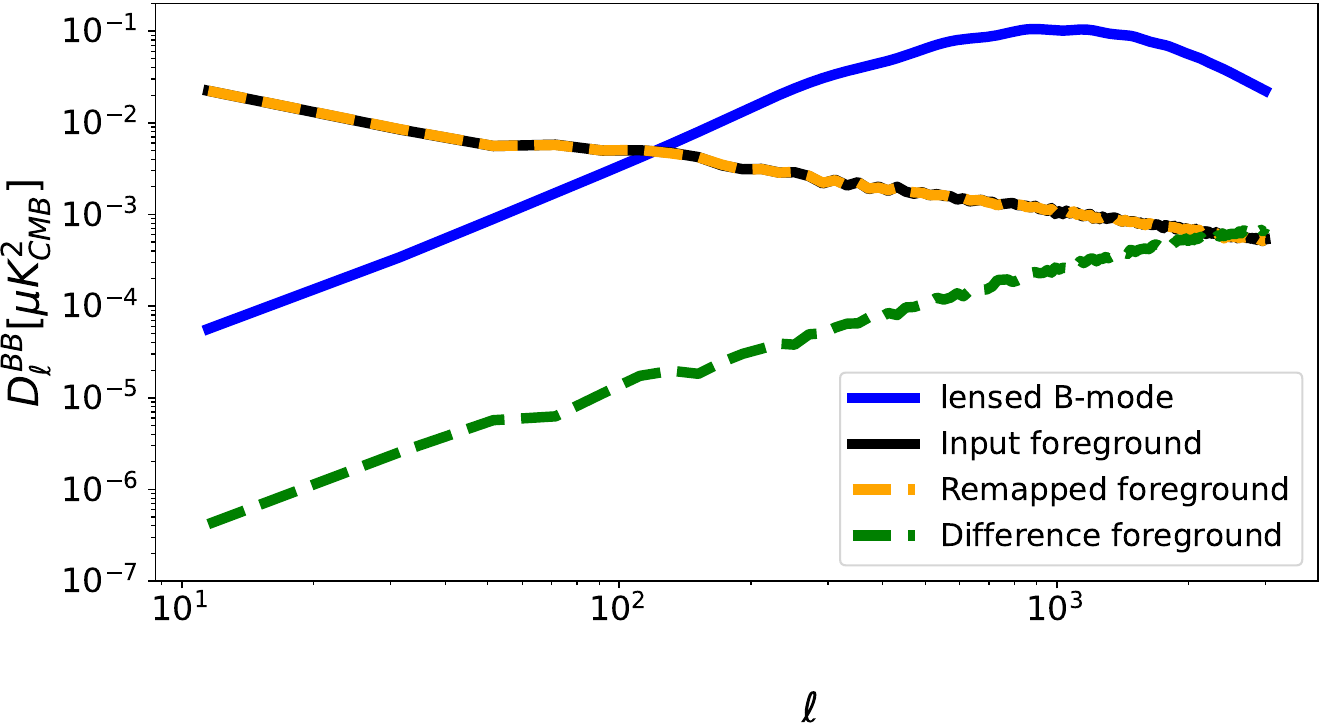}
	\caption{The plot compares the B-mode power spectra of the remapped foreground, the input foreground, and their difference. The difference foreground is derived by subtracting the remapped foreground from the input foreground at the map level. The results show negligible differences on large scales, while the differences become more pronounced as the multipoles increase. It is important to note that the power of the foreground difference correlates with the intensity of the input foreground.}
	\label{fig:remap_fg_cl}
\end{figure}

Using the delensed B-modes, we construct the lensing template by subtracting the delensed B-modes from the observed B-modes,
\begin{equation}\label{EQ:remap_template}
	\begin{aligned}
		B^\text{temp} &= \mathcal{W}^EB^\text{obs} - B^\text{del} \\
				&= \mathcal{W}^E[B^\text{lens} + B^\text{noise} + B^\text{fg}] - [B^\text{del}_S + B^\text{fg,re} + B^\text{del}_{N}] \\
				&\approx [\mathcal{W}^EB^\text{lens} - B^\text{del}_S] + [\mathcal{W}^EB^\text{noise} - B^\text{del}_N]\\
                &\quad + [\mathcal{W}^EB^\text{fg} - B^\text{fg,re}] \\
				&= B^\text{temp}_S + B^\text{df} + B^\text{df,fg},
	\end{aligned}
\end{equation}
Here, we utilize a filtered observed map, taking into account the similarity between the filtered input foreground and the remapped foreground. 
The power spectrum of the foreground difference, specifically the lensing B-mode template (\( B^\text{df,fg} \) in Eq. (\ref{EQ:remap_template})), derived from the pure foreground, is shown by the green dashed line in Fig.~\ref{fig:remap_fg_cl}.
The calculations indicate that the contribution from the foreground remains negligible on large scales, consistent with the assertion below Eq.~(\ref{EQ:template}) that foreground E-modes have minimal impact on lensing B-modes at gradient order on large scales. On small scales, however, the discrepancy between the two methods primarily arises from higher-order terms in the inverse-lensing approach, which become increasingly significant as the scale decreases.

Again, we examine these biases by calculating the auto- and cross-power spectra of the lensing template with the observations.
The auto-power spectrum of lensing template, $B^\text{temp} = \mathcal{T}^{-1} B^\text{lens} + B^\text{df} + B^\text{df,fg} $ is given by:  
\begin{equation}\label{EQ:inverse_auto}
	\begin{aligned}
        C^\text{temp} &= \mathcal{T}^{-2}C^\text{lens} + \langle B^\text{df}B^\text{df} \rangle + \langle B^\text{df,fg}B^\text{df,fg} \rangle \\
                &+ 2\mathcal{T}^{-1}\langle B^\text{lens}B^\text{df} \rangle + 2\mathcal{T}^{-1}\langle B^\text{lens}B^\text{df,fg} \rangle \\
                &+ 2\mathcal{T}^{-1}\langle B^\text{df}B^\text{df,fg} \rangle.
	\end{aligned}
\end{equation}
Here, $\mathcal{T}$ represents the transfer function used to compensate for the signal attenuation caused by the Wiener filter. 
The cross power spectrum between $B^\text{temp}$ and the observed B-modes $B^\text{obs} = B^\text{lens}+B^\text{noise}+B^\text{fg}$,  is
\begin{equation}\label{EQ:inverse_cross}
	\begin{aligned}
        C^\text{cross} &= \mathcal{T}^{-1}C^\text{lens} + \langle B^\text{lens}B^\text{df} \rangle + \langle B^\text{noise}B^\text{temp} \rangle \\
        &+ \langle B^\text{fg}B^\text{temp} \rangle + \langle B^\text{lens}B^\text{df,fg} \rangle.
	\end{aligned}
\end{equation}


For Eq.(\ref{EQ:inverse_auto}), the bias term $\langle B^\text{df} B^\text{df} \rangle$
can be subtracted using simulations, while the terms $\langle B^\text{df,fg} B^\text{df,fg} \rangle$
remain small on large scales but grow with increasing multipole (see Fig.~\ref{fig:remap_fg_cl}). This behavior is acceptable, as our analysis primarily focuses on large-scale modes. Importantly, the magnitude of this term depends on the intensity of the foreground.
In this analysis, we focus on the delensing of the observed maps at 93 GHz, where the foreground polarization is relatively weak. 
We emphasize that only observed maps with sufficiently low foreground contamination can be directly used to construct the lensing B-mode template; otherwise, contamination from polarized foregrounds will significantly degrade the delensing performance. In cases where foreground contamination is significant, a foreground cleaning procedure must be applied prior to template construction.
The first cross term, $2\mathcal{T}^{-1} \langle B^\text{lens} B^\text{df} \rangle$, cannot be neglected due to the correlation between $\phi$ reconstruction noise and the lensed B-mode. However, the last two cross terms can be neglected, as the foreground difference \( B^\text{df,fg} \) is minimal, and the CMB, noise, and foreground components arise from distinct physical processes.

For Eq.(\ref{EQ:inverse_cross}), we observe that neither the second nor the third term, as previously discussed, can be ignored. However, both terms can be effectively eliminated through simulation. The fourth cross term has only a minimal impact, while the final terms can be confidently disregarded on large scales.


It is important to note that Eq.(\ref{EQ:remap}) is accurate to a precision of $1\%-2\%$. Given that the remapping operation is non-linear, many high-order terms are omitted when expressing the result as a summation. In the debiasing process for delensing, we obtain the bias terms by subtracting the signal-only simulation from the noisy simulation. This approach effectively incorporates higher-order noise contributions. We have verified that neglecting these higher-order terms during the debiasing—specifically, by simulating each term individually as described in the equations—does not significantly affect our results, though it does lead to a slight increase in the bias for $r$.

\section{Data simulation}\label{sec:sims}
In this section, we present the simulation framework used in our analysis and demonstrate the feasibility of the delensing pipeline using simulated data. 
We generate 500 sets of mock maps, each consisting of lensed CMB maps, instrumental noise maps, a lensing potential map, foreground maps, and large-scale structure (LSS) tracer maps.
For each set, the lensed CMB maps are produced by applying the lensing potential to the unlensed CMB maps. The LSS tracer maps are simulated using the same lensing potential, ensuring a consistent correlation between the lensed CMB maps and the LSS tracers. In the following subsection, we provide a detailed description of the simulation process for each component.

\begin{figure}[htbp]
    \centering
    \subfigure[Mask for lensing reconstruction and delensing]{
        \includegraphics[width=0.4\linewidth]{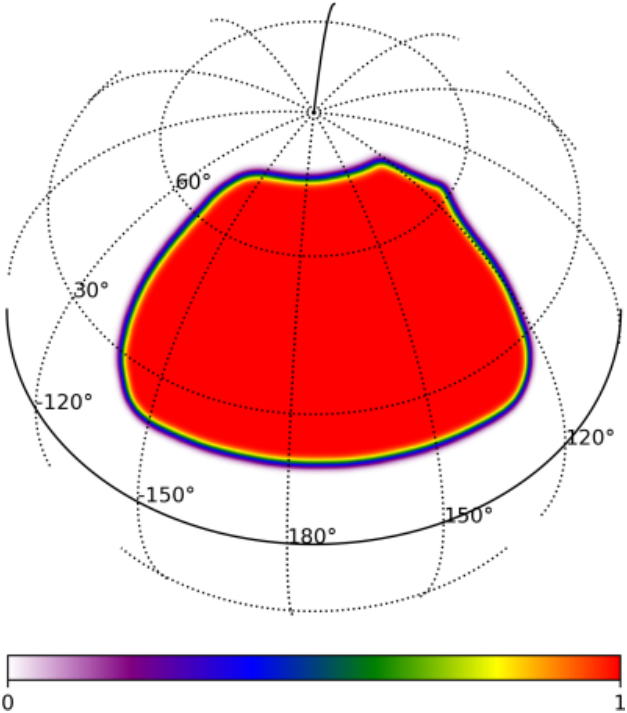}
    }
    \hspace{0.02\linewidth} 
    \subfigure[Mask for pseudo-Cl calculation]{
        \includegraphics[width=0.4\linewidth]{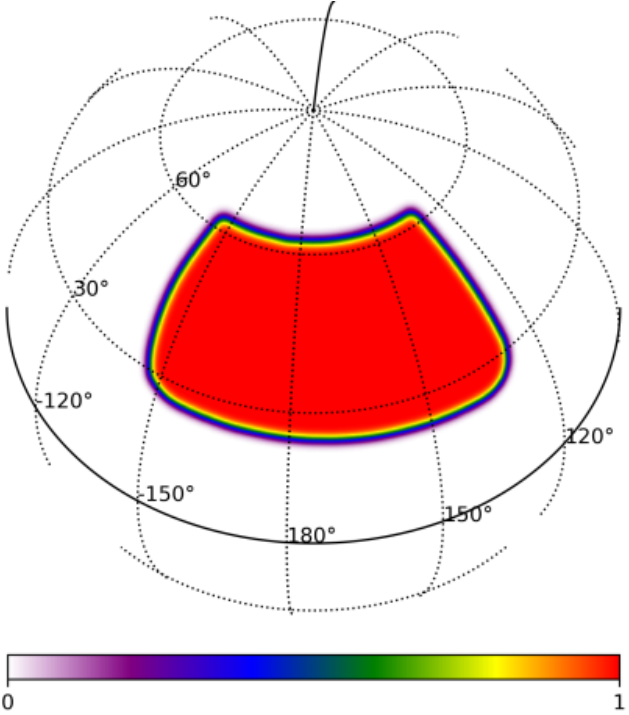}
    }
    \caption{The overlap masks of the LAT and SAT utilized in the simulation. The left panel is the apodized mask used for the delensing procedure, and the right one shows the apodized mask used for the calculation of pseudo-Cl with \texttt{NaMaster} \cite{alonso2023namaster}.}
    \label{fig:masks}
\end{figure}

\subsection{Simulation of lensed CMB maps}\label{sec: sim_cmb}
Our simulations concentrate on future CMB polarization observations aimed at high-precision measurements, incorporating both a small-aperture ground-based $80$\,cm telescope (SAT) and a large-aperture $6$\,m ground-based telescope (LAT). These simulations are designed to assess the effectiveness of our delensing pipelines in reducing the uncertainty of the tensor-to-scalar ratio $r$.   

We generate the unlensed CMB maps through Gaussian realizations derived from a power spectrum computed using the Boltzmann code \texttt{CAMB} \cite{lewis2011camb}. The power spectrum is based on the Planck 2018 best-fit cosmological parameters \cite{aghanim2020planck}, assuming a tensor-to-scalar ratio of $r = 0$.

We generate a lensed CMB map by creating a Gaussian realization from the fiducial lensing potential power spectrum. Using the \texttt{Lenspyx} \cite{carron2020lenspyx,reinecke2023improved} package, we then apply an algorithm to distort the primordial CMB signal based on the generated lensing potential map.

The lensed CMB maps are smoothed using the instrumental beam sizes provided in Table \ref{tab:exp_params}, and all maps are represented in the \texttt{HealPix}\footnote{\url{https://healpix.sourceforge.io/}} \cite{2005ApJ...622..759G,Zonca2019} pixelization scheme with $\texttt{NSIDE}=2048$.

\subsection{Simulation of instrumental noise}
The noise in the sky patch is homogeneous, generated through Gaussian random sampling on the map domain with noise levels specified in Table~\ref{tab:exp_params}.  The noise levels are chosen based on the "goal" configuration of \cite{ade2019simons}, rather than the more conservative "baseline" scenario. This choice is motivated by the fact that improvements from external LSS tracers in delensing are most significant for current experiments, where internal CMB lensing reconstruction remains limited by relatively high noise. For future experiments with ultra-low instrumental noise, such as those discussed in \cite{belkner2024cmb}, internal lensing reconstruction alone may approach the optimal performance, and the additional benefit from external tracers is expected to diminish accordingly. Since the improvement fraction from external data decreases as internal lensing reconstruction improves, our choice aims to provide a modest and realistic forecast of the potential gains.

\begin{table*} 
    \centering
    \caption{The parameters of the experiments. We do not provide the reference multipole range for some frequencies from LAT, as they are not used in this forecast.}
    \label{tab:exp_params}
    \begin{tabularx}{\textwidth}{l *{3}{>{\centering\arraybackslash}X} *{3}{>{\centering\arraybackslash}X}} 
        \hline
        \quad & \multicolumn{3}{c}{SAT} & \multicolumn{3}{c}{LAT} \\
        \cmidrule(lr){2-4}\cmidrule(lr){5-7}
        Freq.  & FWHM  & Noise Level  & $\ell$ Range & FWHM  & Noise Level  & $\ell$ Range \\
        \hline
        27 GHz & 91 arcmin & 25 $\mu$K-arcmin & (20,200) & 7.4 arcmin & 52 $\mu$K-arcmin & \textbackslash \\
        39 GHz & 63 arcmin & 17 $\mu$K-arcmin & (20,260) & 5.1 arcmin & 27 $\mu$K-arcmin & \textbackslash \\
        93 GHz & 30 arcmin & 1.9 $\mu$K-arcmin & (20,320) & 2.2 arcmin & 5.8 $\mu$K-arcmin & (200,4800) \\
        145 GHz & 17 arcmin & 2.1 $\mu$K-arcmin & (20,600) & 1.4 arcmin & 6.3 $\mu$K-arcmin & (200,6000) \\
        225 GHz & 11 arcmin & 4.2 $\mu$K-arcmin & (20,920) & 1.0 arcmin & 15 $\mu$K-arcmin & \textbackslash \\
        280 GHz & 9 arcmin & 10 $\mu$K-arcmin & (20,1100) & 0.9 arcmin & 37 $\mu$K-arcmin & \textbackslash \\
        \hline
    \end{tabularx}
\end{table*}

\subsection{Simulation of foreground emission}\label{sec:sim_fg}
In this paper, two types of foreground simulations are generated for different purposes. One consists of foreground templates from \texttt{PySM3} \cite{zonca2021python,thorne2017python}. \texttt{PySM3} produces full-sky simulations of foreground emissions in both intensity and polarization, based on realistic space observations (e.g., \textit{Planck}, \textit{WMAP}). We generate both Galactic and extragalactic foregrounds, denoted as \( F^{\text{Temp},TEB} \), which are used to quantify the impact of foregrounds on our pipeline using the NILC method (see below and Appendix~\ref{sec:nilc}), and to parameterize the polarized foreground power (see Eq.~(\ref{EQ:fg_model})). The other consists of Gaussian realizations based on Galactic foreground emission models, denoted as \( F^{\text{GS},TEB} \), and will be used for lensing reconstruction, delensing, and parameter fitting.

On the one hand, the simulation $F^{\text{Temp},TEB}$ used to quantify the effect of foregrounds with NILC includes a template from \texttt{PySM3} (with the model specified in brackets). The diffuse Galactic emissions include thermal dust (\texttt{d9} model), synchrotron (\texttt{s4} model), spinning dust (\texttt{a1} model), and free-free emission (\texttt{f1} model), while the Extragalactic emissions include the SZ effect (\texttt{ksz1} model, \texttt{tsz1} model) and the cosmic infrared background (\texttt{cib1} model).  
As discussed in Appendix \ref{sec:nilc}, the NILC method has a limited effect on point sources, necessitating additional processing at the map level for complete removal.
Residual point sources contribute significantly to the power spectrum on small scales. 

Therefore, the Extragalactic foreground components are excluded from the temperature maps used for lensing reconstruction. We do this by generating Gaussian realizations $F^{\text{GS},T}$ of the total power spectrum of the Galactic foreground temperature templates within the corresponding observed sky area. This approach mitigates the impact of point sources on lensing reconstruction and CMB B-mode map-level delensing. A more detailed analysis of residual point sources and their effects is deferred to future work.

On the other hand, we generate Gaussian realization $F^\text{GS}$ for the polarized foreground components based on the emission models.
Here, we primarily consider thermal dust emission and synchrotron emission. A power-law model is used to describe synchrotron emission, while a modified blackbody model is used for thermal dust emission. The total power spectrum in antenna temperature is given by \cite{abitbol2021simons},
\begin{equation}\label{EQ:fg_model}
	\begin{aligned}
		C_{\ell}^{\nu \nu', p p'} |_{FG} = C_{\ell}^{S \times S,\nu \nu', p p'} + C_{\ell}^{D \times D, \nu \nu', p p'} + C_{\ell}^{S \times D,\nu \nu', p p'}.
	\end{aligned}
\end{equation}
Here, $\nu$ and $\nu'$ represent the frequencies of the emission,  and $p \in \{E,B\}$ specifies the polarization channel.
The auto-correlation are modeled as 
\begin{equation}\label{EQ:fg_model1}
	\begin{aligned}
		C_{\ell}^{c \times c,\nu \nu', p p'} = f_c^{\nu}f_c^{\nu'}(\Delta_c)^{log^2(\nu/\nu')/log^2(\nu_{0,c})}C_{\ell}^{c, p p'},
	\end{aligned}
\end{equation}
where $c \in \{S,D\}$ denotes the component index (synchrotron or dust), $f_c^{\nu}$ is the spectra of component $c$ at frequency $\nu$, $\Delta_c$ is a decorrelation parameter that quantifies the decoherence of foregrounds as a function of frequency. Current constraints on $\Delta_D$ suggest it is compatible with zero \cite{sheehy2018no}, although there is some evidence for a non-zero $\Delta_S$ \cite{krachmalnicoff2018s}. However, in this work, we neglect their minor influence and fix both $\Delta_D$ and $\Delta_S$ to zero.

The synchrotron and dust spectra are,
\begin{equation}\label{EQ:fg_model2}
	\begin{aligned}
        f_S^{\nu} = (\nu/\nu_{0,S})^{\beta_s}, \quad f_D^{\nu} = (\nu/\nu_{0,D})^{\beta_d + 1} \frac{e^{h\nu_{0,D}/k_BT_d}-1}{e^{h\nu/k_BT_d }-1} .
	\end{aligned}
\end{equation}
In practice, we choose a pivot scale $\ell_0=80$, and pivot frequencies $\nu_{0,D}=353$\,GHz and $\nu_{0,S}=23$\,GHz.
The scale-dependent factor $C_{\ell}^{c, p p'}$ is parameterized with a power law:
\begin{equation}\label{EQ:fg_model3}
	\begin{aligned}
        \frac{\ell(\ell+1)}{2\pi}C_{\ell}^{c, p p'} = A_c^{pp'}(\frac{\ell}{\ell_0})^{\alpha_c^{pp'}}.
	\end{aligned}
\end{equation}
The cross-correlations between components are modeled as,
\begin{equation}\label{EQ:fg_model4}
	\begin{aligned}
        C_{\ell}^{S \times D, \nu \nu', pp'} = \epsilon_{DS}[f_D^{\nu}f_S^{\nu'} \sqrt{C_{\ell}^{D, p p}C_{\ell}^{S, p' p'}} \\
        + f_S^{\nu}f_D^{\nu'} \sqrt{C_{\ell}^{S, p p}C_{\ell}^{D, p' p'}}].\\
	\end{aligned}
\end{equation}

To match actual foreground observation, we determine the parameters of the foreground model for the observed sky area (see Fig.\ref{fig:masks}) using the foreground templates generated by \texttt{PySM3} \cite{zonca2021python,thorne2017python}. 
We first use the \texttt{d9s4} model \cite{krachmalnicoff2018s,akrami2020planck} in \texttt{PySM3} to generate six foreground templates \( F^{\text{Temp},EB} \) at 27 GHz, 39 GHz, 93 GHz, 145 GHz, 225 GHz, and 280 GHz, and calculate their power spectra $C_{\ell}^{\nu \nu'} |_{FG}$. We then investigate the seven-parameter space $\{A_d, \alpha_d, \beta_d, A_s, \alpha_s, \beta_s, \epsilon_{DS}\}$ using \texttt{Cobaya} \cite{Torrado_2021,torrado2020cobaya}. The peak values of the parameter posterior distribution are then considered a reasonable estimate of the true sky foreground. These values will be used in subsequent foreground simulations. 
Note that the spectral index $\beta_c$ should be $\beta_c(\hat n)$ for strictly speaking, in this work we neglect spatial variations of the foreground SED and treat them as Gaussian random fields.

On one hand, as indicated by \cite{thorne2017python}, the spectral index for dust emission varies on degree scales with a mean of $1.59$ and a standard deviation of $\sigma \sim \mathcal{O}(0.1)$, while \cite{kogut2012synchrotron} fits a synchrotron emission model to a small patch of sky using ten overlapping radio frequency sky surveys along with WMAP 23 GHz data, finding best-fit values of $\beta_s = -3.09 \pm 0.05$. Given that our survey covers a small patch of sky, assuming a constant spectral index model with parameters fitted in the corresponding sky is a reasonable approximation.
On the other hand, the pipeline we presented here can be easily extended to include spatially varying spectral indices. This can be accomplished by incorporating a correction term into the model to account for the spatially varying SED in the angular power spectrum via moment-expansion, as described by \cite{chluba2017rethinking}. This extension would introduce four additional parameters ($B_s$, $B_d$, $\gamma_s$, and $\gamma_d$) to describe the spatial dependence of the emission. 
More importantly, as discussed in \cite{ade2019simons}, the anisotropic and non-Gaussian nature of the foreground may not significantly impact the constraints on \( r \) for SO-like experiments, provided that proper processing is applied. Therefore, we neglect the spatial variation of the SED at this stage, but plan to incorporate it in our future work.

The input values for the B-mode simulation are listed in Table \ref{tab:prior}. The input foreground amplitudes, $A_d$ and $A_s$, for the E-mode simulation are slightly larger than those for the B-mode, and we assume that they share the same remaining parameters. For the dust-synchrotron correlation coefficient \( \epsilon_{DS} \), we obtain a very small fitted value in our case; therefore, we set it to zero for simplicity. 

For the foreground $F^{\text{GS},TEB}$, we simulate B-mode signals for six frequencies, which are used in lensing reconstruction, delensing, and parameter fitting. For the E-mode and temperature signals, only the 93 GHz and 145 GHz frequencies are simulated, as they are used exclusively for lensing reconstruction (and also for delensing in the case of E-modes at 93 GHz).

The simulation of foregrounds $F^{\text{GS},TEB}$ proceeds as follows: for the temperature $F^{\text{GS},T}$, as discussed above, we generate Gaussian realizations based on the total power spectrum of the Galactic foreground templates. For the polarization $F^{\text{GS},EB}$, we first generate Gaussian realizations based on the power spectra of each component, assigning different random seeds to each component to ensure that \(\epsilon_{DS}\) is artificially set to zero. For a single set of simulations, the same random seed is used for different frequency maps of the same component, ensuring that they differ only by a scale factor. The maps from the two components are subsequently combined to create the simulated foreground maps.
To maintain consistency with the CMB maps and instrumental noise maps, we convert the units of the maps from antenna temperature units to thermodynamic temperature units by dividing by a unit conversion factor: 
\[
f_C^{\nu} = e^x \left(\frac{x}{x-1}\right)^2,
\]
where \(x \equiv \frac{h\nu}{k_\text{B} T_\text{CMB}}\). 
The foreground maps are then smoothed using the instrumental beam sizes listed in Table \ref{tab:exp_params}.

Finally, we add the smoothed lensed CMB maps to the smoothed foreground maps and the noise maps, and then multiply them by the sky patch masks to produce the mock observed maps, with a sky coverage of about \(14\%\), as shown in the left panel of Fig.~\ref{fig:masks}.

\begin{table}
	\centering
	\caption{Input value used for B-mode simulation and prior imposed on each parameters for MCMC sampling. $\mathcal{U}(a,b)$ denotes uniform distribution between $[a,b]$.}
	\label{tab:prior}
	\begin{tabular}{lcccr} 
		\hline
		Parameter & Input value & Prior\\
		\hline
		$r$ & 0 &$\mathcal{U}(-0.2,0.2)$ \\
		$A_L$ & 1.000 &$\mathcal{U}(0,1.2)$ \\
		$A_d (\mu K^2)$ & 14.300 &$\mathcal{U}(0,20)$ \\
        $\alpha_d$ & -0.650 &$\mathcal{U}(-1,0)$ \\
        $\beta_d$ & 1.480 &$\mathcal{U}(1,1.8)$ \\
        $A_s (\mu K^2)$ & 2.400 &$\mathcal{U}(0,10)$ \\
        $\alpha_s$ & -0.800 &$\mathcal{U}(-1.4,0)$ \\
        $\beta_s$ & -3.100 &$\mathcal{U}(-3.3, -2.7)$ \\
        $\epsilon_{ds}$ & 0 & $\mathcal{U}(-0.3, 0.3)$ \\
		\hline
	\end{tabular}
\end{table}

\subsection{Simulation of LSS tracers}
We simulate the large-scale structure (LSS) tracer by starting with a 2D projection of the matter overdensity along the line of sight, derived from the 3D matter overdensity field\cite{manzotti2018future}:
\begin{equation}
    \delta^i(\hat{\mathbf{n}})=\int_0^{\infty} d z W^i(z) \delta(\chi(z,\hat{\mathbf{n}}), z),
\end{equation}
where $\delta(\chi(z) \hat{\mathbf{n}}, z)$ corresponds to the dark matter overdensity field at a comoving distance $\chi(z)$ and at a redshift $z$ in the angular direction $\hat{\mathbf{n}}$. 
The kernel $W^i(z)$ acts as the weight function which correlates the 3D matter overdensity field with the 2D LSS tracer fields along the line of sight. 
We can compute the power spectra of two large-scale structure tracer fields $i, j$ as:
\begin{equation}\label{eq:limber_inte}
C_{\ell}^{i j}=\int_0^{\infty} \frac{d z}{c} \frac{H(z)}{\chi(z)^2} W^i(z) W^j(z) P(k=\ell / \chi(z), z),
\end{equation}
where we have used the Limber approximation \cite{limber1953analysis} when calculating the integral. In this equation, $H(z)$ is the Hubble factor at redshift $z, c$ is the speed of light, and $P(k, z)$ is the matter power spectrum evaluated at redshift $z$.  

In this work, we combine the lensing potential reconstructed from the observed CMB with the Cosmic Infrared Background (CIB) and galaxy number density from \emph{Euclid} \cite{amendola2018cosmology}. The models of the large-scale structure (LSS) tracers (the CIB and the galaxy number density) are described in Section \ref{sec: LSS_theory}, while the pipeline for generating LSS tracer realizations is outlined in Section \ref{sec:LSS_realization}. The methodology for combining these tracers is detailed in Section \ref{sec: tracer_comb_exe}.

\subsubsection{LSS tracer models}\label{sec: LSS_theory}
In this section, we present the theoretical models for large-scale structure (LSS) tracers. The simulation of LSS tracers is based on their power spectra, which consist of two main components: (1) the underlying matter distribution, incorporating appropriate biases, and (2) a term accounting for experimental uncertainties, primarily due to shot noise. The procedure for generating realizations from these power spectra will be detailed in Section~\ref{sec:LSS_realization}. We consider three kind of tracers: CMB lensing convergence, Cosmic Infrared Background, as well as galaxy number density. 

\begin{figure}
	\includegraphics[width=\columnwidth]{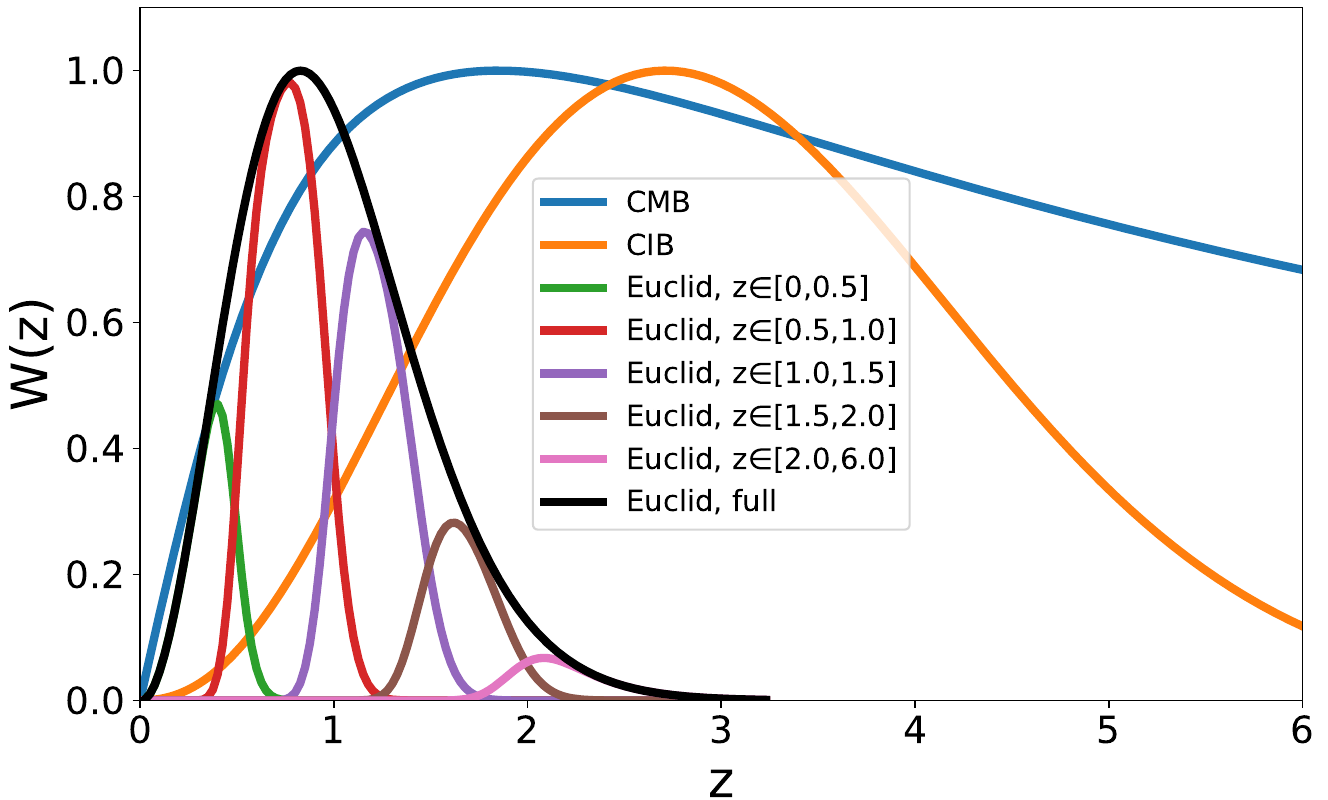}
	\caption{Kernel Comparison: Comparison of the different kernels as a function of redshift for some of the tracers used in this analysis. The function is normalized so that the peak is unity. }
	\label{fig:weight_function}
\end{figure}

\paragraph{\textbf{CMB lensing convergence}}
To begin with, the weight function for the CMB lensing convergence is:
\begin{equation}
W^\kappa(z)=\frac{3 \Omega_{\mathrm{m}}}{2 c} \frac{H_0^2}{H(z)}(1+z) \chi(z) \frac{\chi_*-\chi(z)}{\chi_*},
\end{equation}
where $\chi_*$ is the comoving distance to the last-scattering surface at $z_* \simeq 1090, \Omega_{\mathrm{m}}$ and $H_0$ are the present day values of the Hubble and matter density parameters, respectively. 
We find that although the lensing kernel \( W^\kappa(z) \), as shown in Fig. \ref{fig:weight_function}, extends to very high redshifts, the contribution to the lensing power decreases significantly with increasing redshift due to the limitations of the matter power spectrum. After careful consideration, we have determined that a rather conservative maximum redshift of \( z_{\text{max}} \sim 30 \) is appropriate when calculating the integral in Eq.(\ref{eq:limber_inte}). Our tests indicate that this covers at least $99\%$ of the lensing potential power spectrum.

\
\paragraph{\textbf{Cosmic Infrared Background}}

The observed Cosmic Infrared Background (CIB) is composed of diffuse Extragalactic radiation, primarily generated by unresolved emission from star-forming galaxies, along with shot noise contributions from radio galaxies\cite{toffolatti2014planck}:
\begin{equation}
C_{\text {measured }}^{\nu \times \nu^{\prime}}(\ell)=C_{\mathrm{d}, \text { clust }}^{\nu \times \nu^{\prime}}(\ell)+C_{\mathrm{d}, \text { shot }}^{\nu \times \nu^{\prime}}+C_{\mathrm{r}, \text { shot }}^{\nu \times \nu^{\prime}} .
\end{equation}
the first term $C_{\mathrm{d}, \text { clust }}^{\nu \times \nu^{\prime}}(\ell)$ is the contribution from the clustering of dust star-forming galaxies (DSFGs),  $C_{\mathrm{d}, \text { shot }}^{\nu \times \nu^{\prime}}$ and $C_{\mathrm{r}, \text { shot }}^{\nu \times \nu^{\prime}}$ are the shot noise from DSFGs and radio galaxies respectively, and the clustering of radio galaxies is negligible in the CIB measurement.

The shot noise arises from the finite sampling of a background composed of discrete sources. It is described in \cite{toffolatti2014planck} using two models: the \cite{bethermin2012unified} model for $C_{\mathrm{d}, \text{shot}}^{\nu \times \nu^{\prime}}$ and the \cite{tucci2011high} model for $C_{\mathrm{r}, \text{shot}}^{\nu \times \nu^{\prime}}$. The corresponding flat power spectra are provided in Table 6 and Table 7 of \cite{toffolatti2014planck}.

As for the clustering contribution $C_{\mathrm{d}, \text { clust }}^{\nu \times \nu^{\prime}}$, we use the single-SED model of \cite{hall2010angular}, and its kernel is:
\begin{equation}
W^{\mathrm{CIB}}(z)=b_c \frac{\chi^2(z)}{H(z)(1+z)^2} e^{-\frac{\left(z-z_c\right)^2}{2 \sigma_z^2}} f_{\nu(1+z)},
\end{equation}
for
\begin{equation}
f_\nu= \begin{cases}\left(e^{\frac{h \nu}{k T}}-1\right)^{-1} \nu^{\beta+3} & \left(\nu \leq \nu^{\prime}\right) \\ \left(e^{\frac{h \nu^{\prime}}{k T}}-1\right)^{-1} \nu^{\prime \beta+3}\left(\frac{\nu}{\nu^{\prime}}\right)^{-\alpha} & \left(\nu > \nu^{\prime}\right)\end{cases},
\end{equation}
This corresponds to a gray-body spectrum with $T=34 \mathrm{~K}$. The fiducial model parameters are $z_c=\sigma_z=2$, $\alpha=\beta=2$ and $b_c$ serves as the normalization factor. The power-law transition occurs at $\nu^{\prime} \approx 4955 \mathrm{GHz}$. Considering realistic observations, we set the upper limit of the integration to \( z_{\text{max}} \sim 4.5 \). From Fig. \ref{fig:weight_function}, we observe a significant overlap between the CMB kernel and the CIB kernel. This intuitively suggests a strong correlation between the CIB as a large-scale structure (LSS) tracer and the CMB lensing potential. As noted in \cite{ade2014planck_infrared}, the correlation coefficient can reach up to 80\%.

In this work, we use the CIB map at 353 GHz, i.e., $\nu = 353\,\mathrm{GHz}$. As noted by \cite{yu2017multitracer}, the GNILC CIB map at 353 GHz is found to be most strongly correlated with CMB lensing. Combining it with higher frequencies, such as 545 GHz or 857 GHz, does not significantly enhance the correlation coefficient with CMB lensing. Furthermore, they suggest that, due to the very high signal-to-noise ratio (S/N) of the CIB auto-power spectrum at the measured angular scales, the contribution of instrumental noise to the CIB auto-spectrum is negligible. As a result, we also neglect instrumental noise in our theoretical model.

\

\paragraph{\textbf{Galaxy number density}}
We follow the theoretical modeling of galaxy number density fluctuations as described in \cite{namikawa2024litebird,namikawa2022simons}. The weight function of the galaxy number density fluctuations in the $i$-th redshift bin is given by
\begin{equation}
W^{\mathrm{g}, i}(\chi)=\frac{\mathrm{d} n_{\text {gal }}^i}{\mathrm{~d} \chi}(z(\chi)) b_{\text {gal }}(z(\chi)),
\end{equation}
where $\mathrm{d} n_{\mathrm{gal}}^i / \mathrm{d} \chi$ and $b_{\text {gal }}(z)$ are the normalized redshift distribution of galaxies in the $i$-th bin and the linear galaxy bias, respectively. 
Since the galaxy number density fluctuation is not an unbiased tracer of the matter density fluctuation, it is naive to relate them with a linear bias:
\begin{equation}
    b_\text{gal}(z) = \delta_g(\hat{\mathbf{n}},z) / \delta_m(\hat{\mathbf{n}},z),
\end{equation}
which is just a constant at a fixed redshift. This is only a crude approximation and it is not self-consistent(e.g. $\delta_g(\hat{\mathbf{n}},z)$ will be smaller than $-1$ for $b > 1$) and it is not preserved in time \cite{sigad2000measuring}.
It is reasonable to expect that, at any given time, scale, and for any galaxy type, galaxy biasing is generally nonlinear. In other words, the bias parameter \(b_{\text{gal}}\) should vary as a function of \(\delta_m\) and is typically stochastic, implying that a range of \(\delta_g\) values may correspond to a given \(\delta_m\). \cite{dekel1999stochastic} proposed a comprehensive formalism for galaxy biasing that naturally distinguishes between nonlinearity and stochasticity. For more details, please refer to their work, as a detailed discussion is beyond the scope of this paper.

To fully utilize the redshift information from the galaxy survey, we adopt the following form for the redshift distribution function:
\cite{amendola2018cosmology,abell2009lsst} :
\begin{equation}
\frac{\mathrm{d} n_{\text {gal }}^i}{\mathrm{~d} \chi}(z)=\frac{\beta H(z)}{\Gamma[(\alpha+1) / \beta]} \frac{z^\alpha}{z_0^{\alpha+1}} \exp \left[-\left(\frac{z}{z_0}\right)^\beta\right] p_{\text{gal}}^i(z,\sigma_z),
\end{equation}
where $\alpha, \beta$, and $z_0$ are parameters determining the distribution and $\Gamma[(\alpha+1) / \beta]$ is the Gamma function. The parameter $z_0$ is related to the mean redshift $z_{\mathrm{m}}$ as
\begin{equation}
z_{\mathrm{m}}=\frac{\Gamma[(\alpha+2) / \beta]}{\Gamma[(\alpha+1) / \beta]} z_0,
\end{equation}
$H(z)$ is the Hubble parameter, simply from the conversion between $z$ and $\chi$. The function $p_{\text{gal}}^i\left(z, \sigma_z\right)$ specifies the redshift distribution of galaxies in the $i$-th redshift bin, accounting for the modification of the redshift distribution due to photometric redshift errors. We assume that the photometric redshift estimates follow a Gaussian distribution with an rms fluctuation of $\sigma(z)$. As a result, the top-hat cut in photometric redshift becomes a smoothly overlapping distribution in actual redshift \cite{hu2004measuring}. Consequently, $p_{\text{gal}}^i\left(z, \sigma_z\right)$ takes the following form:
\begin{equation}
p_{\text {gal }}^i\left(z, \sigma_z\right)=\frac{1}{2}\left[\operatorname{erfc}\left(\frac{z_{i-1}-z}{\sqrt{2} \sigma(z)}\right)-\operatorname{erfc}\left(\frac{z_i-z}{\sqrt{2} \sigma(z)}\right)\right],
\end{equation}
where $\sigma(z)=\sigma_z(1+z), z_{i-1}$ and $z_i$ are the minimum and maximum photometric redshift of the $i$ th bin, and the function $\operatorname{erfc}(x)$ is the complementary error function defined by
\begin{equation}
\operatorname{erfc}(x) \equiv \frac{2}{\sqrt{\pi}} \int_x^{\infty} \mathrm{d} z \mathrm{e}^{-z^2},
\end{equation}

Additionally, we need to account for the power of shot noise. When observing the distribution of galaxies, we sample a continuous underlying matter field using a finite number of discrete points (the galaxies). This discreteness introduces uncertainty, as we are not observing all possible points in space but only the locations of the galaxies. Since the galaxy number density follows a Poisson distribution with a mean of $n_g^i$ in the $i$-th redshift bin, the corresponding variance (shot noise) is given by: 
\begin{equation}
P_{SN}^i = \frac{1}{n_g^i \left[sr^{-1}\right]},
\end{equation}
where $n_g^i$ is the number of galaxies per steradian in the $i$-th redshift bin. This value is determined by the total number density $n_g^{\text{tot}}$ and its distribution $p_{\text{gal}}^i\left(z, \sigma_z\right)$ within each redshift bin.

In Fig.~\ref{fig:weight_function}, we observe that the overlap between the CMB kernel and the galaxy kernel occurs primarily at low redshifts and spans a relatively small region. However, given the high number density of \emph{Euclid}, which reduces shot noise, and the fact that structures at different redshifts contribute differently to the generation of large-scale B-modes due to the geometric properties of the lensing kernel, we still anticipate that using galaxies as LSS tracers can significantly enhance the effectiveness of delensing.

We observe that the kernel overlap of a tracer with the CMB lensing potential varies as a function of redshift. Therefore, the optimal strategy is to assign different weights to galaxies at different redshifts based on both their cross-correlation with \(\kappa\) and their auto-spectrum. This approach is well-illustrated by a simple example provided in \cite{manzotti2018future}; see also \ref{app:z_binning} for a brief illustration.

In this paper, we simulate the \emph{Euclid} survey by following the approaches outlined in \cite{yu2017multitracer} and the weak-lensing survey model described in \cite{amendola2018cosmology}. We adopt the parameterization \( \alpha = 2 \), \( \beta = 1.5 \), \( z_0 = 0.9/\sqrt{2} \), \( \sigma_z = 0.05 \), and a redshift-dependent galaxy bias defined as \( b_{\text{gal}}(z) = \sqrt{1+z} \).
 We also assuming an effective surface galaxy density of approximately 30 arcmin$^{-2}$ \cite{laureijs2011euclid}. For our analysis, we artificially divide the redshift range into five tomographic bins: $[0, 0.5, 1.0, 1.5, 2.0, 6.0]$. 

\subsubsection{Generate LSS tracer realization from theoretical models}\label{sec:LSS_realization}
With the explicit expression of the weight function, we calculate all the auto- and cross-spectra of the LSS tracers. Based on these spectra, we generate map realizations for each tracer, following the method outlined in \cite{baleato2022delensing}.

In this work, we focus on three types of tracers: the CIB, \emph{Euclid} galaxy number density, and an internal reconstruction. Their spherical harmonic coefficients are denoted as $I_{\ell m}$, $g_{\ell m}$, and $\kappa_{\ell m}^{\text{rec}}$, respectively. Given a map of the true convergence with spherical harmonic coefficients $\kappa_{\ell m}$, the correlations between the true convergence and the different tracers are described as follows:
\begin{equation}\label{eq:LSS_def}
    \begin{aligned}
        & \kappa_{\ell m}^\text{rec}=\kappa_{\ell m}+n_{\ell m} ,\\
        & g_{\ell m}=A_{\ell}^{g \kappa} \kappa_{\ell m}+u_{\ell m} ,\\
        & I_{\ell m}=A_{\ell}^{I \kappa} \kappa_{\ell m}+A_{\ell}^{g I} u_{\ell m}+e_{\ell m},
    \end{aligned}
\end{equation}
where $n_{\ell m}$, $u_{\ell m}$, and $e_{\ell m}$ are coefficients of the noise, and as such are presumed to be uncorrelated with each other. The coefficients $\kappa_{\ell m}$ represent the true convergence for the particular realization for which we wish to generate correlated tracers. The factors \( A_l^{g \kappa} \) and \( A_l^{I \kappa} \) serve as the "bias" parameters that correlate the CIB and galaxy fields with the true convergence.
 These can be solved as:
\begin{equation}\label{eq:factor_A}
    \begin{aligned}
    A_{\ell}^{g \kappa} & =\frac{C_{\ell}^{g \kappa}}{C_{\ell}^{\kappa \kappa}} ,\\
    A_{\ell}^{I \kappa} & =\frac{C_{\ell}^{I \kappa}}{C_{\ell}^{\kappa \kappa}} ,\\
    A_{\ell}^{g I} & =\frac{C_{\ell}^{g I}-A_{\ell}^{g \kappa} A_{\ell}^{I \kappa} C_{\ell}^{\kappa \kappa}}{C_{\ell}^{u u}},
    \end{aligned}
\end{equation}
and the noise spectra
\begin{equation}\label{eq:LSS_noise}
    \begin{aligned}
    & C_{\ell}^{n n}=N_{\ell}^{\kappa \kappa}, \\
    & C_{\ell}^{u u}=C_{\ell}^{g g}-\left(A_{\ell}^{g \kappa}\right)^2 C_{\ell}^{\kappa \kappa} ,\\
    & C_{\ell}^{e e}=C_{\ell}^{I I}-\left(A_{\ell}^{I \kappa}\right)^2 C_{\ell}^{\kappa \kappa}-\left(A_{\ell}^{g I}\right)^2 C_{\ell}^{u u},
    \end{aligned}
\end{equation}
where $N_{\ell}^{\kappa \kappa}$ is the internal reconstruction noise.
We can generate Gaussian realization from these noise power spectra.
The solution above can be generalized to any number of tracers as described in \cite{baleato2022delensing}.

\subsubsection{Pipeline of LSS tracer simulation}
We summarize our pipeline for the LSS tracer simulation as follows:

\begin{itemize}
    \item Calculate the auto- and cross-spectra of all LSS tracers based on the theoretical models described in Section \ref{sec: LSS_theory}, using the matter power spectrum and tracer model parameters as inputs. The matter power spectrum is computed using \texttt{pyccl} \cite{chisari2019core}, incorporating nonlinear evolution, with the cosmology defined by the Planck 2018 best-fit parameters. In this work, we consider a total of seven tracers: the internally reconstructed convergence \(\kappa^{\text{rec}}\), the CIB intensity \(I\), and the \emph{Euclid} survey divided into five redshift bins \(g^i\) (\(i = 1, \ldots, 5\)).
    \item Compute the scaling factors \(A_\ell^{g \kappa}\) and \(A_\ell^{I \kappa}\) using Eq.\,(\ref{eq:factor_A}), as well as the noise power spectra using Eq.\,(\ref{eq:LSS_noise}).
    \item Generate Gaussian realizations based on the noise power spectra. Scale the input true convergence \(\kappa\) by the factor \(A_\ell\), and add the noise realizations to it following Eq.\,(\ref{eq:LSS_def}).
\end{itemize}

\begin{figure*}
	\includegraphics[width=\textwidth]{./fig_revise/pipeline2_revise.pdf}
	\caption{The flowchart illustrates the delensing pipeline, which is divided into five major components: map simulation (top), lensed CMB map processing (middle left), lensing reconstruction (middle right), implementation of two delensing methods (lower section), and parameter constraint (bottom). Rectangular boxes represent data products—cyan for temperature and polarization fields of the microwave sky and red for fields correlated with large-scale structure (mass tracers). Processing steps and methods are indicated by boxes with rounded corners. We omit the spherical harmonic transformation in the plot for clarity. Note: The impact of foregrounds on the pipeline, assessed using the NILC method, is not shown in this diagram.}
	\label{fig:pipeline}
\end{figure*}

\section{Delensing implementation and Results}\label{sec:result}

\subsection{Motivation of the pipeline}\label{sec:motivation}
We perform NILC on the observed $\Theta$, $E$, and $B$ maps, as described in Appendix \ref{sec:nilc}, to quantify the effect introduced by the foregrounds. The harmonic transformations mentioned therein are executed using \texttt{healpy} / \texttt{HEALPix} package. A well-tested accompanying Python package to implement the NILC pipeline, is available at \texttt{openilc}\footnote{\url{https://github.com/dreamthreebs/openilc}}.
The results of NILC and detailed discussion are also listed there.

In this work, we just exclude the contribution of Extragalactic foregrounds in the temperature maps used solely for lensing reconstruction in the following sections. As discussed in Appendix~\ref{sec:nilc}, the NILC method leaves some non-Gaussian extragalactic foreground residuals, which may lead to a small temperature reconstruction bias. This bias can be specifically removed through various mitigation methods. We expect this exclusion to have a limited impact on the effectiveness of CMB B-mode delensing, though we acknowledge that it may slightly overestimate the results. A detailed investigation of the influence of Extragalactic point sources will be addressed in future studies.

For the purposes of this paper, we do not include any map-level foreground cleaning procedure in our pipeline. Before proceeding to describe the implementation and results, we will first explain the rationale behind this decision and its feasibility in detail by addressing the three main components of the pipeline:
\begin{itemize}
\item The lensing potential reconstruction, can be performed effectively without the need for NILC, as long as the large-scale modes are deprojected (in this paper $\ell < 200$) and the foreground power is properly accounted for in the filtering process, with the price of a little larger reconstruction noise.
\item For constructing the lensing B-mode template (LT), the presence of foregrounds at 
$\ell<200$ will inevitably degrade the construction performance. However, as discussed in \cite{manzotti2017cmb}, the key factor determining delensing efficiency is the accuracy of the lensing estimate, not the contamination in the CMB maps that are to be delensed. More importantly, as discussed in Section \ref{sec:bias}, the contribution of foregrounds to LT construction is negligible on large scales.
\item For the parameter $r$ constraint, we can simultaneously parameterize the foreground components in the model to avoid significant biases from the foreground. Even if the spatially varying SED of the foreground is considered, we can introduce four additional parameters in the model to properly account for it, as described by \cite{chluba2017rethinking}.
\end{itemize}
Therefore, in our baseline approach, we omit the NILC step and instead apply the aforementioned methods to mitigate the foreground impact. The advantage of this approach is that it avoids the computational cost required by NILC, with minimal degradation in performance.

\subsection{Lensing potential reconstruction}\label{phirec}
\subsubsection{Internal lensing reconstruction}

We reconstruct the lensing potential using observed maps at 93 GHz and 145 GHz from the LAT. The signal-to-noise ratio (S/N) of the reconstructed potential depends on both the multipole range and the noise level: a higher S/N is achieved with a larger maximum multipole and lower noise. Considering this, we select these two frequency bands due to their relatively lower foreground contamination and noise.
To reduce contamination from diffuse foreground residuals and noise residuals while maintaining the S/N, we exclude modes with \(\ell < 200\), as most lensing information comes from smaller angular scales.
Additionally, as highlighted by \cite{lizancos2021impact,teng2011cosmic}, lensing biases caused by the overlap between the B-modes used for reconstructing \(\phi\) and those intended for delensing (i.e., the observed B-modes) can be significantly reduced by overlapping B-mode deprojection. This aligns with the analysis presented in Section~ \ref{sec:bias}.

The reconstruction is performed strictly following the pipeline described in \ref{sec:phirec}. Note that for the polarization lensing reconstruction, we use the noise level accounted for the instrumental noise only in filtering, which introduces negligible sub-optimality since we have excluded the multipoles on large scales. We separately reconstruct the minimum-variance (MV) estimators \(\hat{\phi}\) from the LAT observations at 93 GHz and 145 GHz for \(200 < \ell < 6000\), with the maximum multipole for reconstruction set to \(L_\text{max} = 3071\). These are then combined using the inverse-variance weights:

\begin{equation}\label{EQ:phi_com}
	\begin{aligned}
        \hat{\phi}^\text{MV}_{LM} &= \sum_{\nu_i} \omega_{\nu_i, L} \hat{\phi}^{\text{MV}, \nu_i}_{LM}, \\
        \omega_{\nu_i, L} &= \frac{\left[N^{(0),\nu_i}_L\right]^{-1}}{\sum_{\nu_i} \left[N^{(0),\nu_i}_L\right]^{-1}}.
	\end{aligned}
\end{equation}
Here, \(\hat{\phi}^{\text{MV}, \nu_i}_{LM}\) represents the MV estimator derived from the observed maps at frequency \(\nu_i\) (where $\nu_i$ $\in$ $\{93 \text{GHz}$, $145 \text{GHz}\}$), and \(N^{(0),\nu_i}_L\) denotes the corresponding reconstruction noise power spectrum.

\begin{figure}
	\includegraphics[width=\columnwidth]{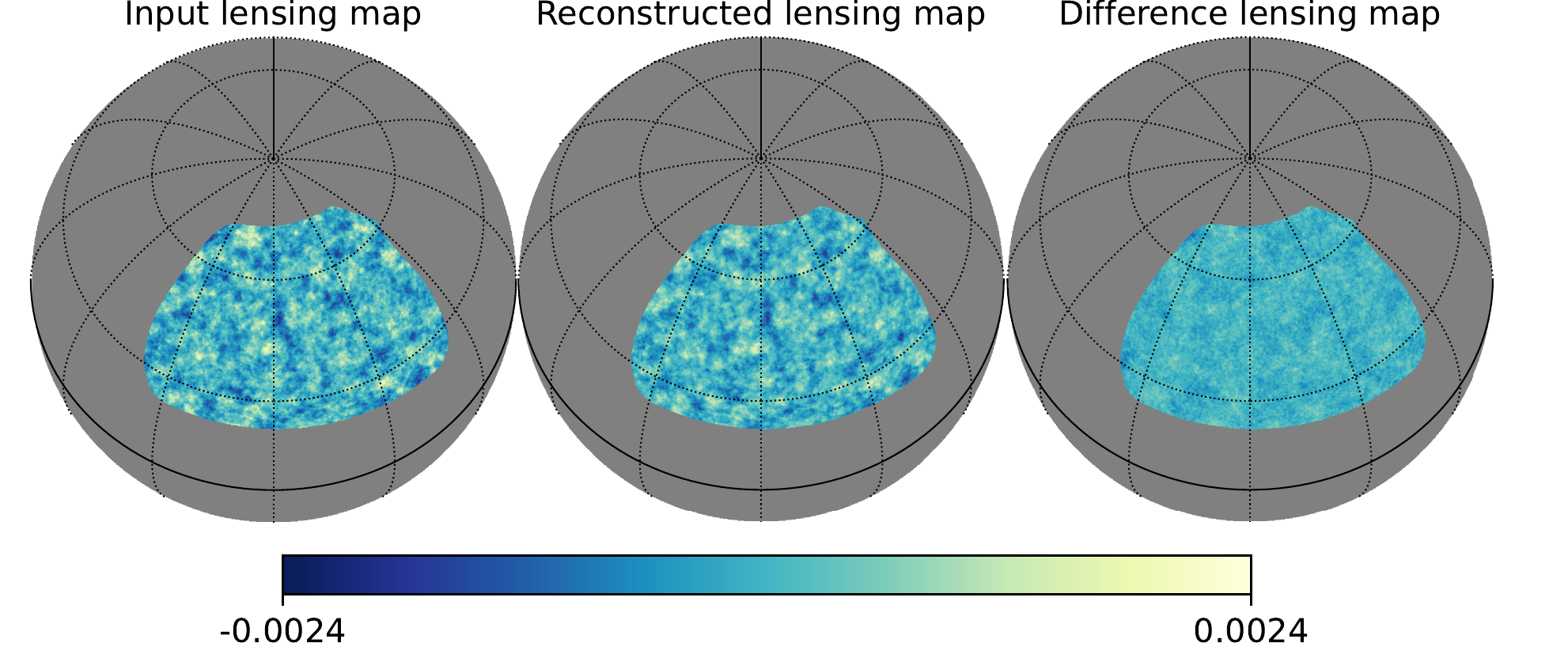}
	\caption{Reconstructed lensing map. The left panel is the input lensing map; the middle panel is the Wiener-filtered reconstructed map from the MV quadratic estimator, the right panel is the difference between them. In order to highlight the lens structure, we plot the Wiener -filtered lensing deflection angle amplitude $\hat \alpha^\text{WF} = \sqrt{L(L+1)} \hat \phi_{LM}^\text{MV} C^{\phi\phi,\text{fid}}_L/[C^{\phi\phi,\text{fid}}_L + N^{(0),\text{ana}}_L]$, with multipole range $20<L<2000$. Note that the edges of the map, which are not used in subsequent analysis, have been trimmed for clarity.}
	\label{fig:phi_rec}
\end{figure}

The final reconstructed lensing potential, \(\hat{\phi}^\text{MV}_{LM}\), is shown in Figure \ref{fig:phi_rec}. To highlight the lensing structures, we plot the Wiener-filtered deflection angle amplitude, defined as 
\[
\hat{\alpha}^\text{WF} = \sqrt{L(L+1)} \frac{ C^{\phi\phi,\text{fid}}_L}{C^{\phi\phi,\text{fid}}_L + N^{(0),\text{ana}}_L}\hat{\phi}^\text{MV}_{LM}.
\]
The left panel presents the input data, while the middle panel shows the reconstructed deflection angle from the MV estimator combining 93 GHz and 145 GHz. The right panel displays their difference. It is evident that the reconstructed deflection successfully captures most of the features of the input data, and the homogeneous distribution of brightness and darkness in the difference map confirms the effectiveness of the reconstruction.

\subsubsection{Combination with external tracers }\label{sec: tracer_comb_exe}
Using the simulated CIB maps and \emph{Euclid} galaxy number density maps, we obtained the combined tracer through a linear combination of \(I^i\):
\begin{equation}
    I = \sum_i c^i I_i,
\end{equation}
where $I_i$ in this context is the MV quadratic estimator (Eq. (\ref{EQ:phi_com})), the CIB and the galaxy number density from \emph{Euclid}, and the coefficients \(c^i\) are determined by maximizing the correlation coefficient between the combined tracer and the true convergence \(\kappa\). 
The derivation is detailed in \cite{sherwin2015delensing}, and a brief summary is provided in \ref{sec:external_rec}.
The inclusion of LSS tracers enhances the signal-to-noise ratio (S/N) of the reconstruction, particularly on small angular scales. This improvement significantly increases the effectiveness of delensing, thereby tightening the constraints on \(r\).

\begin{figure}
	\includegraphics[width=\columnwidth]{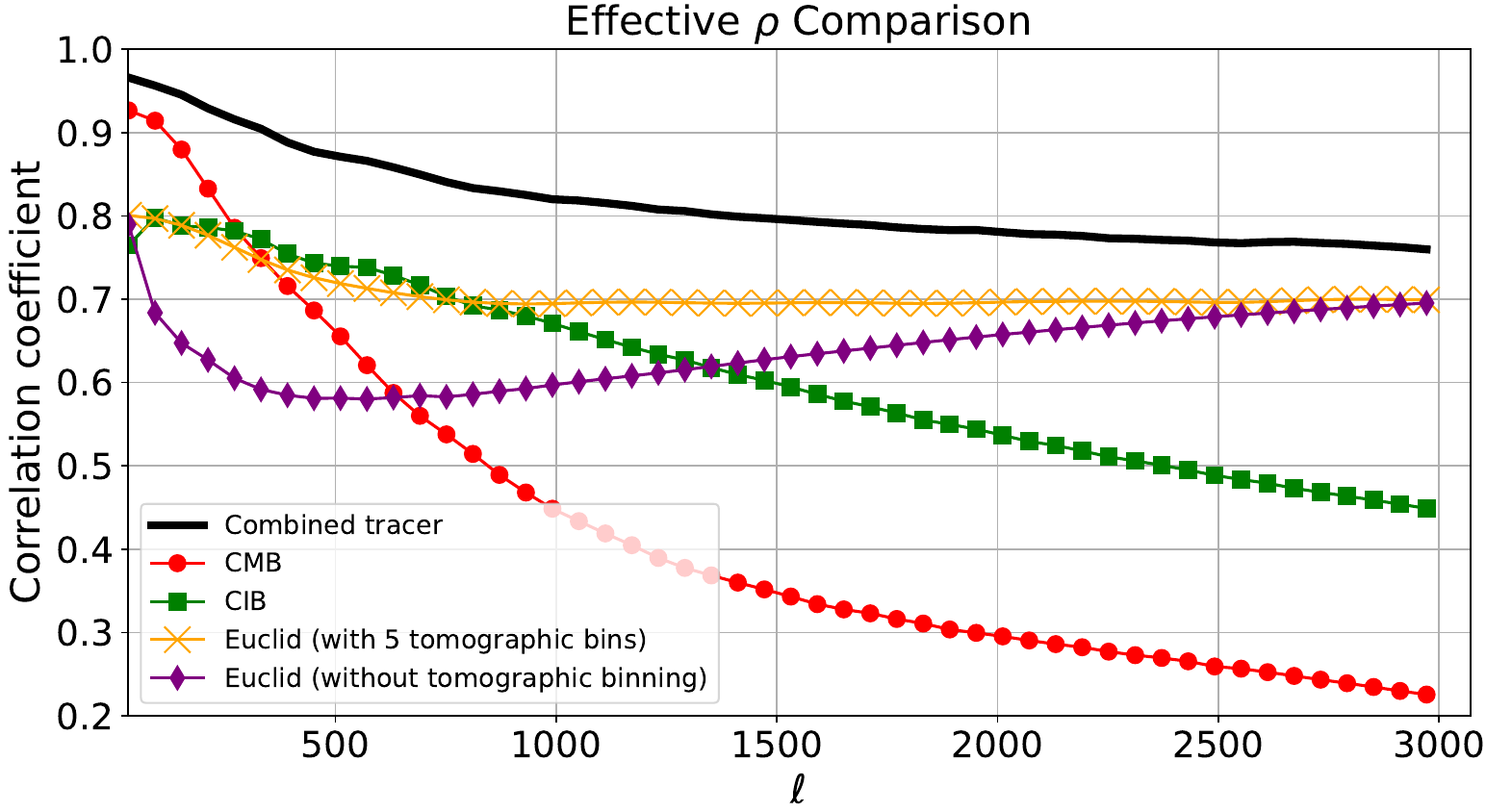}
	\caption{The correlation coefficient $\rho$ of each tracer, as well as the combined tracer (black), with the true convergence.}
	\label{fig:tracer_rhos}
\end{figure}

In Fig.~\ref{fig:tracer_rhos}, we present the correlation coefficient \(\rho\) between the tracers and the true convergence. Our results demonstrate that combining the CIB and \emph{Euclid} with internal lensing reconstruction does not significantly improve the correlation on large scales, where internal reconstruction already achieves a high signal-to-noise ratio (S/N). However, as anticipated, the external tracers enhance the correlation on smaller angular scales.

\begin{figure}
	\includegraphics[width=\columnwidth]{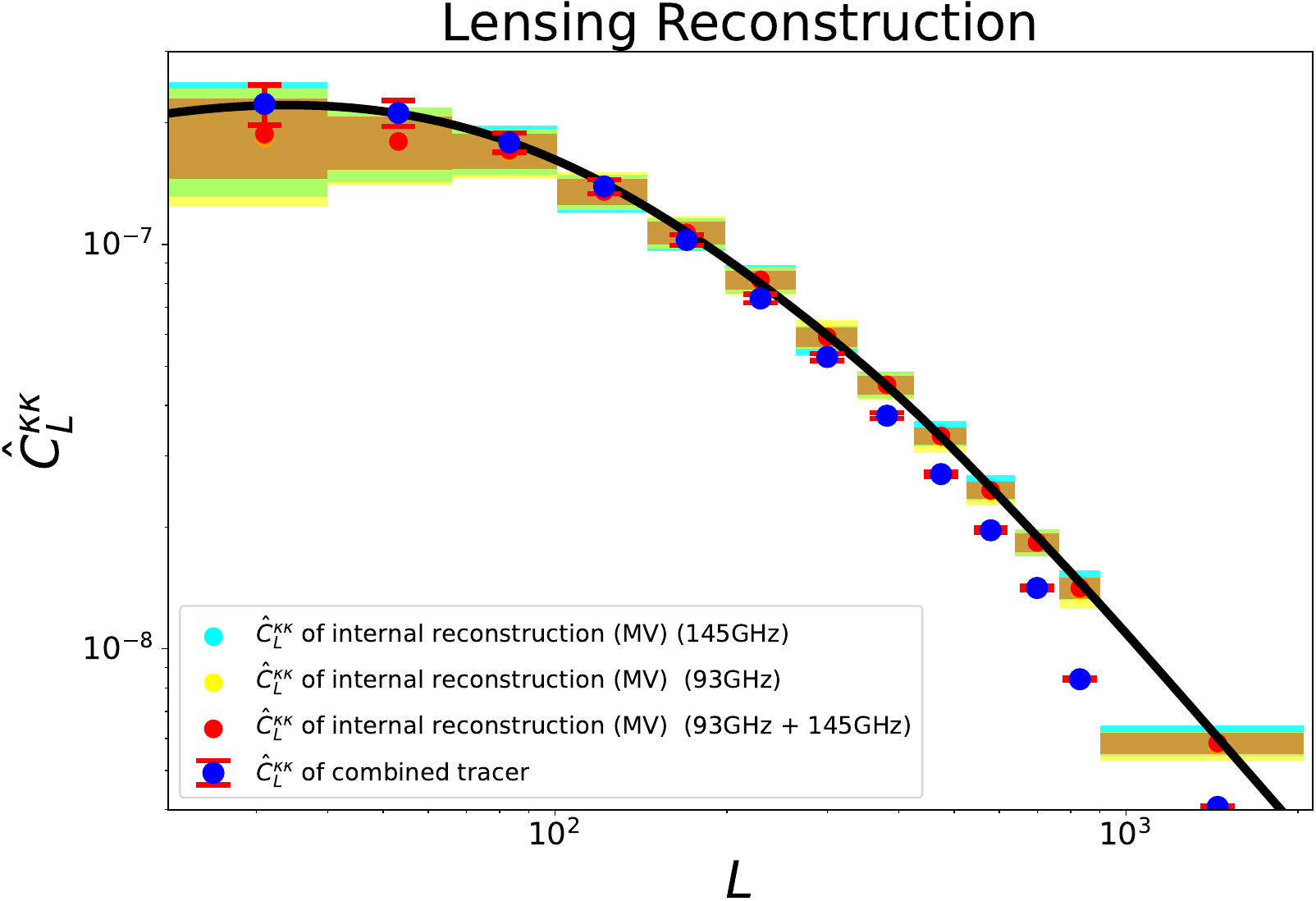}
	\caption{The lensing power spectra of the internal reconstruction (MV) (93 GHz and 145 GHz) and the combined tracer from one simulation are shown. The internal reconstruction (MV) from 93 GHz and 145 GHz was combined using inverse-variance weighting, and we refer to this \(\kappa^{\text{rec}}\) as the result of the internal reconstruction, which is then combined with external tracers. 
    It should be noted that the multi-tracer combined tracer (internal + external) is not an unbiased estimate of the lensing potential due to the filter included in the combination coefficients. An obvious suppression is observed for the combined tracer at \(L > 200\), which is caused by the filter. This suppression is designed to minimize the lensing B-mode residuals in the delensing procedure.
    The estimate \(\hat{C}_L^{\kappa\kappa}\) is derived from 100 simulation sets for the mean-field calculation and 400 simulation sets for the variance calculation.
    The error bars shown in the figure are the standard deviation scaled by a factor of 2 for clarity.}
	\label{fig:lensing_power}
\end{figure}

In Fig.~\ref{fig:lensing_power}, we present the power spectra of the lensing reconstruction. It is evident that incorporating external tracers reduces the variance of the reconstruction (from the red shaded region to the red error bars), particularly on small angular scales, as anticipated.

It is important to note that the combined tracer is no longer an unbiased estimate of the true convergence. This arises due to a filtering effect incorporated into the diagonal and non-diagonal elements of the coefficient \(c^i\). 
To illustrate this, consider a simple case where only two tracers are combined: \(I = I_1 + I_2\), with \(I_i = A_i \kappa + n_i\), where \(A_i\) is the scale factor and \(n_i\) represents independent noise. In this case, the coefficients \(c^i\) become:

\begin{equation}
    \begin{aligned}
        c_1 &= (C_{II}^{-1})_{11}C^{\kappa I_1} + (C_{II}^{-1})_{12}C^{\kappa I_2} \\
        &= \frac{1}{A_1} \frac{C^{\kappa\kappa}}{C^{\kappa\kappa} + \frac{N_1}{A_2^2} + \left(\frac{A_2^2}{A_1^2} \frac{N_1}{N_2}\right) C^{\kappa\kappa}}, \\
        c_2 &= (C_{II}^{-1})_{22}C^{\kappa I_2} + (C_{II}^{-1})_{21}C^{\kappa I_1} \\
        &= \frac{1}{A_2} \frac{C^{\kappa\kappa}}{C^{\kappa\kappa} + \frac{N_2}{A_1^2} + \left(\frac{A_1^2}{A_2^2} \frac{N_2}{N_1}\right) C^{\kappa\kappa}}.
    \end{aligned}
\end{equation}
Here, \(1/A_i\) can be interpreted as the normalization factor, and \(C^{\kappa\kappa}/(C^{\kappa\kappa} + {N_i}/{A_j^2})\) (\((i, j) \in \{(1, 2), (2, 1)\}\)) as the usual Wiener filter. Additionally, we observe an excess term in the denominator, which arises from the correlation between the tracer to be added and the tracers existed. 
Therefore, the term
\[
\frac{C^{\kappa\kappa}}{C^{\kappa\kappa} + \frac{N_i}{A_j^2} + \left(\frac{A_j^2}{A_i^2} \frac{N_i}{N_j}\right) C^{\kappa\kappa}}
\]
can be recognized as the modified Wiener filter, designed to minimize the noise of the combined tracer.
Consequently, no further filtering is needed in the delensing procedure when using this combined tracer.

\subsection{Lensed CMB maps processing}\label{subsec:mapproc}
Before proceeding with the delensing procedure, we need to process the lensed QU maps. We initially combine the observed maps from SAT at 93 GHz and from LAT at 93 GHz with an inverse-variance weight:
\begin{equation}
		\begin{aligned}
			&\chi_{\ell m} = \sum_i \omega_{i, \ell} \chi_{i, \ell m}, \\
			&\omega_{i, \ell} = \frac{N_{i, \ell}^{-1}}{\sum_i N_{i, \ell}^{-1}} .
		\end{aligned}
\end{equation}
Here, $\omega_{i, \ell}$ represents the weights of the $i$-th experiment, $N_{i, \ell}$ is the sum of the noise power spectrum of the $i$-th experiment after beam deconvolution and the foreground power spectrum, $\chi_{i, \ell m}$ denotes the harmonic coefficients of fields of the $i$-th experiment after beam deconvolution, and $\chi_{\ell m}$ represents the combined alm.
We initially perform Wiener filter on the combined observed QU maps to enhance the signal-to-noise ratio (S/N) as:
\begin{equation}
	\begin{gathered}
		Q_{\ell m} \Rightarrow \frac{C_{\ell}^{EE}}{C_{\ell}^{EE} + N_{\ell}^{EE}} Q_{\ell m}, \\
		U_{\ell m} \Rightarrow \frac{C_{\ell}^{EE}}{C_{\ell}^{EE} + N_{\ell}^{EE}} U_{\ell m}. \\
	\end{gathered}
\end{equation}
where $C_{\ell}^{EE}$ and $N_{\ell}^{EE}$ are obtained from ensembles of signal-only and noise-only simulations, respectively.
Next, we perform Wiener filter on the combined maps, as mentioned above, and the combined E-mode noise power spectrum is given by:
\begin{equation}
	N_{\ell}^{EE} = \sum_i \omega_{i, \ell}^2 N_{i, \ell}^{EE} .
\end{equation}
We only use observations at 93 GHz from both SAT and LAT when constructing the lensing B-mode template, as we anticipate minimal improvement from enhancing the signal-to-noise ratio (S/N) of the maps to be delensed (e.g. further combining with observation at 145 GHz). Instead, we expect that achieving a higher S/N in the mass estimate will play a pivotal role in improving the efficiency of delensing, as discussed in \cite{manzotti2017cmb}.

\subsection{Delensing}\label{sec:delensing}
In this paper, we construct the lensing B-mode template using two delensing methods, as discussed in Section \ref{sec:method}.

Regarding the Gradient-order lensing template method, we proceed by converting the spin-2 fields $Q + iU$ to spin-1 and spin-3 fields, and transform the spin-0 lensing potential field to spin-$-1$ and spin-1 fields using \texttt{CMBlensplus} \cite{namikawa2021CMBlensplus}. It's important to note that only E-modes should be included in the QU fields. Following this conversion, we compute the gradient lensing template according to Section \ref{sec: method_intro}. The resulting QU represents the lensing effect on E-mode, from which the lensing B-mode can be separated using harmonic transformation and EB leakage  is corrected with an "E-mode Recycling" method described in \cite{liu2019methods}.

Regarding the Inverse-lensing method, we proceed by estimating the inverse deflection angle $ \mathbf{d^{inv}} $ from the filtered lensing potential map $\hat \phi^\text{MV}$ and from combined tracer. This is done by iteratively solving Eq.(\ref{hat_n}) using \texttt{CMBlensplus} \cite{namikawa2021CMBlensplus}. Subsequently, we remap the filtered observed QU maps to obtain the delensed QU maps. These are then subtracted from the observed QU maps to derive the noisy lensing template QU maps. The lensing template B map can be separated through harmonic transformation from QU fields, and EB leakage should also be corrected. 

We compute the auto and cross pseudo-$C_{\ell}$ of lensing B-mode template (LT) and observed SAT B-modes ($C_{\ell}^{LT}$ and $C_{\ell}^{LT \times \text{SAT}, \nu}$) using the \texttt{NaMaster} \cite{alonso2023namaster} code. As for the corresponding bias terms analyzed in Section \ref{sec:bias}, we input the required QU maps and the potential maps to the same pipeline as described above to simulate all the bias terms, their auto-pseudo-$C_{\ell}$ and cross-pseudo-$C_{\ell}$ spectra are then also calculated using the \texttt{NaMaster} code. 
As mentioned in Section \ref{sec:bias}, we can replace the baseline simulation with a signal-only simulation in the pipeline, causing some bias terms to vanish. This enables us to isolate these terms by taking the difference between the baseline and signal-only simulation results. We found that both debiasing methods have similar effects, allowing us to choose either one flexibly.
The average of these terms over 500 simulations serves as an estimate of the biases, which is then used for debiasing.
Note that the resulting lensing B-mode template (LT) is a filtered version of the lensed B-mode due to the suppression on signal from Wiener filter. So after debiasing, the auto- and cross- pseudo-$C_{\ell}$ are both filtered version and we then further compute the transfer function by signal-only simulation to compensate for the suppression on power-spectrum level. The transfer function for $C_{\ell}^{LT}$ is calculated as the ratio of the average power of the lensing B-mode template from signal-only simulation to the average lensed BB power. Besides, the transfer function for $C_{\ell}^{LT \times \text{obs}}$ is similarly calculated by replacing the numerator with the cross-spectrum. We use these transfer functions to rescale $C_{\ell}^{LT \times \text{SAT,}\nu}$ and $C_{\ell}^\text{LT}$ when fitting parameters.

\subsection{Result of Lensing B-mode Template Construction}

\begin{figure}
	\includegraphics[width=\columnwidth]{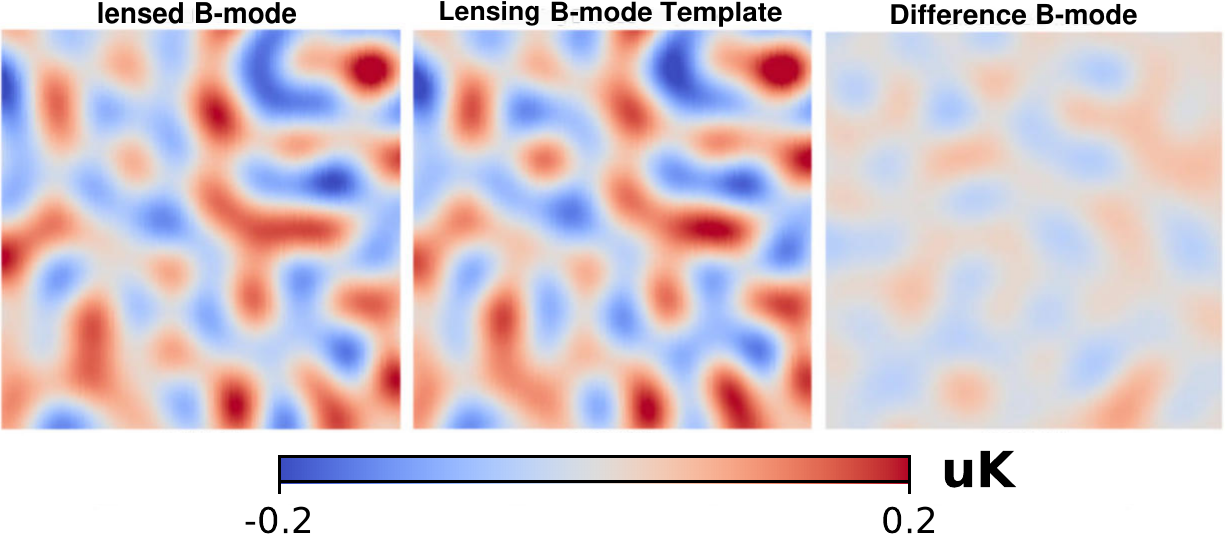}
	\caption{The input lensed B-mode map (left), the lensing B-mode template (noise subtracted) constructed using the gradient-order method (middle), and their difference (right) are shown. The shaded regions in the difference map represent the reduction of the lensing B-mode.}
	\label{fig:LT_temp}
\end{figure}

We plot one of the Lensing B-mode Templates (LT) from our simulation sets for \( \ell < 200 \) in Figure \ref{fig:LT_temp}. The LT map shown here is constructed using the Gradient-order template method with only the internal lensing reconstruction. After subtracting the noise part from the noisy lensing template on map level, we divide it by the square root of the transfer function, as calculated in Section \ref{sec:delensing}, to compensate for signal suppression caused by filtering. 
It is evident that most features of the lensed B-mode are effectively captured by the LT, and the difference map further confirms their similarity.

\begin{figure*}
	\includegraphics[width=\textwidth]{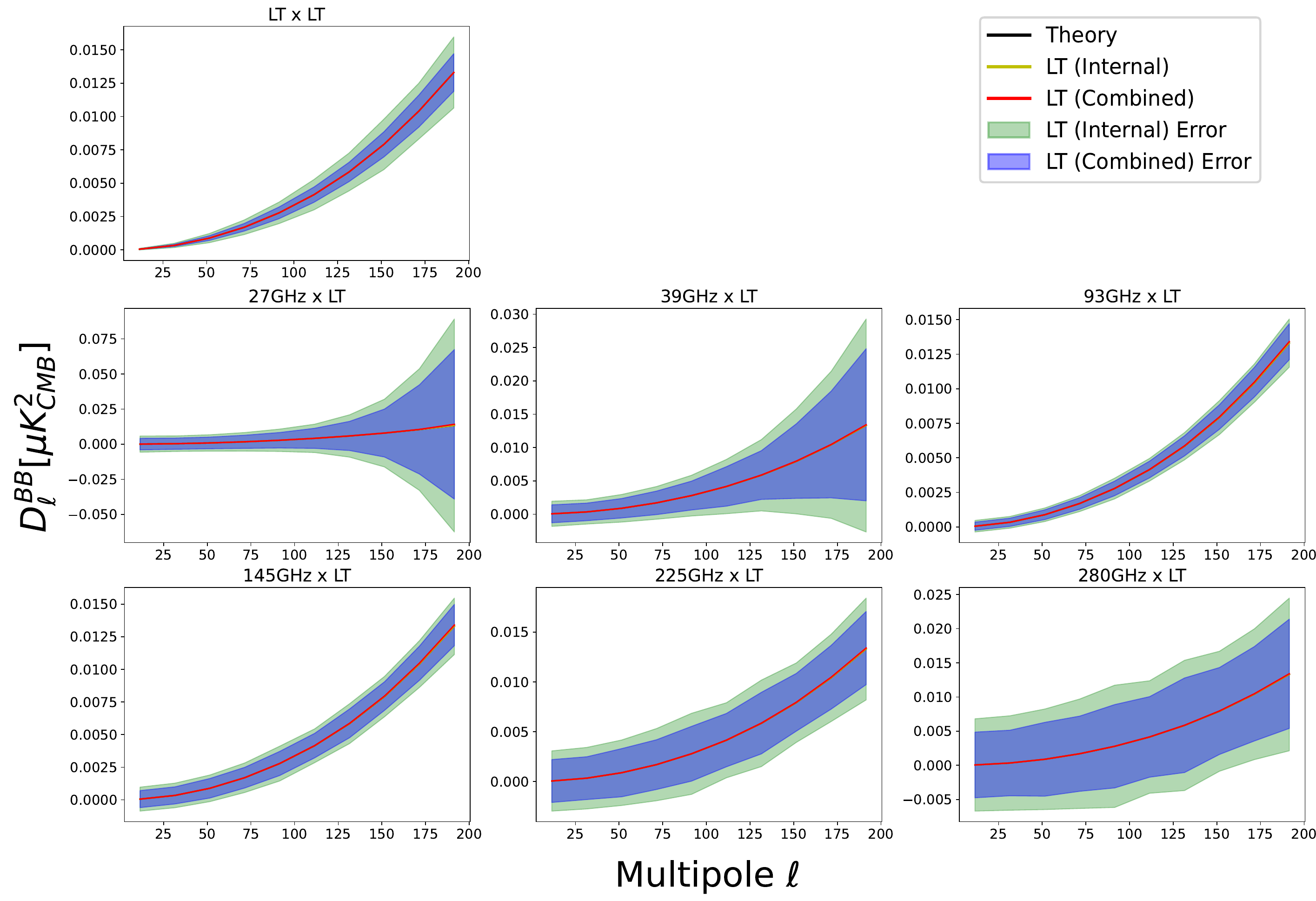}
	\caption{The cross-power spectra between the SAT observed maps at six frequencies and the lensing B-mode template (LT) are shown. All spectra have been debiased using simulations. The LT is generated using the gradient-order method, based either on internal lensing reconstruction or on a combined tracer. The black lines represent the theoretical predictions derived from our models with best-fit parameters. The green and blue shaded areas indicate the \( 1\sigma \) standard deviation computed from 500 observation simulations for the LT obtained from internal lensing reconstruction and from the combined tracer, respectively. The yellow and red lines denote the mean LT auto- and cross-power spectra for internal lensing reconstruction and the combined tracer, respectively. As shown, all the mean spectra are in good agreement with the theoretical predictions, and a significant reduction in the uncertainty is observed when comparing the LT derived from the combined tracer to that from internal lensing reconstruction. }
	\label{fig:all_cl_data}
\end{figure*}

The auto-power spectra of the Lensing B-mode Template (LT), the SAT observed maps at six frequencies, and their cross-power spectra are shown in Figure \ref{fig:all_cl_data}. The LT is constructed using the Gradient-order template method with either the internal lensing reconstruction or the combined tracer. The last row shown here has been debiased using the estimate and rescaled with the transfer function for better comparison.
As shown, both the auto-power spectrum of the LT and its cross-power spectra with the SAT observed maps at each frequency agree well with the theoretical predictions. Furthermore, employing a combined tracer effectively reduces the variance of the delensing procedure, as evidenced in the last row, where the variance involving the LT derived from the combined tracer is reduced compared to that from the internal lensing reconstruction.

\subsection{Results on the $r$ constraint}\label{sec:posterior}
\subsubsection{Baseline result}\label{sec: baseline_fit}
We investigate the parameter space using \texttt{Cobaya} \cite{Torrado_2021,torrado2020cobaya}, which employs a Markov chain Monte Carlo approach. To improve the constraint on the tensor-to-scalar ratio \( r \), rather than subtracting the lensing B-mode template (LT) from the observation (map-level delensing), we treat it as an additional pseudo-channel and calculate the auto and cross power spectra to form the theoretical model (cross-spectral method delensing). Both methods lead to effectively the similar uncertainty, as pointed out by \cite{hertig2024simons}. Our map channels are \( B^\text{SAT,27GHz} \), \( B^\text{SAT,39GHz} \), \( B^\text{SAT,93GHz} \), \( B^\text{SAT,145GHz} \), \( B^\text{SAT,225GHz} \), \( B^\text{SAT,280GHz} \), and \( B^{LT} \), effectively seven channels.
Then the dataset vector can be:
$\mathbf{\hat{X}}_{\ell}=[C_{\ell}^{\text{SAT},\nu \nu'},$ $C_{\ell}^{\text{LT} \times \text{SAT},\nu},C_{\ell}^{\text{LT}}]$, 
which includes 7 auto-power spectra and 21 cross-power spectra. We parameterize each elements of $\mathbf{\hat{X}}_{\ell}$ with the baseline model (nine-parameter):
\begin{align}
    &\begin{aligned}
    C_{\ell}^{\text{SAT},\nu \nu'} &= rC_{\ell}^\text{tens}|_{r=1} + A_LC_{\ell}^\text{lens} 
     + N_{\ell}^{BB,\nu \nu'} \\ &\quad + C_{\ell}^{\nu \nu'}|_\text{FG (7 params.)},
    \end{aligned}\\
    &C_{\ell}^{\text{LT} \times \text{SAT},\nu} = {A_L}C_{\ell}^\text{lens}, \\
    &C_{\ell}^\text{LT} = A_LC_{\ell}^\text{lens} + N_{\ell}^\text{temp}.
\end{align}
where parameter $A_L$ scales the intensity of the lensing B-mode. $N_{\ell}^{BB,\nu \nu'}$ is the power spectrum of instrumental noise,
\begin{align}
    N_{\ell}^{BB,\nu \nu'} &= 
    \begin{cases} 
        N_{\ell}^{\text{SAT},BB,\nu}, & \text{if } \nu = \nu' \\
        0, & \text{if } \nu \neq \nu'.
    \end{cases}
\end{align}

We use the band-power spectrum from $\ell_{\text{min}} = 20$ to $\ell_{\text{max}} = 200$, with a binning width of $\Delta_\ell = 20$. The priors on the nine parameters are imposed as uniform distributions listed in Table \ref{tab:prior} . We use Gaussian likelihood
to give parameter posterior distribution. 
The Gaussian approximation log-likelihood used in this work is given by:
\begin{equation}
	\begin{gathered}
        -2 \ \text{ln} \ \mathcal{L}(C_{\ell}|\hat{C_{\ell}}) = (\mathbf{X}_{\ell}-\mathbf{\hat X}_{\ell})^T \mathbf{M_{\ell}^{-1}} (\mathbf{X}_{\ell}-\mathbf{\hat X}_{\ell}) +  \text{ln} \ |\mathbf{M_{\ell}}|,
	\end{gathered}
\end{equation}
where $\mathbf{M_{\ell}}$ is the covariance matrix, and the vector $\mathbf{\hat X}_{\ell}$ contains the band-power estimates from the observed cut-sky maps, calculated using the \texttt{NaMaster} \cite{alonso2023namaster} code.

We also test the Hamimeche \& Lewis (HL) likelihood, which effectively captures the non-Gaussian shape of the likelihood. It is particularly recommended for experiments focusing on small sky areas and large-scale observations, where a non-Gaussian likelihood approximation is essential.
The Hamimeche \& Lewis likelihood is expressed as: 
\begin{equation}
	\begin{aligned}
        -2 \ln \mathcal{L}(C_{\ell} | \hat{C}_{\ell}) &= \frac{2l + 1}{2} \mathrm{Tr} \left[ \left( \mathbf{C}_{f_{\ell}}^{-1/2} \mathbf{C}_{g_{\ell}} \mathbf{C}_{f_{\ell}}^{-1/2} \right)^2 \right] \\ 
        &= \mathbf{X}_{g_{\ell}}^T \mathbf{M}_{f_{\ell}}^{-1} \mathbf{X}_{g_{\ell}},
	\end{aligned}
\end{equation}
where $\mathbf{X}_{g_{\ell}}=\text{vecp}(C_{g_{\ell}})$ is the vector of distinct elements of the transformed power spectra $C_{gl}$. For a comprehensive review, we refer the reader to reference \cite{hamimeche2008likelihood}.

The covariance matrix $\mathbf{M_{\ell}}$ used in the likelihood is derived from the baseline of 500 simulations to account for the effects of masking, cosmic variance, noise, foregrounds, and other factors. 
The simulation of the mock maps is detailed in Section \ref{sec:sims}.
Delensing is performed on these simulations to generate the lensing B-map templates. Subsequently, all power spectra contained in the aforementioned $\mathbf{\hat X}_{\ell}$ are calculated for each simulation using the \texttt{NaMaster} \cite{alonso2023namaster} code, as described in Section \ref{sec:delensing}. 
The covariance matrix is then estimated from these 500 realizations of $\mathbf{\hat X}_{\ell}$, with the elements of the matrix:
\begin{equation}
    \mathbf{M}_{\ell,ij} = \langle \mathbf{\mathbf{\hat{X}}}_{\ell,i} \mathbf{\mathbf{\hat{X}}}_{\ell,j} \rangle - \langle \mathbf{\mathbf{\hat{X}}}_{\ell,i} \rangle \langle \mathbf{\mathbf{\hat{X}}}_{\ell,j} \rangle,
\end{equation}
where the average is taken over 500 simulations, and the subscript $i,j$ represents the elements of $\mathbf{\hat X}_{\ell}$.



We present the posterior distributions for the parameters using the nine-parameter model with simulated data in Fig.~\ref{fig:posterior_gs_temp}. These results correspond to the LT constructed using the gradient-order method and parameter constraints obtained with a Gaussian likelihood.
The degeneracy between $r$ and $A_L$ is gradually canceled with the inclusion of the lensing B-mode template, as shown in the contour.
The summarized results from the Gaussian likelihood and HL likelihood analyses are shown in Table~\ref{tab:posterior_gs_all} and Table~\ref{tab:posterior_hl_all}, respectively.

The uncertainty in $r$ decreases from approximately $1.8 \times 10^{-3}$ to $1.0 \times 10^{-3}$ after incorporating the LT constructed with the internal lensing reconstruction, as shown in Fig.~\ref{fig:LT_temp}, as an independent observation channel into the likelihood. This represents a reduction of about 40\%. Furthermore, we find that the uncertainty in $r$ reduces by approximately 60\% when using the LT constructed with the combined tracer, whether from the Gradient-order method or the Inverse-lensing method. However, the latter introduces slightly more uncertainty in $r$, which is expected, as the Inverse-lensing method involves more high-order noise terms in the lensing template, as described in Section~\ref{sec:bias2}.

As expected, the inclusion of the LT in the likelihood has minimal impact on the constraints for the seven foreground parameters. This further suggests that the delensing procedure does not introduce any significant additional bias from the foreground into the parameter estimates.

\begin{figure}
	\includegraphics[width=\columnwidth]{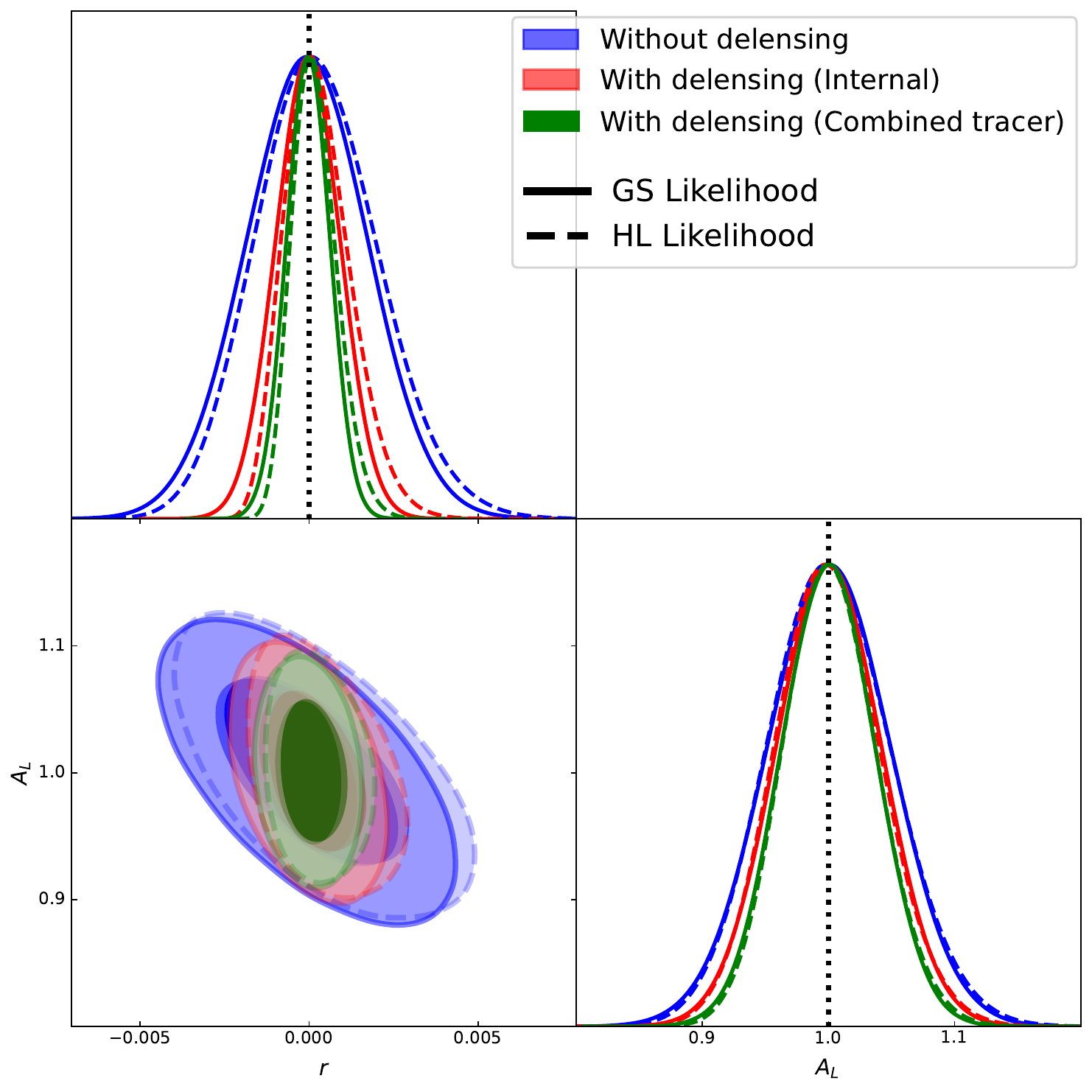}
        \caption{Marginalized posterior distributions for $r$ and $A_L$, obtained by sampling the HL likelihood (dashed lines) and its Gaussian approximation (solid lines). Results are presented for three cases for each likelihood: without delensing, delensing with internal lensing reconstruction, and delensing with the combined tracer.}
	\label{fig:hl_gs_comparison}
\end{figure}

A comparison of the marginalized posterior distributions between Gaussian likelihood and HL likelihood is given in Fig.~\ref{fig:hl_gs_comparison}. 
We observe some skewness toward 0 ($\gamma_1 > 0$) in the $r$ posterior derived from the HL likelihood. We attribute this to the physical constraint that the tensor-to-scalar ratio must be non-negative, whereas our prior allows for a negative \( r \).
During the progress of MCMC sampling, the power spectra matrix for some of the samples with $r < 0$ will no longer be positive-definite, which actually violates the requirements of the HL likelihood (see \cite{hamimeche2008likelihood} for details), leading to the suppression of these samples.
Even though, we observe no significant deviation between the results derived from the Gaussian likelihood and the HL likelihood. This consistency suggests that the adopted binning strategy is effective, providing sufficient degrees of freedom for the likelihood to approximate a Gaussian distribution, in accordance with the central limit theorem.


\begin{figure*}
	\includegraphics[width=\textwidth]{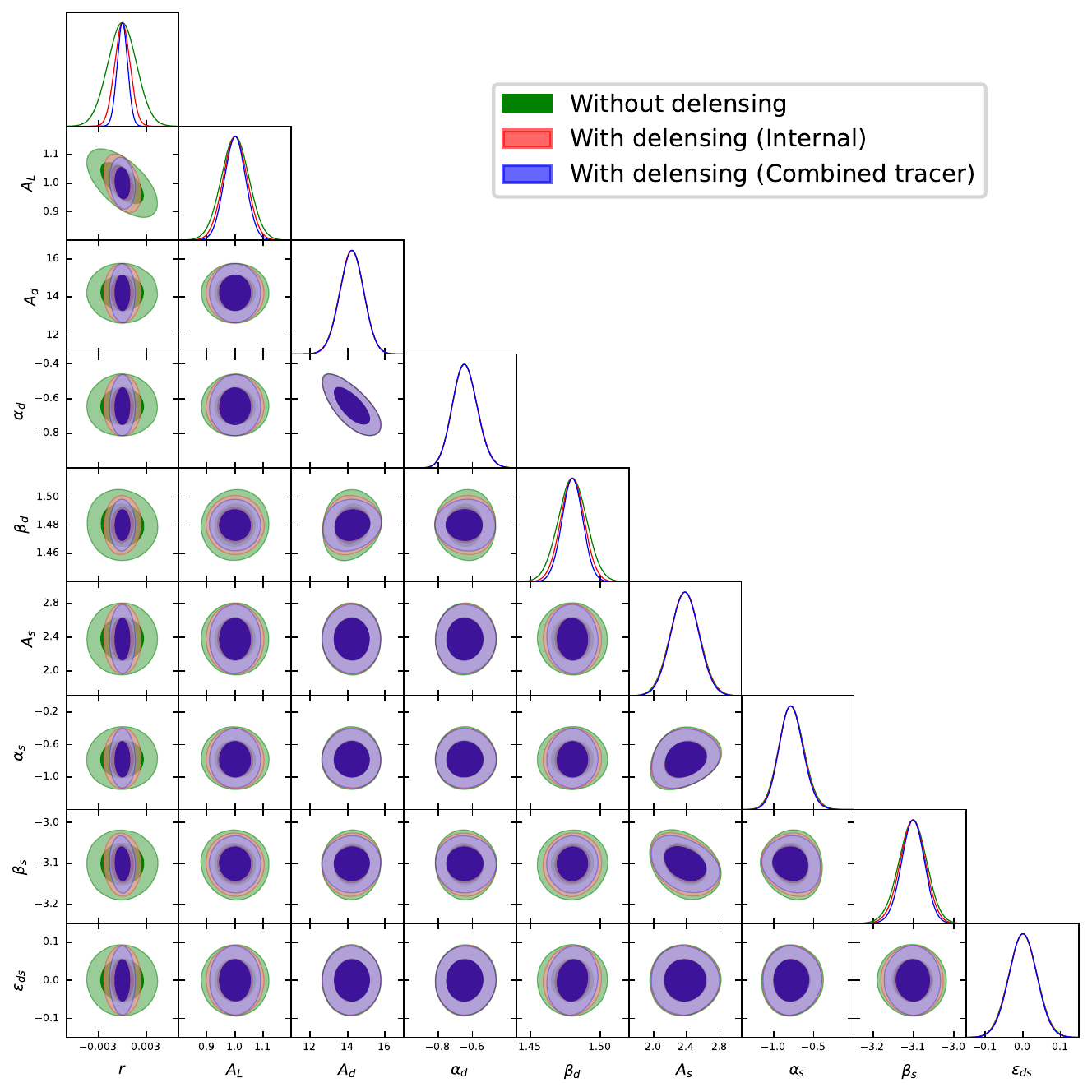}
        \caption{Posterior distributions of the baseline model parameters are shown for the baseline simulation data set with delensing (red and blue) and without delensing (green). A Gaussian approximation was used when performing the likelihood analysis. The lensing template (LT) was constructed using the gradient-order method, either with internal lensing reconstruction (red) or with a combined tracer (blue). The posterior distributions without incorporating the LT are shown in green. It is evident that the uncertainty in $r$ is reduced by including the LT (red line) in the likelihood analysis.}
	\label{fig:posterior_gs_temp}
\end{figure*}

\begin{table*} 
	\centering 
	\caption{The mean and $1\sigma$ standard deviation of each parameter using Gradient-order method/Inverse-lensing method, with the lensing proxy from internal reconstruction/combined tracer. Gaussian approximation likelihood was used in likelihood analysis.}
	\label{tab:posterior_gs_all}
	\begin{adjustbox}{max width=\textwidth} 
	\begin{tabular}{lcccccc} 
		\hline
		& & & \multicolumn{2}{c}{Gradient-order Method} & \multicolumn{2}{c}{Inverse-lensing Method} \\
		\cmidrule(lr){4-5} \cmidrule(lr){6-7} 
		Parameter & Input value & Before adding LT & After adding LT (Internal) & After adding LT (Combined tracer) & After adding LT (Internal) & After adding LT (Combined tracer) \\
		\hline
		$r (\times 10^3)$ & 0 & $-0.055 \pm 1.800$ & $-0.038 \pm 0.958$ & $-0.030 \pm 0.649$ & $-0.033 \pm 0.984$ & $-0.028 \pm 0.660$ \\
		$A_L$ & 1 & $1.000 \pm 0.049$ & $1.000 \pm 0.042$ & $1.000 \pm 0.037$ & $1.000 \pm 0.042$ & $1.000 \pm 0.037$ \\
		$A_d (\mu K^2)$ & 14.300 & $14.215 \pm 0.640$ & $14.215 \pm 0.637$ & $14.222 \pm 0.633$ & $14.219 \pm 0.638$ & $14.223 \pm 0.633$ \\
            $\alpha_d$ & -0.650 & $-0.642 \pm 0.073$ & $-0.642 \pm 0.072$ & $-0.643 \pm 0.072$ & $-0.642 \pm 0.072$ & $-0.643 \pm 0.072$ \\
            $\beta_d$ & 1.48 & $1.480 \pm 0.010$ & $1.480 \pm 0.009$ & $1.480 \pm 0.007$ & $1.480 \pm 0.009$ & $1.480 \pm 0.008$ \\
            $A_s (\mu K^2)$ & 2.400 & $2.378 \pm 0.175$ & $2.378 \pm 0.170$ & $2.378 \pm 0.167$ & $2.378 \pm 0.170$ & $2.379 \pm 0.168$ \\
            $\alpha_s$ & -0.800 & $-0.775 \pm 0.157$ & $-0.778 \pm 0.152$ & $-0.778 \pm 0.149$ & $-0.778 \pm 0.153$ & $-0.778 \pm 0.149$ \\
            $\beta_s$ & -3.100 & $-3.102 \pm 0.035$ & $-3.101 \pm 0.031$ & $-3.101 \pm 0.028$ & $-3.101 \pm 0.031$ & $-3.101 \pm 0.028$ \\
            $\epsilon_{ds}(\times 10^2)$ & 0 & $0.003 \pm 3.811$ & $0.003 \pm 3.729$ & $0.012 \pm 3.684$ & $0.019 \pm 3.730$ & $0.016 \pm 3.678$ \\
		\hline
	\end{tabular}
	\end{adjustbox}
\end{table*}


\begin{table*} 
	\centering 
	\caption{The same as Table.\ref{tab:posterior_gs_all}, but Hamimeche \& Lewis approximation likelihood was used in likelihood analysis.}
	\label{tab:posterior_hl_all}
	\begin{adjustbox}{max width=\textwidth} 
	\begin{tabular}{lcccccc} 
		\hline
		& & & \multicolumn{2}{c}{Gradient-order Method} & \multicolumn{2}{c}{Inverse-lensing Method} \\
		\cmidrule(lr){4-5} \cmidrule(lr){6-7} 
		Parameter & Input value & Before adding LT & After adding LT (Internal) & After adding LT (Combined tracer) & After adding LT (Internal) & After adding LT (Combined tracer) \\
		\hline
		$r (\times 10^3)$ & 0 & $0.259 \pm 1.829$ & $0.287 \pm 0.984$ & $0.178 \pm 0.668$ & $0.295 \pm 1.010$ & $0.182 \pm 0.680$ \\
		$A_L$ & 1 & $1.001 \pm 0.049$ & $1.000 \pm 0.042$ & $1.001 \pm 0.037$ & $1.000 \pm 0.042$ & $1.001 \pm 0.037$ \\
		$A_d (\mu K^2)$ & 14.300 & $14.392 \pm 0.645$ & $14.379 \pm 0.640$ & $14.382 \pm 0.638$ & $14.380 \pm 0.640$ & $14.381 \pm 0.637$ \\
            $\alpha_d$ & -0.650 & $-0.656 \pm 0.072$ & $-0.655 \pm 0.072$ & $-0.655 \pm 0.072$ & $-0.655 \pm 0.072$ & $-0.655 \pm 0.071$ \\
            $\beta_d$ & 1.48 & $1.480 \pm 0.010$ & $1.480 \pm 0.008$ & $1.480 \pm 0.007$ & $1.480 \pm 0.008$ & $1.480 \pm 0.007$ \\
            $A_s (\mu K^2)$ & 2.400 & $2.421 \pm 0.175$ & $2.419 \pm 0.171$ & $2.420 \pm 0.168$ & $2.420 \pm 0.172$ & $2.419 \pm 0.168$ \\
            $\alpha_s$ & -0.800 & $-0.816 \pm 0.154$ & $-0.814 \pm 0.149$ & $-0.814 \pm 0.146$ & $-0.815 \pm 0.150$ & $-0.815 \pm 0.147$ \\
            $\beta_s$ & -3.100 & $-3.102 \pm 0.034$ & $-3.102 \pm 0.031$ & $-3.102 \pm 0.028$ & $-3.102 \pm 0.031$ & $-3.102 \pm 0.028$ \\
            $\epsilon_{ds}(\times 10^2)$ & 0 & $-0.020 \pm 3.781$ & $-0.031 \pm 3.705$ & $-0.009 \pm 3.656$ & $-0.020 \pm 3.699$ & $-0.017 \pm 3.639$ \\
		\hline
	\end{tabular}
	\end{adjustbox}
\end{table*}

\subsubsection{A generalized modeling result}\label{sec: generalized_fit}
We propose a more general model that additionally accounts for imperfections in the construction of the lensing B-mode template, parameterized by $A_L'$. Given the complexity of the LT construction process, we expect the bias in the LT to be encapsulated by $A_L'$, with $A_L' = A_L = 1$ representing a perfect construction in standard \( \Lambda \)CDM. Here, $A_L$ describes the lensed B-modes in the observation. Thus, deviations of $A_L$ from 1 ($A_L \neq 1$) will capture discrepancies between the observation and the theoretical model, while deviations of $A_L'$ from 1 ($A_L' \neq 1$) will reflect biases introduced by the pipeline.

Then the model now includes ten parameters as follows:
\begin{align}
    &\begin{aligned}
    C_{\ell}^{\text{SAT},\nu \nu'} &= rC_{\ell}^\text{tens}|_{r=1} + A_LC_{\ell}^\text{lens} 
     + N_{\ell}^{BB,\nu \nu'} \\ &\quad + C_{\ell}^{\nu \nu'}|_\text{FG (7 params.)},
    \end{aligned}\\
    &C_{\ell}^{\text{LT} \times \text{SAT},\nu} = \sqrt{A_L A_L^{'}}C_{\ell}^\text{lens}, \\
    &C_{\ell}^\text{LT} = A_L^{'}C_{\ell}^\text{lens} + N_{\ell}^\text{temp}.
\end{align}
This reduces to our baseline model with \( A_L \approx A_L' \), which serves as a good approximation in most cases, as discussed above.

\begin{figure*}
	\includegraphics[width=\textwidth]{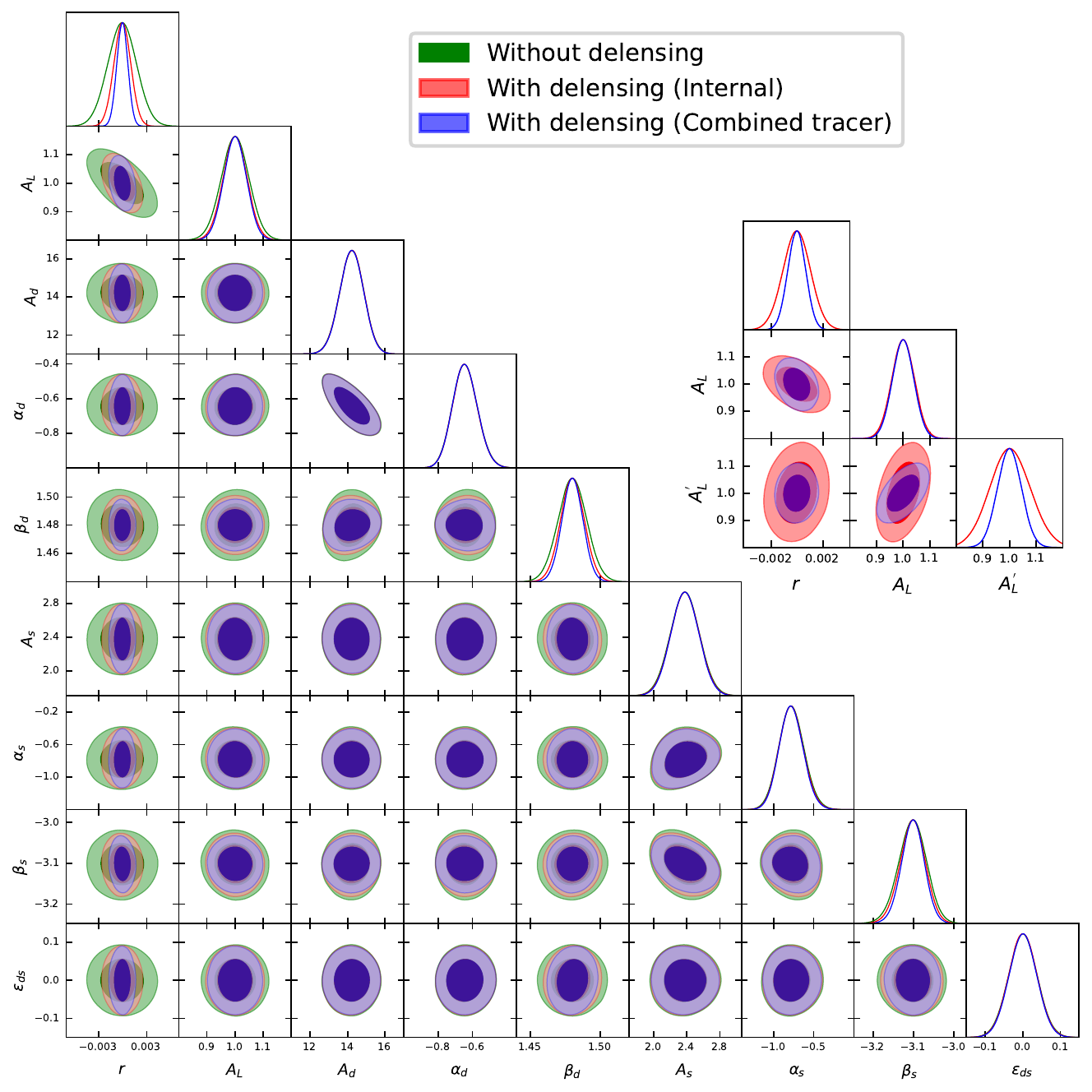}
        \caption{The posterior distributions of the generalized model parameters are shown for the baseline simulation data set, with delensing (red and blue) and without delensing (green). A Gaussian approximation was used in the likelihood analysis. The lensing template (LT) was constructed using the gradient-order method, either with internal lensing reconstruction (red) or with a combined tracer (blue). The posterior distributions of $r$, $A_L$, and $A_L^{'}$ for the delensing cases are shown exclusively in the upper right.}
	\label{fig:posterior_gs_temp_general}
\end{figure*}

\begin{table*} 
	\centering 
	\caption{The mean and $1\sigma$ standard deviation of each parameter using Gradient-order method/Inverse-lensing method, with the lensing proxy from internal reconstruction/combined tracer. Gaussian approximation likelihood was used in likelihood analysis. Two models described in Section \ref{sec: baseline_fit} and Section \ref{sec: generalized_fit} were considered.}
    \label{tab:posterior_gs_general}
	\begin{adjustbox}{max width=\textwidth} 
	\begin{tabular}{lcccccc} 
		\hline
		& & & \multicolumn{2}{c}{Generalized Model} & \multicolumn{2}{c}{Baseline Model} \\
		\cmidrule(lr){4-5} \cmidrule(lr){6-7} 
		Parameter & Fiducial value & Before adding LT & After adding LT (Internal) & After adding LT (Combined tracer) & After adding LT (Internal) & After adding LT (Combined tracer)\\
		\hline
		$r (\times 10^3)$ & 0 & $-0.055 \pm 1.800$ & $-0.042 \pm 1.048$ & $-0.042 \pm 0.686$  & $-0.038 \pm 0.958$ & $-0.030 \pm 0.649$\\
		$A_L$ & 1 & $1.000 \pm 0.049$ & $1.000 \pm 0.042$ & $1.000 \pm 0.040$  & $1.000 \pm 0.042$ & $1.000 \pm 0.037$\\
		$A_L^{'}$ & 1 & N/A & $1.001 \pm 0.074$ & $1.000 \pm 0.045$  & N/A & N/A\\
		$A_d (\mu K^2)$ & 14.300 & $14.216 \pm 0.640$ & $14.215 \pm 0.636$ & $14.222 \pm 0.634$  & $14.215 \pm 0.637$ & $14.222 \pm 0.633$\\
        $\alpha_d$ & -0.650 & $-0.642 \pm 0.073$ & $-0.642 \pm 0.072$ & $-0.642 \pm 0.072$  & $-0.642 \pm 0.072$ & $-0.643 \pm 0.072$\\
        $\beta_d$ & 1.48 & $1.480 \pm 0.010$ & $1.480 \pm 0.009$ & $1.480 \pm 0.007$  & $1.480 \pm 0.009$ & $1.480 \pm 0.007$\\
        $A_s (\mu K^2)$ & 2.400 & $2.378 \pm 0.175$ & $2.378 \pm 0.170$ & $2.378 \pm 0.167$  & $2.378 \pm 0.170$ & $2.378 \pm 0.167$\\
        $\alpha_s$ & -0.800 & $-0.775 \pm 0.157$ & $-0.777 \pm 0.152$ & $-0.779 \pm 0.148$  & $-0.778 \pm 0.152$ & $-0.778 \pm 0.149$\\
        $\beta_s$ & -3.100 & $-3.102 \pm 0.035$ & $-3.102 \pm 0.031$ & $-3.101 \pm 0.028$  & $-3.101 \pm 0.031$ & $-3.101 \pm 0.028$\\
        $\epsilon_{ds}(\times 10^2)$ & 0 & $0.003 \pm 3.811$ & $0.015 \pm 3.734$ & $0.025 \pm 3.677$ & $0.003 \pm 3.729$ & $0.012 \pm 3.684$\\
		\hline
	\end{tabular}
	\end{adjustbox}
\end{table*}

Posterior distributions of the generalized model parameters are shown in Fig.\ref{fig:posterior_gs_temp_general} and summarized in Table.~\ref{tab:posterior_gs_general}. The LT is constructed using the Gradient-order method, and a Gaussian approximated likelihood is used for the likelihood analysis. The prior for \( A_L^{'} \) is set to be the same as \( A_L \), as listed in Table \ref{tab:prior}.
The triangle plot on the upper right demonstrates that the inclusion of the LT from the combined tracer effectively tightens the constraint on $A_L^{'}$ compared to the one derived from internal tracers, with almost no tightening effect on the constraint of $A_L$. This behavior is expected, as \( A_L \) scales the lensed B-modes from observations, with its uncertainty primarily determined by the primordial B-modes, instrumental noise, and foregrounds.
The auto- and cross-power spectra of LT jointly constrain the lensing B-mode template ($A_L^{'}$), which subsequently helps constrain the observed lensed B-mode ($A_L$), ultimately leading to a tighter constraint on $r$.
Compared to the results derived from baseline model, we find that introducing the parameter $A_L^{'}$ results in a slight increase in the uncertainty on $r$. However, this addition allows for the capture of biases due to lensing, thereby mitigating the bias on \( r \) (although this effect is not evident in this case due to the large parameter space).

To highlight the advancement of the generalized model, we also conduct a toy test to evaluate the performance of these two models in \ref{app:model_toy}, where we artificially introduce bias into both the lensed B-mode power and the lensing B-mode template. In this case, we find that the baseline model results in a significantly larger bias on \( r \) compared to the generalized model, as the approximations made in the former become less effective as the bias increases. This demonstrates the potential of the generalized model to better account for the bias in the lensing power, thereby mitigating the bias on \( r \) when the delensing procedure is applied.

\subsection{Limitations of current forecasts}
Although we have attempted to produce realistic forecasts in terms of Galactic foreground and lensing reconstruction, one of the limitations of these forecasts is the simplicity of the foreground model. We plan to refine the foreground emission model in our future work.
Furthermore, we neglect Extragalactic foregrounds in the temperature maps used for reconstructing the lensing potential to avoid the influence of residual point sources. This makes our forecast somewhat optimistic, but it may not significantly affect our results. According to \cite{baleato2022impact}, combining with external tracers will reduce the impact of point source contamination in the internal lensing reconstruction, thereby largely alleviating biases and the degradation of delensing efficiency. A more comprehensive approach to handling residual point sources and assessing their impact requires further investigation.


Finally, the LSS tracer simulations are based solely on Gaussian realizations with the corresponding models in this paper, which may require more realistic considerations (e.g., non-Gaussianity) if we aim to be compatible with real observations. N-body simulations model dark matter halos by simulating the gravitational interactions between a large number of particles. These halos act as the building blocks for the formation of galaxies and other cosmic structures. Several major projects, such as the Millennium Run \cite{springel2005simulations}, Bolshoi \cite{klypin2011dark}, and the Outer Rim Simulation \cite{heitmann2019outer}, have been conducted to simulate the evolution of large-scale structures in the universe and have been widely used. By applying appropriate physical models to these dark matter halos, one can generate a variety of Extragalactic foregrounds, including the Sunyaev-Zeldovich (SZ) effect, the Cosmic Infrared Background (CIB), and galaxy distributions. This approach has been utilized in previous studies \cite{omori2024agora,sehgal2010simulations}, which have generated realistic sky survey maps for use in cosmological analysis. Although these simulation-based approaches are powerful and more realistic, they also require substantial computational resources and careful modeling choices, which are beyond the scope of this work focused on building and validating a baseline delensing pipeline. We plan to incorporate such realistic non-Gaussian simulations in future studies to further test and refine our pipeline under observational conditions.

\section{Conclusions}\label{sec:conclusion}

In this paper, we comprehensively address various aspects aimed at improving the constraints on \(r\) for future ground-based CMB B-mode observations. We provide an in-depth analysis of both the delensing residuals arising from the methods themselves and the delensing noise biases, which stem from Galactic foregrounds, instrumental noise, and the lensing reconstruction process.

We use the NILC method to examine the foreground components in the LAT observations and primarily assess the influence of the foreground residuals. Our analysis suggests that the point source residuals in temperature maps may require further processing, whereas the diffuse foreground residuals are acceptable, although there is no significant improvement compared to some of the observed maps at specific frequencies. We thus argue that our pipeline is feasible without a prior foreground cleaning process, saving the computational and time cost while enabling a more precise detection of \( r \) with minimal degradation.

We perform internal lensing reconstruction using simulated data from large-aperture ground-based telescopes at both 93 GHz and 145 GHz, utilizing both temperature and polarization information. 
Our results indicate that this approach is feasible and can reduce the cost associated with foreground removal. These are then combined with the simulated LSS tracers to enhance the quality of mass reconstruction.

The lensing B-mode template (LT) is constructed using simulated data from both small-aperture and large-aperture ground-based telescopes at 93 GHz. 
For parameter constraints, we extend the likelihood function to include all auto- and cross-spectra between the lensing B-mode template and the observed B-mode. Our baseline results show that the uncertainty in \(r\) is reduced by 40\% when using the LT from internal lensing reconstruction. This reduction increases to 60\% when using the LT constructed from the combined tracer. Furthermore, we observe consistent results between the Gaussian approximation and the Hamimeche \& Lewis approximation in the likelihood analysis. We also propose a generalized model that simultaneously accounts for the intensity of the lensed B-mode from observations and the lensing B-mode template from delensing. This model has the potential to capture the biases introduced during the delensing procedure, thereby mitigating the bias on $r$, at the cost of a slight increase in uncertainty.

There remains some room for improvement in our pipeline. First, we have neglected the impact of foreground non-Gaussianity on the delensing results in this study, which warrants further investigation. Second, since the temperature still contributes significantly to the internal lensing reconstruction for SO-like experiments, finding the optimal combination of foreground removal methods and lensing reconstruction techniques to mitigate point source contamination in the temperature field remains an important area of study.
Lastly, the LSS tracer models employed in this study requires further refinement to ensure compatibility with real observational data. More accurate and realistic LSS tracers may be based on N-body simulations with corresponding emission models. We will consider these factors in our future works.

\appendix

\section{Needlet Internal linear Combination}\label{sec:nilc}
Component separation with an ILC method has been widely used by CMB observation, in either pixel domain (PILC) \cite{fernandez2016exploring} or harmonic domain (HILC) \cite{kim2009cmb}. However, considering that the variation of the foreground intensity across the sky, and the instrumental noise is dominant on small scales, we here perform the ILC method in spherical needlets (NILC) \cite{basak2012needlet}, which allows us to localize our statistics both in the pixel domain and in the harmonic domain.

\subsection{Basic theory of NILC}
We here describe the method in brief, we recommend the reader to reference \cite{basak2012needlet} for thorough details.
Considering an observation with CMB, instrumental noise foreground components, and assuming their independence:
\begin{equation}
    X_{\ell m}^{\mathrm{obs}, c} = b_\ell^c X_{\ell m}^{\mathrm{CMB}} + b_\ell^c X_{\ell m}^{\mathrm{FG}, c} + X_{\ell m}^{\mathrm{N}, c},
\end{equation}
where $X \in \{ \Theta, E,B\}$ is the harmonic coefficient of a field, $b_\ell^c$ is the beam function and $c$ is the frequency channel. We then perform a filter $h_\ell^{j}$ in harmonic space to extract observation at scale $j$,
\begin{equation}
    X_{\ell m}^{c, j} = h_\ell^{j} X_{\ell m}^{c},
\end{equation}
this correspond to expanding the observed harmonic coefficient with spherical needlets at scale $j$, pixel $k$:
\begin{equation}
    \beta_{jk}^{X, c} = \beta_{jk}^{\mathrm{CMB}} + \beta_{jk}^{\mathrm{FG}, c} + \beta_{jk}^{\mathrm{N}, c},
\end{equation}
with corresponding spherical needlets given by:
\begin{equation}
    \begin{aligned}
        \beta_{jk}^{\mathrm{CMB}} &= \sqrt{\lambda_j} \sum_{\ell m} h_\ell^j b_\ell X_{\ell m}^{\mathrm{CMB}} Y_{\ell m}(\xi_j), \\
        \beta_{jk}^{\mathrm{FG}, c} &= \sqrt{\lambda_j} \sum_{\ell m} h_\ell^j b_\ell^c X_{\ell m}^{\mathrm{FG}, c} Y_{\ell m}(\xi_j), \\
        \beta_{jk}^{\mathrm{N}, c} &= \sqrt{\lambda_j} \sum_{\ell m} h_\ell^j \frac{b_\ell}{b_\ell^c} X_{\ell m}^{\mathrm{N}, c} Y_{\ell m}(\xi_j).
    \end{aligned}
\end{equation}

Since all the observations actually measure signal $bX^{\mathrm{CMB}}$ with some error $bX^{\mathrm{FG}} + X^{\mathrm{N}}$, consists in averaging all these measurements giving a specific weight $w_i$ to each of them, we naturally questing for a solution of the form in needlet space:
\begin{equation}
    \beta_{jk}^{\mathrm{NILC}} = \sum_{c=1}^{n_c} w_{jk}^c \beta_{jk}^{X, c},
\end{equation}
the weight $w_{jk}^c$ is determined by minimizing the variance of the reconstructed CMB, subject to the unbiasedness constraint ($\sum_{c=1}^{n_c} w_{jk}^c=1$). Using the method of Lagrange multipliers to enforce the constraint, the solution is given by:
\begin{equation}
    \hat{\omega}_{jk}^c = 
    \frac{
        \sum_{c'} \left[\hat{\mathbf{R}}_{jk}^{-1}\right]^{cc'} a^{c'}
    }{
        \sum_c \sum_{c'} a^c \left[\hat{\mathbf{R}}_{jk}^{-1}\right]^{cc'} a^{c'}
    },
\end{equation}
where $\left[\hat{\mathbf{R}}_{jk}^{-1}\right]^{cc'}$ is the inverse of the cross frequency covariance matrix:
\begin{equation}
        \hat{\mathbf{R}}_{jk}^{cc'} = \frac{1}{n_k} \sum_{k'} \omega_j(k,k') \beta_{jk}^{c}\beta_{jk}^{c'},
\end{equation}
where $\omega_j(k,k')$ selects the domain of $n_k$ pixels around the $k$-th pixel over which we perform our averaging to estimate the covariance.
Finally, the cleaned CMB map is given by:
\begin{equation}
        \hat X_\text{NILC}(\hat{\mathbf{n}}) = \sum_{\ell m}\sum_{jk} \beta_{jk}^{\mathrm{NILC}} \sqrt{\frac{4\pi}{\lambda_j}} h_{\ell}^j Y_{\ell m}(\xi_j)Y_{\ell m}(\hat{\mathbf{n}}).
\end{equation}

\subsection{The NILC results and some discussion}
\begin{figure*}
	\includegraphics[width=\textwidth, height=0.95\textheight, keepaspectratio]{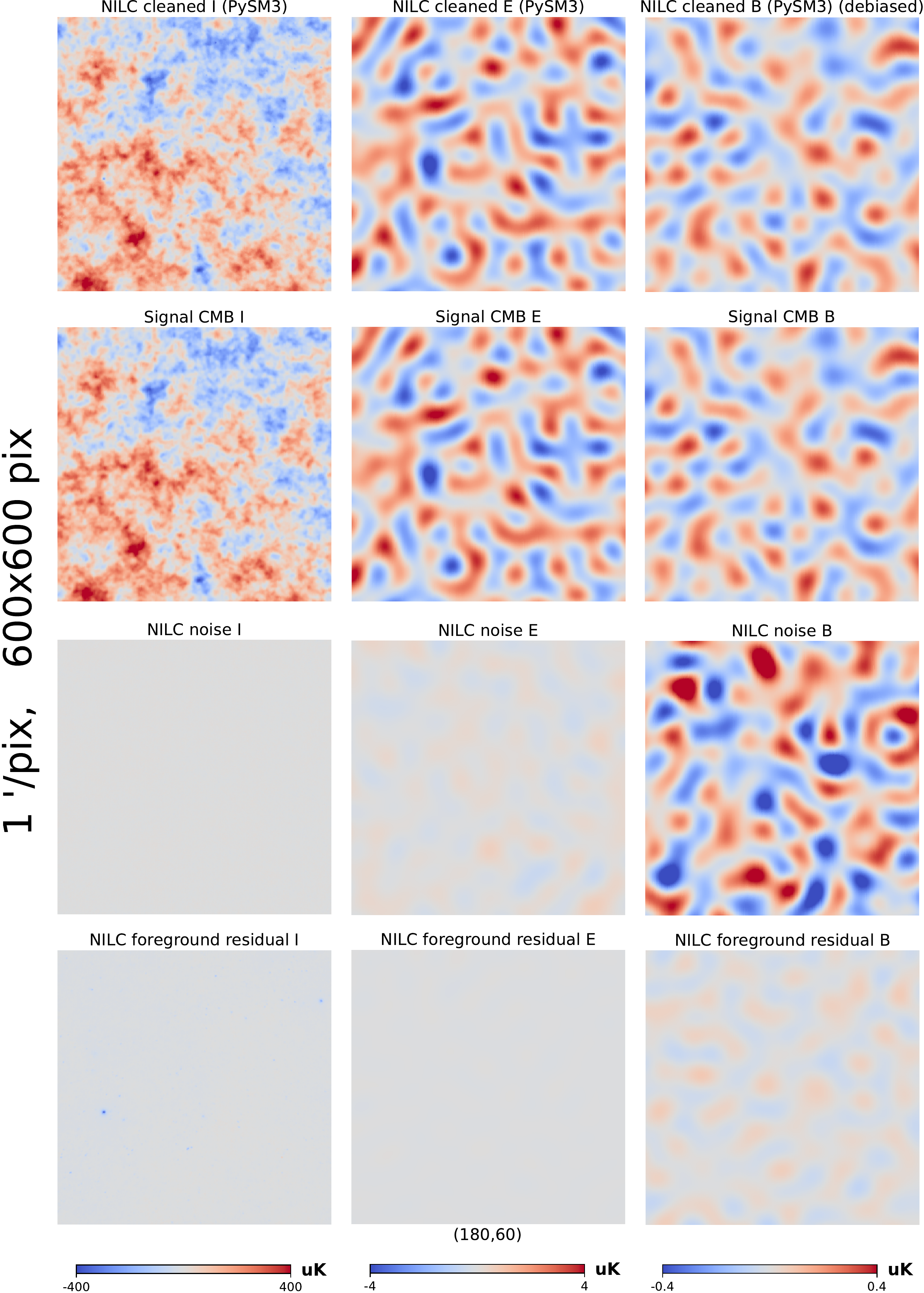}
	\caption{NILC result maps for I,E,B. From top to bottom, the rows represent the cleaned maps given by NILC, the input CMB maps, the noise maps, and the foreground residual maps. The noise maps and foreground residual maps illustrate the noise and foreground components remaining in the cleaned maps, respectively. Note that the cleaned map has been de-biased by subtracting the residuals at the map level. The maximum multipole of the polarization is set to $\ell \sim300$.}
	\label{fig:nilc_map}
\end{figure*}

\begin{figure}[htbp]
    \centering
    \subfigure[NILC result: power spectra of temperature]{
        \includegraphics[width=1\linewidth]{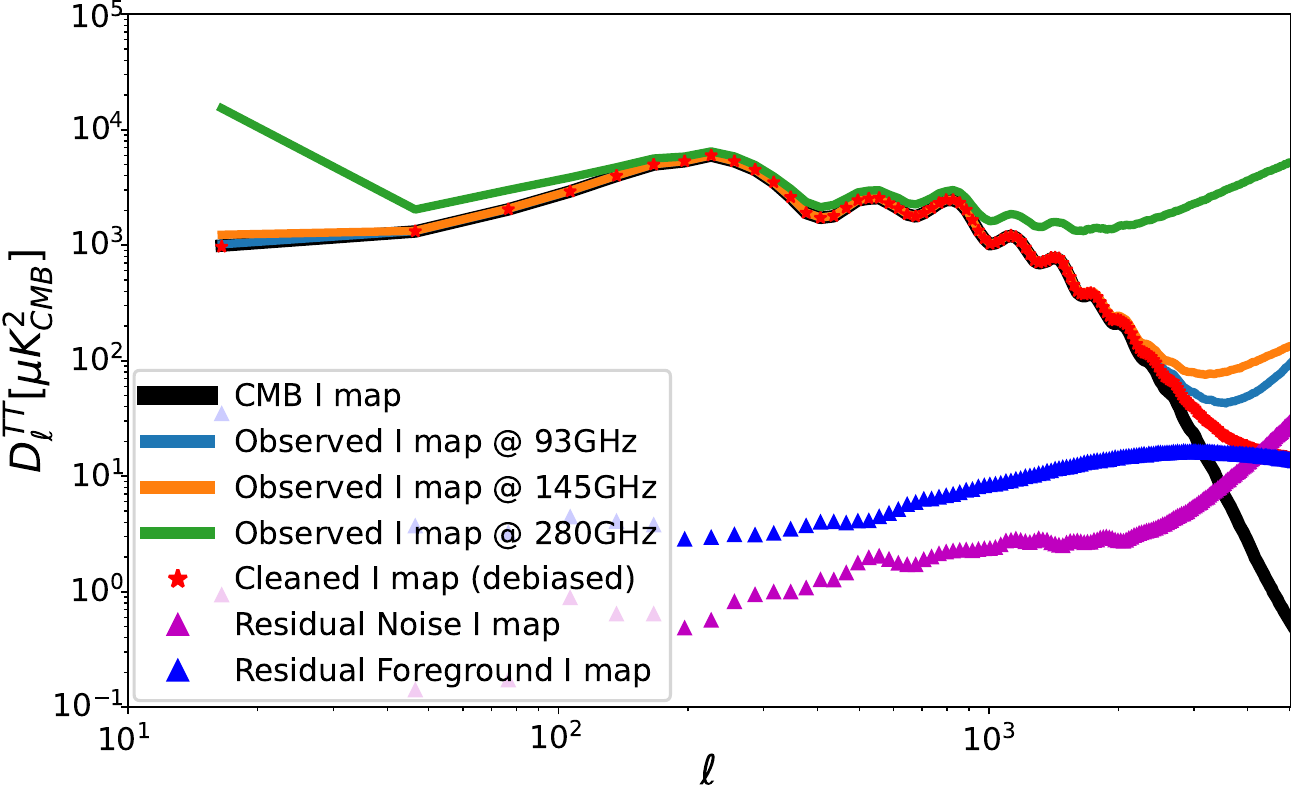}
    }
    \vspace{0.02\linewidth} 
    \subfigure[NILC result: power spectra of E-mode polarization]{
        \includegraphics[width=1\linewidth]{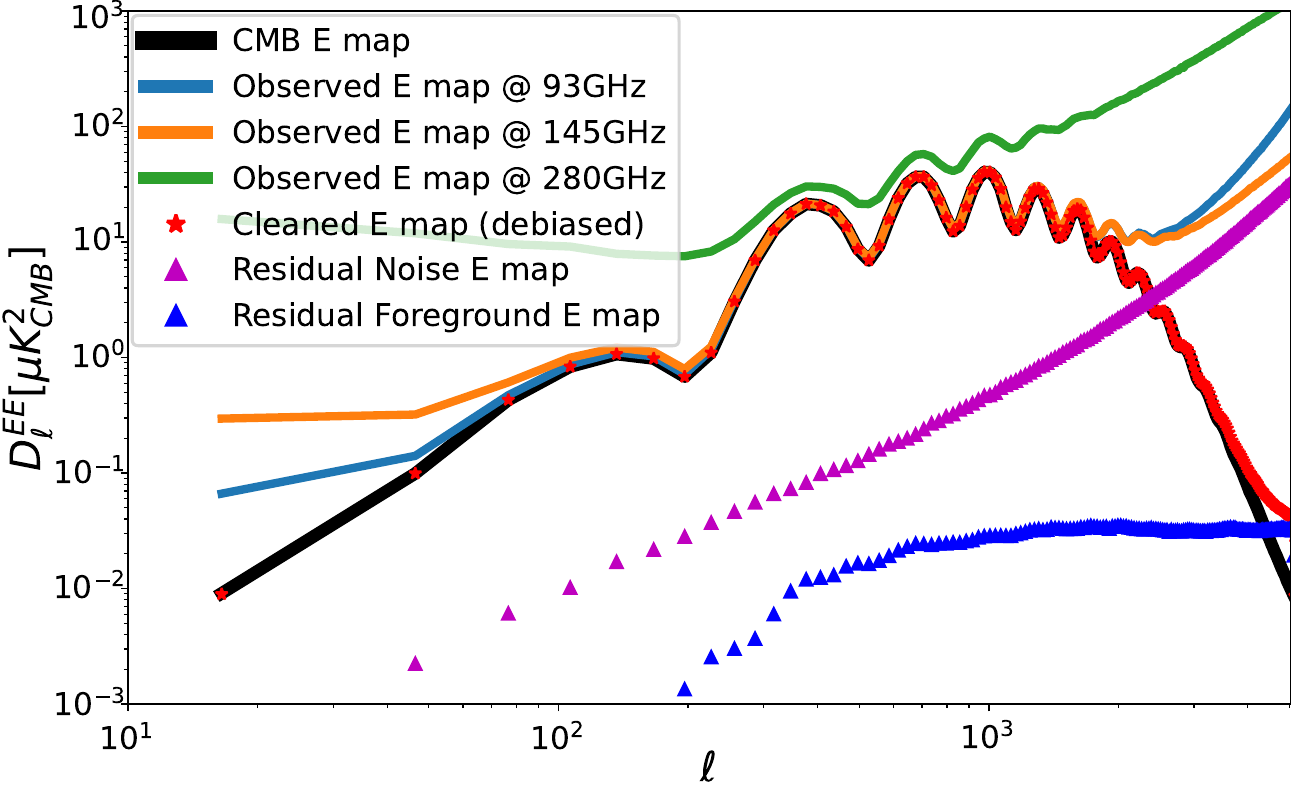}
    }
    \vspace{0.02\linewidth} 
    \subfigure[NILC result: power spectra of B-mode polarization]{
        \includegraphics[width=1\linewidth]{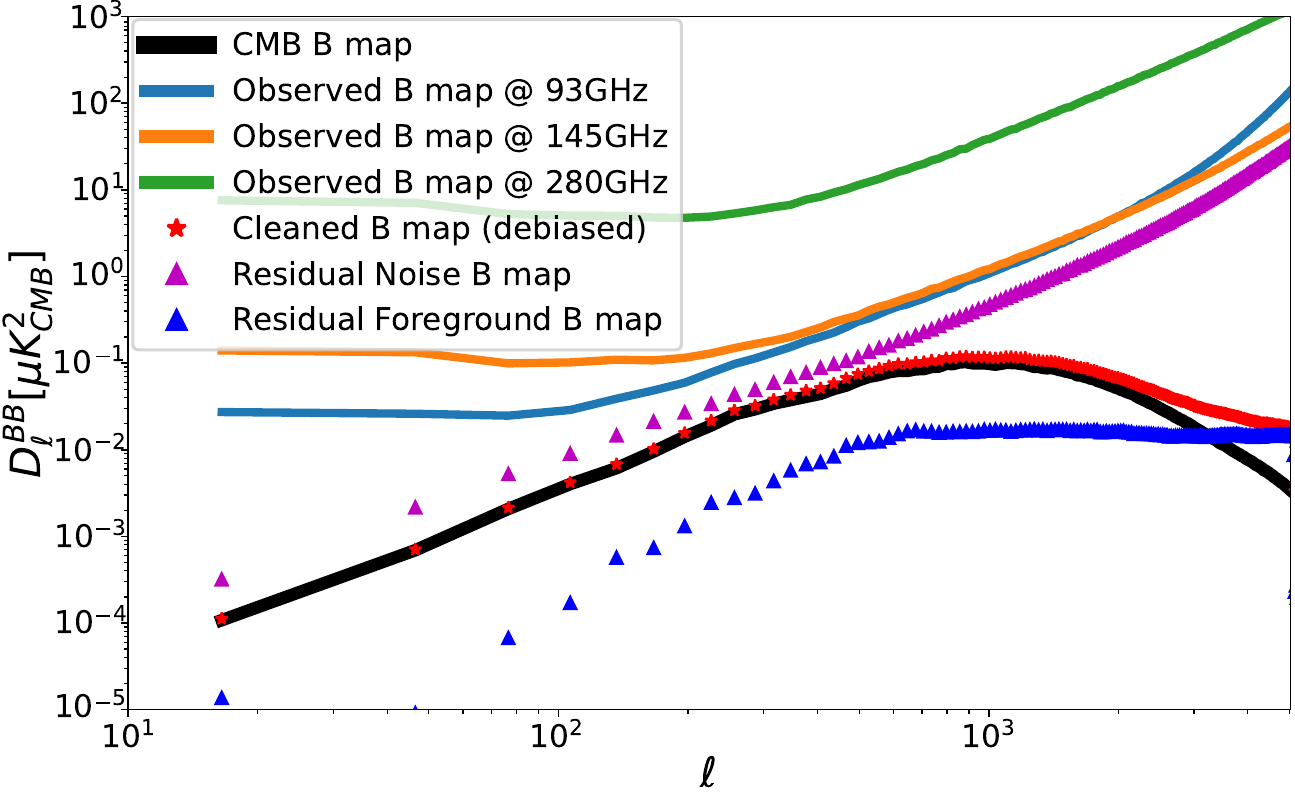}
    }
    \caption{The power spectra of the NILC result. The power of the cleaned maps have been de-biased by subtracting the NILC noise. We see that the cleaned map power spectra closely align with those of the signals on large scales.}
    \label{fig:nilc_cls}
\end{figure}

We present the results of the NILC method in Fig.~\ref{fig:nilc_map}. The similarity between the signal maps and the cleaned maps underscores the effectiveness of the foreground removal. We present the NILC residual power with different foreground components in Fig.~\ref{fig:nilc_cls}.
In the cleaned CMB maps, the diffuse foreground has been significantly suppressed and has become subordinate to the signal. 
However, we observe some evident point source residuals in the NILC residual temperature map, which mainly come from the thermal Sunyaev-Zel'dovich effect and the CIB in our case. This point source residual is expected, as the NILC method is known to have limited effectiveness in mitigating compact sources, as discussed in \cite{zhang2022efficient}.

As shown, for $\ell>200$, the majority of the residuals originate from the thermal Sunyaev-Zel'dovich effect and CIB, while the kinetic Sunyaev-Zel'dovich effect is subdominant. In contrast, thermal dust, synchrotron emission, spinning dust, and free-free emission contribute relatively little to the residuals. We suggest that the majority of point source residuals originating from the Extragalactic foreground require further processing. 

On the one hand, different foreground removal methods exhibit varying performance with respect to point sources \cite{sailer2021optimal}. De-projection of the tSZ or CIB can effectively eliminate one foreground, but often significantly amplifies the other. De-projection of both tSZ and CIB, however, leads to a substantial increase in residual noise. The ILC method suppresses total residual noise, although neither of the foregrounds is completely eliminated. Finally, masking or in-painting with the matched-filter can also address some of the point sources.

On the other hand, several lensing quadratic estimators (QEs) have been developed to mitigate foreground contamination, such as the bias-hardened estimator \cite{namikawa2013bias,osborne2014extragalactic,sailer2020lower,calabrese2024atacama} and the shear-only estimator \cite{qu2023cmb,schaan2019foreground}. The former is based on the standard QE \cite{okamoto2003cosmic} and nullifies foreground biases through a linear combination, while the latter reconstructs the lensing potential using only the quadrupole ($m=2$, corresponds to shear), which ensures negligible response to foregrounds. However, both methods alleviate foreground biases at the cost of increasing reconstruction noise compared to the standard QE.

Therefore, the optimal combination of foreground removal methods and lensing reconstruction techniques remains an important area of study, with ongoing correlative research in this field \cite{schaan2019foreground,lizancos2021impact}.

\begin{figure}
	\includegraphics[width=\columnwidth]{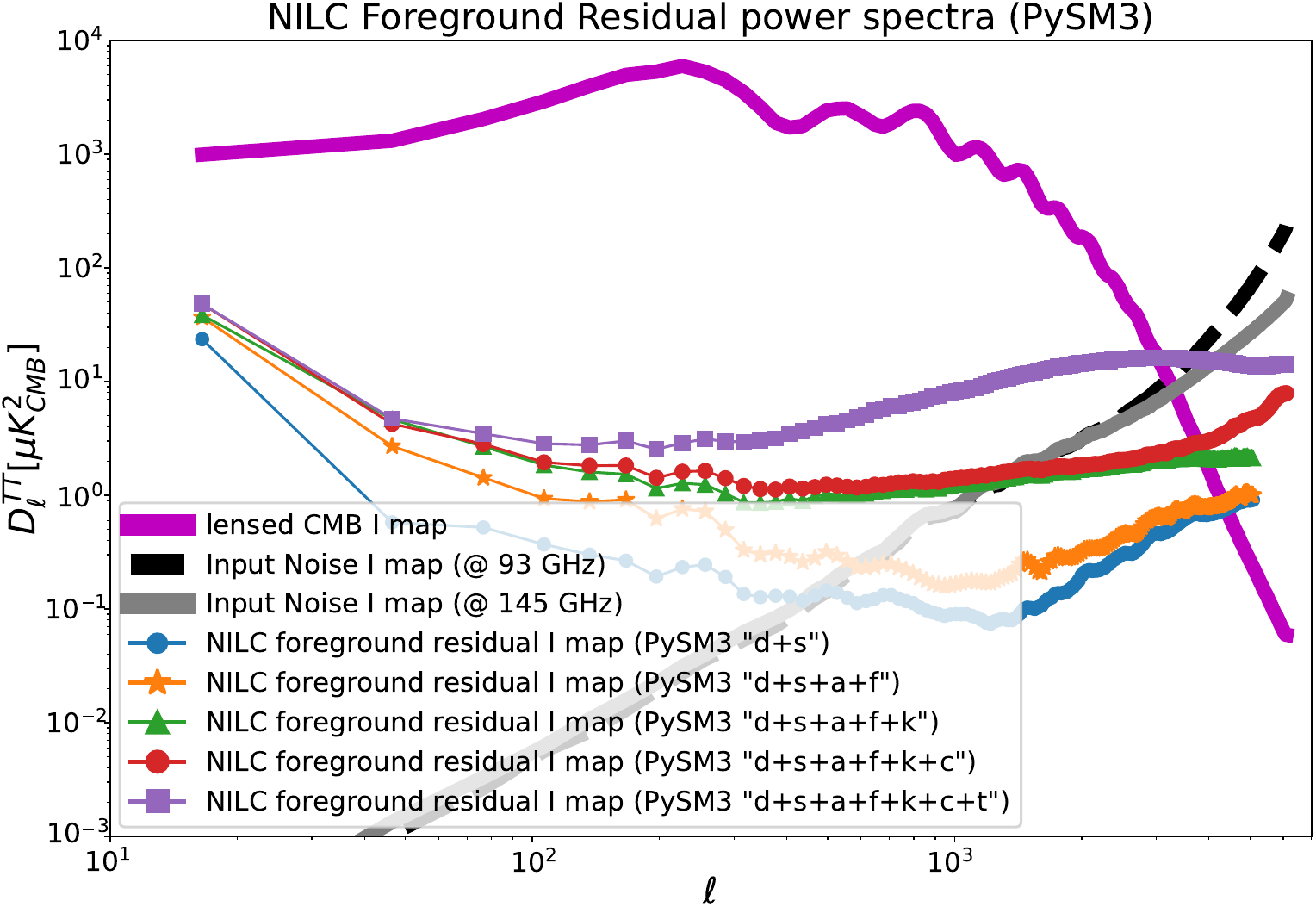}
	\caption{NILC temperature residual maps with different foreground components are included. The results are obtained by averaging over 50 simulations, each with varying CMB and noise realizations while keeping the foreground components fixed. The dotted lines represent the power of the residuals with varying levels of foreground complexity as input, where 'd' indicates thermal dust emission, 's' represents synchrotron emission, 'a' stands for spinning dust, 'f' denotes free-free emission, 'c' corresponds to the CIB, 'k' refers to the kinetic-SZ effect, and 't' represents the thermal-SZ effect. The instrumental noise curves have been beam-corrected.}
	\label{fig:nilc_comp_res}
\end{figure}

\section{Validation of the simulation debiasing}\label{sec:check_debias}
\subsection{Varying Cosmology}
In this work, component maps are simulated based on various models. 
The simulation of CMB and lensing (and LSS tracers) is based on the \(\Lambda\)CDM model with parameters given by the Planck 2018 best-fit \cite{aghanim2020planck}. 
The simulation of instrumental effects is described by the white noise level and the Gaussian beam. 
The simulation of foregrounds is based on models provided by PySM3 \cite{zonca2021python,thorne2017python} and the parameterization described in Section \ref{sec:sim_fg}.
In realistic observations, the difference between the simulated and observed data may introduce additional systematic effects. The impact of these effects on the delensing procedure should be carefully assessed to avoid introducing large bias on $r$ constraint.

Since instrumental effects are experiment-specific—depending on factors such as time-stream simulations \cite{quad2009second} or sign-flipping techniques \cite{ade2014bicep2,van2012measurement}, and foreground simulations are highly model-dependent, we do not attempt to quantify their impact in this work. A comprehensive treatment of these effects is beyond the scope of this study.
In this section, we briefly investigate the impact of cosmological parameter uncertainties on delensing performance and $r$ constraint, arising from the mismatch between the assumed cosmology and the true underlying universe.

\begin{table*} 
	\centering 
	\caption{The 68\% interval given by Planck 2018 "TT,TE,EE+lowE+lensing+BAO" and the five varying parameter cases used for validation the simulation.}
	\label{tab:vary_cosmos_params}
	\begin{adjustbox}{max width=\textwidth} 
	\begin{tabular}{lccccccc} 
		\hline
		&  \multicolumn{1}{c}{Planck 2018 "TT,TE,EE+lowE+lensing+BAO"} & \multicolumn{5}{c}{Varying Parameter Cases} \\
		\cmidrule(lr){2-2} \cmidrule(lr){3-7} 
        Params.  & 68\% interval & Exact Scale & Underest. Large Scale  & Underest. Small Scale & Overest. Small Scale  & Overest. Large Scale  \\
        \hline
        $\Omega_b h^2$ & $0.02242 \pm 0.00014$ & $0.02242$ & $0.02256$ & $0.02256$ & $0.02228$ & $0.02228$ \\
        $\Omega_c h^2$ & $0.11933 \pm 0.00091$ & $0.11933$ & $0.11842$ & $0.11842$ & $0.12024$ & $0.12024$ \\
        $100 \theta_\text{MC}$ & $1.04101 \pm 0.00029$ & $1.04101$ & $1.04130$ & $1.04130$ & $1.04072$ & $1.04072$ \\
        $\tau$ & $0.0561 \pm 0.0071$ & $0.0561$ & $0.0632$ & $0.0632$ & $0.0490$ & $0.0490$ \\
        $ln(10^10 A_s)$ & $3.047 \pm 0.014$ & $3.047$ & $3.033$ & $3.033$ & $3.061$ & $3.061$ \\
        $n_s$ & $0.9665 \pm 0.0038$ & $0.9665$ & $0.9703$ & $0.9627$ & $0.9703$ & $0.9627$ \\
        \hline
		\hline
	\end{tabular}
	\end{adjustbox}
\end{table*}

We consider varying six fundamental parameters for the base-\(\Lambda\)CDM model: \{$\Omega_b h^2$, $\Omega_c h^2$, $100 \theta_\text{MC}$, $\tau$, $\ln(10^{10} A_s)$, $n_s$\}, within the 68\% confidence interval as provided in the "TT,TE,EE+lowE+lensing+BAO" case of Table 2 in \cite{aghanim2020planck}, where $\Omega_b h^2$ is the physical baryon density, $\Omega_c h^2$ is the physical cold dark matter density, $\theta_\text{MC}$ is the angular acoustic scale (approximate), $\tau$ is the optical depth to reionization, $A_s$ is the amplitude of the primordial scalar power spectrum and $n_s$ represents the scalar spectral index. For each parameter, we assign four values between the 68\% interval lower bound and upper bound.
The six-dimensional parameter space consists of \( 4^6 \) mesh points. We compute the lensing potential power spectrum and the lensed CMB power spectra at each of these points using \texttt{CAMB} \cite{lewis2011camb}, resulting in \( 4^6 \) curves, illustrated as the shaded region in Fig.~\ref{fig:vary_cosmos_power}. 

In addition, we select five representative parameter combinations, listed in Table~\ref{tab:vary_cosmos_params}, that yield the maximum and minimum lensing potential power on both large ($\ell<300$) and small ($\ell>300$) angular scales, as shown in the thick colored lines in Fig.~\ref{fig:vary_cosmos_power}. These combinations also correspond to extrema in the lensed B-mode power on large scales. For instance, in the upper-left panel, the "Overestimate Small Scale Case" lies near the upper edge of the shaded region for \(\ell > 300\), whereas the "Overestimate Large Scale Case" dominates the upper bound for \(\ell < 300\). These selections are motivated by the scale-dependent nature of the lensing effect, which manifests differently across angular scales.

\subsection{Effect on the Pipeline}
Given the five sets of parameter combinations, we generate lensed CMB map simulations following the same procedure described in Section~\ref{sec: sim_cmb}, with the only modification being the use of different power spectra. The simulations of instrumental noise and foregrounds are kept consistent with those used in the main analysis. For each parameter set, we produce 50 realizations of the simulated maps.
We treat the “Exact Case” as representing the true underlying cosmology, while the remaining four parameter sets are interpreted as our imperfect estimates of the real universe. These reflect possible mismatches between the true cosmos and our assumed cosmological model within the current parameter constraints.

Subsequently, we apply the same analysis pipeline as outlined in the main text, with the delensing procedure restricted to internal lensing reconstruction for simplicity. The four sets of mismatched simulations are used to estimate the lensing template bias for the “Exact Case”. The flowchart of this validation procedure is presented in Fig. \ref{fig:vary_cosmos_flowchart}. In Fig.~\ref{fig:vary_cosmos_temp_diff}, we present the fractional difference between the debiased lensing template power spectra and the theoretical lensing B-mode power spectrum of the “Exact Case”. 

Our results indicate that, within the current uncertainties of the base cosmological parameters (see Table~\ref{tab:vary_cosmos_params}), the mismatch in cosmology used for bias estimation introduces a bias residual of approximately 3\% to 4\%. This level of bias is relatively small compared to the contributions from instrumental noise and foreground residuals.

In the likelihood evaluation, we retain the same covariance matrix used in the main analysis. This is justified by the fact that the contribution to the covariance from the variation in the lensing B-mode power spectrum—induced by mismatches in cosmological parameters—is negligible compared to that from the dominant components: the CMB signal, instrumental noise, and foreground residuals. However, to properly reflect the differences in assumed cosmologies, we use distinct bias estimates derived from the five simulation sets. Additionally, the theoretical power spectra entering the likelihood function are replaced by those computed under each of the five different cosmological parameter combinations.

The parameter constraint results are summarized in Table~\ref{tab:vary_cosmos_fit}, and the corresponding posterior distributions are shown in Fig.~\ref{fig:vary_cosmos_fit}. As expected, the mismatched cosmologies introduce negligible changes to the parameter uncertainties. More importantly, under the current level of cosmological parameter uncertainties (see Table~\ref{tab:vary_cosmos_params}), the bias in the inferred tensor-to-scalar ratio $r$ due to mismatched debiasing is found to be $\Delta r \sim 1 \times 10^{-4}$, which is significantly smaller than the statistical uncertainty ($ \sigma(r) \sim 1 \times 10^{-3}$). For future CMB experiments targeting a detection of $\sigma(r) \sim \mathcal{O}(10^{-4})$—such as CMB-S4 \cite{abazajian2016cmb}—the bias induced by mismatches in cosmological parameters is expected to diminish, as constraints on these parameters become increasingly precise with improved instrumental sensitivity.

We thus conclude that the simulation-based debiasing procedure adopted in our pipeline remains robust and reliable under the current constraints on cosmological parameters, as summarized in Table~\ref{tab:vary_cosmos_params}. A more comprehensive investigation of the validity and potential limitations of this debiasing approach is left for future work.

\begin{figure*}
	\includegraphics[width=\textwidth, height=0.95\textheight, keepaspectratio]{./fig_revise/vary_cosmos_power.jpg}
	\caption{The power spectra of the lensing potential, lensed TT, lensed EE, and lensed BB fields are computed using the parameters from a six-dimensional parameter space with \( 4^6 \) mesh points. We show the ratio of the power spectra between the values at the mesh points and those at the best-fit parameters. The thick solid black lines delineate the upper and lower bounds of the ratio regions, and we have arbitrarily chosen five sets of parameter combinations for the simulations. For the lensed EE power, and the power below \( \ell \sim 20 \) is significantly affected by variations in the optical depth \( \tau \) due to reionization.}
	\label{fig:vary_cosmos_power}
\end{figure*}

\begin{figure*}
	\includegraphics[width=\textwidth, height=0.95\textheight, keepaspectratio]{./fig_revise/vary_cosmos_flowchart}
	\caption{Flowchart illustrating the validation test of the delensing de-biasing procedure under cosmology mismatch. The process begins with constructing a six-dimensional parameter grid based on the 68\% confidence intervals of Planck 2018 best-fit parameters. Five representative cosmologies are selected from this grid to generate simulated lensed CMB maps. These simulations are then used to assess the impact of mismatched bias estimates on the delensing procedure and parameter constraints, assuming the "Exact Case" as the true cosmology. }
	\label{fig:vary_cosmos_flowchart}
\end{figure*}

\begin{figure*}
	\includegraphics[width=\textwidth, height=0.95\textheight, keepaspectratio]{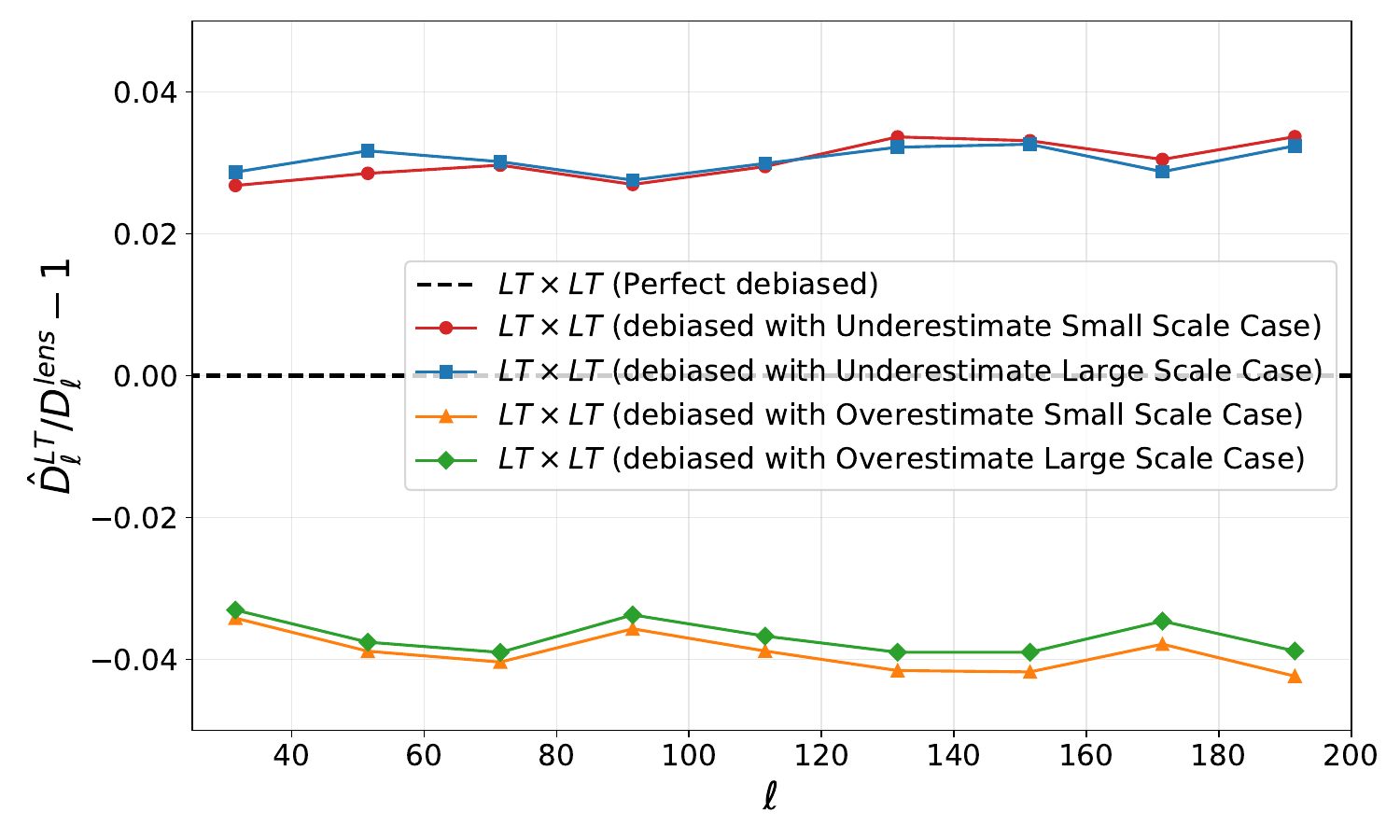}
	\caption{The fractional differences between the debiased lensing template power spectra and the theoretical lensing B-mode power spectrum of the “Exact Case” are shown. We consider five sets of simulations used to debias the lensing template derived from the “Exact Case” simulation. The “Perfect Debiasing” case refers to using the same simulation for both bias estimation and correction, which naturally leads to a negligible residual bias. The remaining four cases involve debiasing the “Exact Case” template using bias estimates from mismatched simulations, each corresponding to a different cosmological model. Under the current 68\% confidence interval constraints on cosmological parameters (see Table~\ref{tab:vary_cosmos_params}), such mismatches introduce a residual bias at the level of approximately 3\% to 4\%.}
	\label{fig:vary_cosmos_temp_diff}
\end{figure*}

\begin{figure*}
	\includegraphics[width=\textwidth]{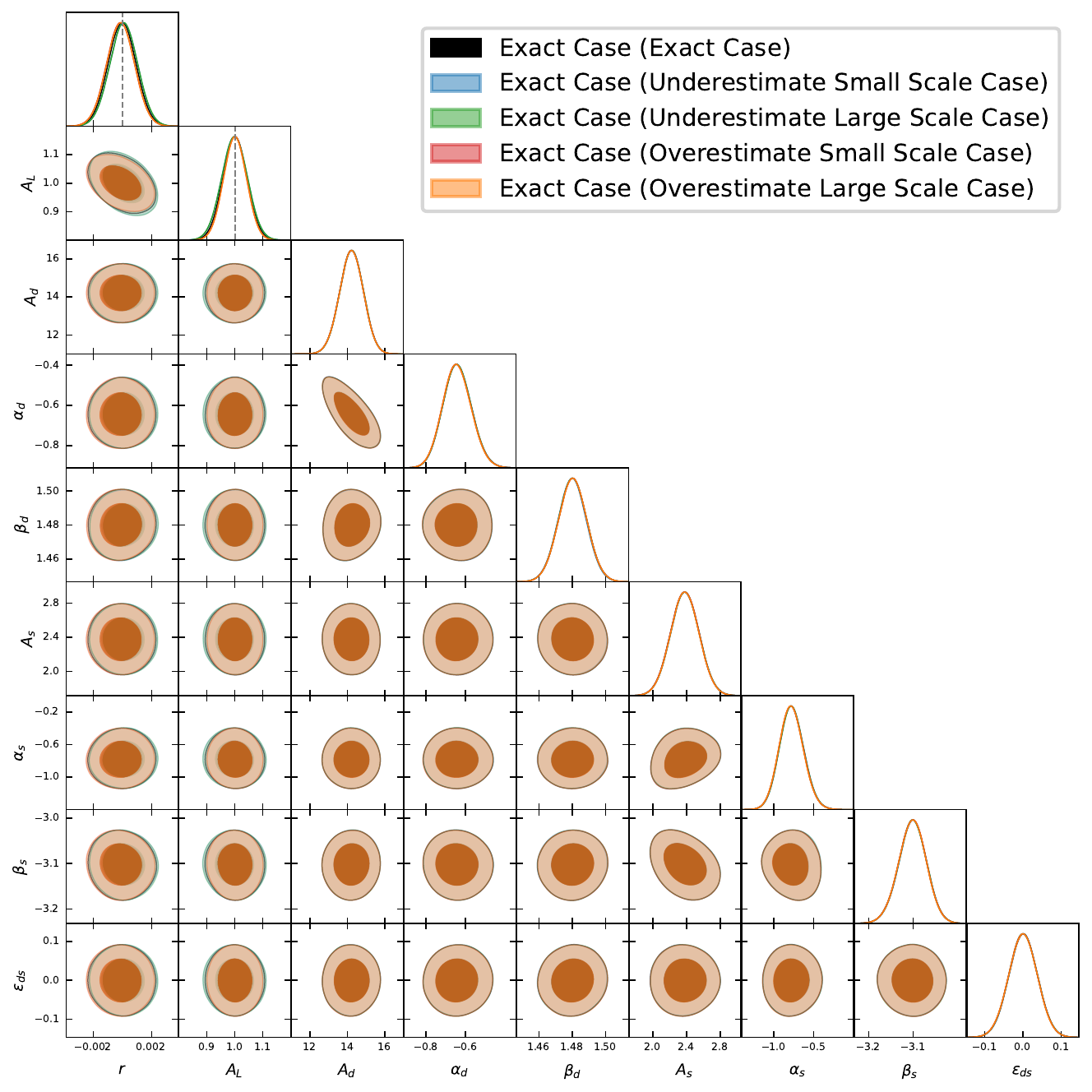}
        \caption{The posterior distributions of the baseline model parameters are shown for the “Exact Case” simulation dataset, where the bias in the lensing template is estimated using different cosmological simulations. A Gaussian approximation is adopted in the likelihood analysis. The labels in parentheses in the legend indicate the specific simulation set used for estimating the lensing template bias.}
	\label{fig:vary_cosmos_fit}
\end{figure*}

\begin{table*} 
	\centering 
	\caption{The mean and 1$\sigma$ standard deviation of each parameter for the “Exact Case” simulation are shown. The parameter constraints are obtained using a Gaussian-approximated likelihood. The difference between the third and subsequent columns lies in the choice of simulations used to estimate and subtract the bias: each case uses a different one of the five simulation sets for debiasing the lensing template in the “Exact Case” analysis.}
	\label{tab:vary_cosmos_fit}
	\begin{adjustbox}{max width=\textwidth} 
	\begin{tabular}{lcccccc} 
		\hline
		& & & \multicolumn{2}{c}{Underestimate Case} & \multicolumn{2}{c}{Overestimate Case} \\
		\cmidrule(lr){4-5} \cmidrule(lr){6-7} 
		Parameter & Input value & Exact Case & Small Scale & Large Scale & Small Scale & Large Scale \\
		\hline
		$r (\times 10^3)$ & 0 & $-0.036 \pm 0.954$ & $0.073 \pm 0.954$ & $0.081 \pm 0.957$ & $-0.145 \pm 0.955$ & $-0.133 \pm 0.953$ \\
		$A_L$ & 1 & $1.000 \pm 0.043$ & $0.999 \pm 0.046$ & $0.999 \pm 0.046$ & $1.001 \pm 0.041$ & $1.001 \pm 0.041$ \\
		$A_d (\mu K^2)$ & 14.300 & $14.214 \pm 0.636$ & $14.218 \pm 0.638$ & $14.219 \pm 0.637$ & $14.219 \pm 0.638$ & $14.212 \pm 0.635$ \\
        $\alpha_d$ & -0.650 & $-0.642 \pm 0.072$ & $-0.641 \pm 0.072$ & $-0.643 \pm 0.072$ & $-0.643 \pm 0.072$ & $-0.643 \pm 0.072$ \\
        $\beta_d$ & 1.48 & $1.480 \pm 0.008$ & $1.480 \pm 0.008$ & $1.480 \pm 0.008$ & $1.480 \pm 0.008$ & $1.480 \pm 0.008$ \\
        $A_s (\mu K^2)$ & 2.400 & $2.378 \pm 0.170$ & $2.380 \pm 0.170$ & $2.380 \pm 0.170$ & $2.377 \pm 0.170$ & $2.378 \pm 0.170$ \\
        $\alpha_s$ & -0.800 & $-0.778 \pm 0.152$ & $-0.774 \pm 0.152$ & $-0.775 \pm 0.152$ & $-0.787 \pm 0.152$ & $-0.780 \pm 0.152$ \\
        $\beta_s$ & -3.100 & $-3.102 \pm 0.031$ & $-3.102 \pm 0.031$ & $-3.102 \pm 0.031$ & $-3.102 \pm 0.031$ & $-3.102 \pm 0.031$ \\
        $\epsilon_{ds}(\times 10^2)$ & 0 & $-0.022 \pm 3.736$ & $-0.048 \pm 3.719$ & $-0.017 \pm 3.740$ & $-0.014 \pm 3.736$ & $-0.025 \pm 3.730$ \\
		\hline
	\end{tabular}
	\end{adjustbox}
\end{table*}

\section{Lensing Potential Reconstruction}\label{sec:phirec}
\subsection{Internal Lensing Reconstruction}
CMB internal lensing reconstruction is anticipated to provide the optimal estimate of the true convergence, thanks to the high-redshift origin of the CMB. This reconstruction not only contributes to constraining matter-related cosmological parameters (e.g. $\sigma_8$, $\Omega_m$, $\Sigma m_{\nu}$) \cite{aghanim2020planck}, but also serves as a crucial prerequisite for CMB delensing.

In this paper, we reconstruct the lensing potential $\hat{\phi}$ from pairs of filtered observed fields, following the implementation described in \cite{carron2017maximum,aghanim2020planck}.

Below, we provide a brief outline of the procedure for reconstructing \( \phi \) in real space, based on the Planck 2018 baseline \cite{aghanim2020planck}. For further details, we recommend that readers refer to the original sources.

\begin{enumerate}
    \item Filter of the CMB map.

    To down-weight noise-dominated modes and deconvolve the window function caused by the finite beam size and pixelization, we first apply a Wiener filter to the observed maps
    \begin{equation}
        \begin{pmatrix}
            \Theta^{\text{WF}} \\                                          
            E^{\text{WF}} \\   
            B^{\text{WF}} \\                                           
        \end{pmatrix}
        = C^{\text{fid}} \mathcal{T}^{\dagger} \text{Cov}^{-1}
        \begin{pmatrix}
            \Theta^{\text{dat}} \\
            _{2}P^{\text{dat}} \\
            _{-2}P^{\text{dat}} \\
        \end{pmatrix},
    \end{equation}

    where $C^\text{fid}$ is the fiducial CMB spectrum,  $\mathcal{T} = \mathcal{B} \mathcal{Y}$ is the transfer function from multipole to pixel sky, $\mathcal{B}$ is the window function accounting for beam and pixel convolution effect, and $\mathcal{Y}$ contains the spin-weighted spherical harmonic functions to map from multipoles to the sky. $\text{Cov} = \mathcal{T} C^\text{fid} \mathcal{T}^{\dagger }+ N$ is the pixel-space covariance.
    We also calculate the inverse variance weighted CMB maps
    \begin{equation}
        \bar X (\hat{\mathbf{n}}) = [\mathcal{B} ^{\dagger} \text{Cov}^{-1} X^\text{dat}](\hat{\mathbf{n}}),
    \end{equation}
    where $X=\{\Theta,E,B\}$ in real space.
    \item Construction of the quadratic lensing estimator.
    
     A Gaussian likelihood is sufficient to describe the \( \Lambda \)CDM CMB lensing potential \( \phi \), according to \cite{carron2017maximum}, where an optimal Bayesian estimate is obtained by maximizing the log-posterior, implemented through iteration starting from a quadratic estimator (QE). In this paper, we focus on the QE, as we do not expect significant improvement from the Bayesian estimate for our experimental configuration. The spin-1 real-space (unnormalized) lensing displacement QE is given by
    \begin{equation}
        {_{1}}\hat{d}(\hat{\mathbf{n}}) = - \sum_{s=0,\pm2} {_{-s}}\bar{X}(\hat{\mathbf{n}}) \big[\sharp_s X^{\text{WF}}\big](\hat{\mathbf{n}}).
    \end{equation}
    Here, $\sharp$ denotes the spin-raising operator, and the subscript $s$ represents the spin of the fields. 
    The estimator takes as input the gradient of the Wiener-filtered maps and the real-space inverse-variance filtered maps. Using the following relation, we derive a raw estimator for the lensing potential:
    \begin{equation}
        _{\pm1}\hat d (\hat{\mathbf{n}}) \equiv \left(\sum_{LM} \frac{\hat g_{LM} \pm i\hat c_{LM}}{\sqrt{L(L+1)}}\right) {_{\pm1}Y_{LM}(\hat{\mathbf{n}})},
    \end{equation}
    where the gradient mode $\hat g_{LM}$ is what we want and the curl mode $\hat c_{LM}$ is expected to be extremely weak.
    
    \item Mean-field subtraction and normalization.
    
    It is worth noting that even in the absence of lensing effects, factors such as masks, noise, beam effects, and pixelization can still introduce anisotropies that mimic deflections (see \cite{ade2014planck}).
    Since $\phi$ is considered Gaussian, its one-point correlation function averages to zero across different realizations of $\phi$, leaving behind the spurious lensing effects, we thus estimate their contribution by taking the mean of $\hat{g}_{LM}$ over several Monte Carlo simulations. We denote these effects here and subtract the mean-field bias from the raw estimator $\hat{g}_{LM}$ derived from the data.

    Additionally, for the unbiasedness of the estimator, the response $\mathcal{R}$ must be calculated to serve as a normalization factor. We apply an approximate normalization at the map level, calculated analytically following \cite{Okamoto:2003zw}. Although this semi-analytic normalization is only accurate within $1\%-2\%$, it is sufficient for the purpose of map-level delensing. After these adjustments, the estimator is refined as follows:
    \begin{equation}\label{eq:qe_norm}
        \hat \phi_{LM} = \frac{1}{\mathcal{R}_L^{\phi}} \left(\hat g_{LM} - \langle \hat g_{LM}\rangle_{MC} \right).
    \end{equation}
    
    \item Calculation of the power spectrum of the lensing map and subtraction of additional biases.
    
    For any two of the estimators we get from previous steps, their cross spectrum estimator can be written as:
    \begin{equation}
        \hat C^{\hat \phi_1 \hat \phi_2}_L \equiv \frac{1}{(2L+1)f_\text{sky}} \sum_{M=-L}^L \hat \phi _{1,LM} \hat \phi_{2,LM}^*.
    \end{equation}
    Further calculations reveal that it is not an unbiased estimator, indicating the presence of reconstruction noise power. Therefore, the estimate of the lensing power is given by:
    \begin{equation}\label{eq:qe_cl_debias}
        \hat C^{\phi \phi}_L \equiv \hat C^{\hat \phi_1 \hat \phi_2}_L - \Delta C^{\hat \phi_1 \hat \phi_2}_L|_\text{RDN0} - \Delta C^{\hat \phi_1 \hat \phi_2}_L|_\text{N1} ,
    \end{equation}
    where, $\Delta C^{\hat \phi_1 \hat \phi_2}_L|_\text{RDN0}$ comes from disconnected signal from Gaussian fluctuation, and it exists  even if the CMB maps do not have lensing in them, and it is of the zeroth order of  $C^{\phi \phi}_L$.
    $\Delta C^{\hat \phi_1 \hat \phi_2}_L|_\text{N1}$ comes from the connected part of 4-point function, which arises from the effects from other $\ell$ modes when we estimate a certain $\phi_L$, and it is of the first order of $C^{\phi \phi}_L$. The $N^{(1)}$ bias shows up as excess power $\left(\sim 10 \%\right.$ of $C_L^{\phi \phi}$ at $\left.L \sim 1000\right)$ on small scales.
    \item Binning, and application of a multiplicative correction.
    
    To emphasize and compare the statistical features of $\hat{C}^{\phi \phi}_L$ and $C^{\phi \phi, \text{fid}}_L$, we define a band-power estimate as follows:
    \begin{equation}
        \hat C^{\kappa \kappa}_{L_b} = \left(\sum_L \mathcal{B}^L_b \hat C^{\kappa \kappa}_L \right) \left( \frac{\sum_L \mathcal{B}^L_b C^{\kappa \kappa, \text{fid}}_L}{\sum_L \mathcal{B}^L_b \langle\hat C^{\kappa \kappa}_L\rangle_\text{MC}}\right),
    \end{equation}
    please refer to \cite{aghanim2020planck} for an explicit expression of the binning function $\mathcal{B}^L_b$.
\end{enumerate}

All these steps can be implemented with \texttt{Plancklens} \cite{aghanim2020planck}, which is a Python package allowing us to calculate the MV $\phi$ estimators from filtered maps on curved sky fast.

\subsection{Combination with external tracers}\label{sec:external_rec}
As mentioned in the Introduction, under current conditions of CMB experiments, combining external LSS tracers, such as CIB and galaxy number density, can significantly improve the signal-to-noise ratio (S/N) of lensing reconstruction, thereby enhancing the delensing efficiency.
In this work, we combine external LSS tracers following the methodology described in \cite{sherwin2015delensing}, where a detailed derivation is provided in their Appendix. 

The lensing potential \(\phi\) can be estimated as a linear combination of tracers $I_i$:
\begin{equation}
    I = \sum_i c^i I_i.
\end{equation}
By maximizing the correlation coefficient \( \rho = \frac{C^{\kappa I}}{\sqrt{C^{\kappa \kappa} C^{II}}} \) between the combined tracer and the true convergence \( \kappa \), the explicit form of the coefficients \( c^i \) is determined:
\begin{equation}
    c_i = \left(C_{II}^{-1}\right)_{ij} C^{\kappa I^j},
\end{equation}
with \(C_{II}\) being the covariance matrix of the LSS tracers, which is calculated using the nine tracer realizations in each simulation. The “effective” correlation \(\rho^2_\ell\) of the combined tracer with gravitational lensing is then given by:
\begin{equation}
    \rho_{\ell}^2 = \sum_{i, j} \frac{C_{\ell}^{\kappa I^i} \left(C_{\ell}^{II}\right)_{ij}^{-1} C_{\ell}^{\kappa I^j}}{C_{\ell}^{\kappa \kappa}}.
\end{equation}

Note that the gain from adding a new tracer is not solely proportional to its correlation with the CMB lensing but also depends on its correlation with the existing set of tracers. The correlation coefficient \(\rho\) serves as a direct indicator of the quality of our lensing reconstruction. We observe a significant improvement in the correlation when combining the quadratic estimator from CMB with CIB at 353\,GHz and galaxy number density from \emph{Euclid}, as discussed in Section \ref{sec: tracer_comb_exe}.

\section{Some algorithms of delensing effect}\label{sec:algorithms}
Here, we highlight an algorithm for the delensing effect, which could be useful for separating the components of the delensing output.

We begin by assuming that a lensed field is remapped using a potential field $\phi = \phi_1 + \phi_2$, which is artificially divided into two components (e.g., signal and noise), corresponding to the inverse-lensing angles $\mathbf{d^{inv}_1}$ and $\mathbf{d^{inv}_2}$, respectively. Assuming that the total inverse-lensing angle satisfies $\mathbf{d^{inv}} \approx \mathbf{d^{inv}_1} + \mathbf{d^{inv}_2}$, we acknowledge that this approximation neglects the high-order term of $\mathbf{d^{inv}}$ (when solving Eq.(\ref{hat_n}) with Newton-Raphson method, the gradient operation is linear, while the iteration is not). However, we have verified that this approximation has negligible influence under reasonable lensing (or delensing) scenarios.
Then we can write the delensed field as:

\begin{equation}
\begin{aligned}
	X^{de}(\hat{\mathbf{n}}) &\approx  \tilde X (\hat{\mathbf{n}} + \mathbf{d^{inv}_1} + \mathbf{d^{inv}_2}) \\
	& \approx \tilde X(\hat{\mathbf{n}} ) + \nabla \tilde X (\hat{\mathbf{n}}) ( \mathbf{d^{inv}_1} + \mathbf{d^{inv}_2})\\ & \quad + \frac{1}{2} \nabla^2 \tilde X (\hat{\mathbf{n}}) (\mathbf{d^{inv}_1} + \mathbf{d^{inv}_2})^2 + \mathcal{O} (\mathbf{d^{inv}} ^2)\\
	& \approx \left[\tilde X(\hat{\mathbf{n}})  +  \nabla \tilde X(\hat{\mathbf{n}}) \mathbf{d^{inv}_1} +  \frac{1}{2} \nabla^2 \tilde X(\hat{\mathbf{n}}) \mathbf{d^{inv}_1}^2\right]\\
	& \quad + \left[ \tilde X(\hat{\mathbf{n}}) + \nabla \tilde X(\hat{\mathbf{n}}) \mathbf{d^{inv}_2}  + \frac{1}{2} \nabla^2 \tilde X(\hat{\mathbf{n}}) \mathbf{d^{inv}_2}^2\right] \\ &\quad - \tilde X(\hat{\mathbf{n}} )+ \mathcal{O} (\mathbf{d^{inv}} ^2)\\
	& \approx \tilde X(\hat{\mathbf{n}} + \mathbf{d^{inv}_1}) + \tilde X(\hat{\mathbf{n}} + \mathbf{d^{inv}_2}) - \tilde X(\hat{\mathbf{n}} ) + \mathcal{O} (\mathbf{d^{inv}} ^2).
\end{aligned}
\end{equation}

However, if we remap a sum of two lensed field $\tilde X = \tilde X_1 + \tilde X_2$ with a single potential, it will be easy finding that this just a linear algorithm, so:
\begin{equation}
	\begin{aligned}
		X^{de}(\hat{\mathbf{n}}) &= \tilde X(\hat{\mathbf{n}}+\mathbf{d^{inv}}) \\
        &=  X_1(\hat{\mathbf{n}} + \mathbf{d^{inv}}) +  X_2(\hat{\mathbf{n}} + \mathbf{d^{inv}}) \\
        &= X_1^\text{de}(\hat{\mathbf{n}} ) +  X_2^\text{de}(\hat{\mathbf{n}} ).
	\end{aligned}
\end{equation}

Above all, we can summary here the delensing algorithms :
\begin{equation}\label{EQA3}
	\begin{aligned}
		&X \star \left(\phi_1 + \phi_2\right) \approx  X \star \phi_1 + X \star \phi_2 - X + \mathcal{O} (\mathbf{d^{inv}} ^2), \\
		&(X_1 + X_2) \star \phi =  X_1 \star \phi + X_2 \star \phi.
	\end{aligned}
\end{equation}
Here, we use $\star$ to represent the delensing operation, $X_i (i=1,2)$ represents one of  $\{ \Theta, Q \pm iU \}$ among the CMB field, instrumental noise, or the observed CMB field, and $\phi_i (i=1,2)$ represents the lensing potential or its reconstruction noise.

As for the gradient order template method, we find the lensing template map is straightforward and just obeys a linear algorithm as :
\begin{equation}
	\begin{aligned}
		&\nabla[(Q_1+Q_2)\pm i(U_1+U_2)](\hat{\mathbf{n}})\nabla[\phi_1(\hat{\mathbf{n}})+\phi_2(\hat{\mathbf{n}})] \\
		&= \nabla[Q_1\pm iU_1](\hat{\mathbf{n}})\nabla[\phi_1(\hat{\mathbf{n}})] + \nabla[Q_2\pm iU_2](\hat{\mathbf{n}})\nabla[\phi_2(\hat{\mathbf{n}})] \\
		&\quad + \nabla[Q_2\pm iU_2](\hat{\mathbf{n}})\nabla[\phi_1(\hat{\mathbf{n}})] + \nabla[Q_1\pm iU_1](\hat{\mathbf{n}})\nabla[\phi_2(\hat{\mathbf{n}})].
    \end{aligned}
\end{equation}

\section{Redshift binning in galaxy survey}\label{app:z_binning}
The institutional example provided by \cite{manzotti2018future} is briefly summarized here. For a more detailed discussion, please refer to the original source.

Let's split a single survey $I$ into two non-overlapping redshift bins $I_1$ and $I_2$ with $I=I_1+I_2$. For the full survey the effective cross-correlation is equal to
\begin{equation}
\rho_{\text {full }}^2=\frac{\left(C_{\ell}^{\kappa I}\right)^2}{C_{\ell}^{\kappa \kappa} C_{\ell}^{I I}}=\frac{\left(C_{\ell}^{\kappa I_1}+C_{\ell}^{\kappa I_2}\right)^2}{C_{\ell}^{\kappa \kappa}\left(C_{\ell}^{I_1 I_1}+C_{\ell}^{I_2 I_2}\right)},
\end{equation}
while for the split survey it will be
\begin{equation}
\rho_{\mathrm{split}}^2=\frac{\left(C_{\ell}^{\kappa I_1}\right)^2}{C_{\ell}^{\kappa \kappa} C_{\ell}^{I_1 I_1}}+\frac{\left(C_{\ell}^{\kappa I_2}\right)^2}{C_{\ell}^{\kappa \kappa} C_{\ell}^{I_2 I_2}}.
\end{equation}



Now it can be shown that \( \rho_{\text{split}} \geq \rho_{\text{full}} \) since:
\begin{equation}
\rho_{\text{split}}^2 - \rho_{\text{full}}^2 \propto \left(C_{\ell}^{I_1 I_1} C_{\ell}^{\kappa I_2} - C_{\ell}^{I_2 I_2} C_{\ell}^{\kappa I_1}\right)^2.
\end{equation}
Thus, \( \rho^2 \) is always larger in the tomographic case, and the two are equal only when \( \frac{C^{\kappa I_i}}{C^{I_i I_i}} \) is the same across all redshift bins. In this scenario, a single optimal weight would suffice for the entire survey.
While binning improves the efficiency of galaxy surveys, it is less effective for tracers with poor redshift information, such as the Cosmic Infrared Background (CIB) or radio continuum surveys. Although redshift binning enables better utilization of redshift information, excessive binning can result in a low number density within each bin, leading to increased shot noise. Therefore, the choice of binning strategy requires careful consideration.

\section{A toy test of the generalized model}\label{app:model_toy}
In this section, we assess the potential of the generalized model to capture possible biases, including those arising from discrepancies between the theoretical and observed lensed B-modes, as well as biases artificially introduced during the construction of the lensing B-mode template.

We test the effectiveness of the model using theoretical power spectra, where the fiducial lensed B-mode power is artificially biased by multiplying it by 1.02, $\sqrt{1.02 \times 0.98}$, and 0.98, respectively, to serve as the power in the mock data $\mathbf{\hat{X}}_{\ell}^{\text{toy}} = \left[ C_{\ell}^{\text{SAT},\nu \nu'}, C_{\ell}^{\text{LT} \times \text{SAT},\nu}, C_{\ell}^{\text{LT}} \right]$, while keeping the other power spectra unchanged:
\begin{align}
    &\begin{aligned}
    \hat C_{\ell}^{\text{SAT},\nu \nu'} &= 1.02*C_{\ell}^\text{lens} 
     + N_{\ell}^{BB,\nu \nu'} + C_{\ell}^{\nu \nu'}|_\text{FG},
    \end{aligned}\\
    &\hat C_{\ell}^{\text{LT} \times \text{SAT},\nu} = \sqrt{1.02 * 0.98}C_{\ell}^\text{lens}, \\
    &\hat C_{\ell}^\text{LT} = 0.98 * C_{\ell}^\text{lens} + N_{\ell}^\text{temp}.
\end{align}

We then perform MCMC sampling on this mock data using both the generalized and baseline models, with a Gaussian approximation likelihood to obtain constraints on the parameters. The covariance matrix used in the main text is approximately applied here, and we expect this will not have a significant impact on the results.

\begin{figure}
	\includegraphics[width=\columnwidth]{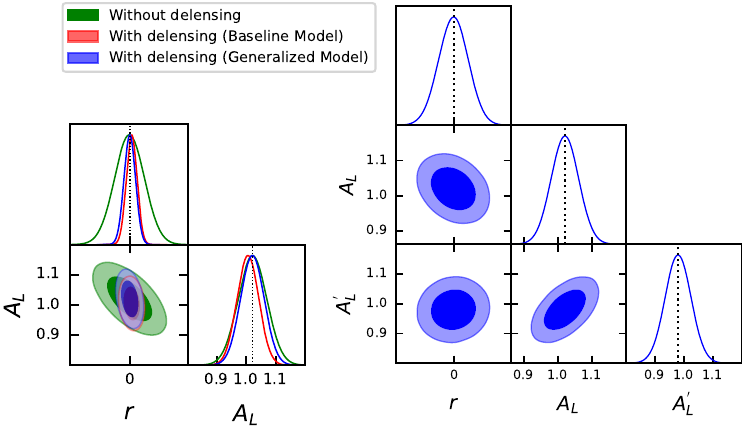}
        \caption{A comparison of the posterior distributions for the parameters \(r\), \(A_L\), and \(A_L'\) from the generalized model and the baseline model is shown here. We focus on these three parameters, as the seven foreground parameters show no significant differences between the two models. The black dashed lines in each subplot represent the fiducial values. From the left triangle plot, we observe that the baseline model fails to constrain both \(r\) and \(A_L\) effectively, with noticeable biases for both. In contrast, the generalized model provides a good constraint on \(r\), \(A_L\), and \(A_L'\), with only a slight increase in uncertainty compared to the baseline model. }
	\label{fig:posterior_fit_toy}
\end{figure}

The posterior distributions of the parameters are shown in Fig.\ref{fig:posterior_fit_toy} and summarized in Table.\ref{tab:posterior_gs_toy}. The posterior of $r$ from the generalized model shows significantly lower bias compared to that from the baseline model, with only a slight increase in uncertainty. We find that the generalized model effectively recognizes $A_L$ and $A_L^{'}$, thereby preventing the lensing B-mode power from leaking into the tensor B-mode power, which would lead to a bias on $r$.

In summary, this toy test demonstrates the potential of the generalized model to better account for the bias in the lensing power, thus mitigating the bias on $r$ when the delensing procedure is considered.

\begin{table*} 
	\centering 
	\caption{The mean and \( 1\sigma \) standard deviation of each parameter using the Gradient-order method, with the lensing proxy from the combined tracer, were obtained. A Gaussian approximation likelihood was used in the likelihood analysis. Two models described in Section \ref{sec: baseline_fit} and Section \ref{sec: generalized_fit} were considered.}
    \label{tab:posterior_gs_toy}
	\begin{adjustbox}{max width=\textwidth} 
	\begin{tabular}{lcccccc} 
		\hline
		& & & \multicolumn{1}{c}{Generalized Model} & \multicolumn{1}{c}{Baseline Model} \\
		\cmidrule(lr){4-4} \cmidrule(lr){5-5} 
		Parameter & Fiducial value & Before adding LT &  After adding LT (Combined tracer) &  After adding LT (Combined tracer)\\
		\hline
		$r (\times 10^3)$ & 0 & $-0.068 \pm 1.808$ & $-0.041 \pm 0.691$ & $0.198 \pm 0.648$  \\
		$A_L$ & 1.020 & $1.020 \pm 0.049$ & $1.020 \pm 0.040$ & $1.005 \pm 0.037$  \\
		$A_L^{'}$ & 0.980 & N/A & $0.980 \pm 0.045$  & N/A\\
		$A_d (\mu K^2)$ & 14.300 & $14.213 \pm 0.640$ & $14.217 \pm 0.633$ & $14.206 \pm 0.636$  \\
        $\alpha_d$ & -0.650 & $-0.642 \pm 0.072$ & $-0.642 \pm 0.072$ & $-0.643 \pm 0.072$  \\
        $\beta_d$ & 1.48 & $1.480 \pm 0.010$ & $1.480 \pm 0.007$ & $1.480 \pm 0.007$  \\
        $A_s (\mu K^2)$ & 2.400 & $2.378 \pm 0.175$ & $2.378 \pm 0.167$ & $2.385 \pm 0.167$  \\
        $\alpha_s$ & -0.800 & $-0.777 \pm 0.157$ & $-0.778 \pm 0.148$ & $-0.771 \pm 0.148$  \\
        $\beta_s$ & -3.100 & $-3.102 \pm 0.035$ & $-3.102 \pm 0.028$ & $-3.101 \pm 0.028$  \\
        $\epsilon_{ds}(\times 10^2)$ & 0 & $0.015 \pm 3.823$ & $0.011 \pm 3.694$ & $0.010 \pm 3.684$ \\
		\hline
	\end{tabular}
	\end{adjustbox}
\end{table*}

\section*{Acknowledgments}

We greatly appreciate Sebastian Belkner for his valuable and insightful discussions and suggestions.  
We also thank Siyu Li, Yongping Li, Yepeng Yan, Zi-Xuan Zhang, and Qian Chen for useful discussions.  
This study is supported by the National Key R\&D Program of China No. 2020YFC2201601 and the National Natural Science Foundation of China No. 12403005.  
We acknowledge the use of several Python packages, including:
\texttt{CAMB} \cite{lewis2011camb},  
\texttt{healpy} and \texttt{HEALPix}\footnote{\url{https://healpix.sourceforge.io/}} \cite{2005ApJ...622..759G,Zonca2019},  
\texttt{Lenspyx} \cite{carron2020lenspyx,reinecke2023improved},  
\texttt{CMBlensplus} \cite{namikawa2021CMBlensplus},  
\texttt{Plancklens} \cite{aghanim2020planck},  
\texttt{PySM3} \cite{zonca2021python,thorne2017python},  
\texttt{pyccl} \cite{chisari2019core},  
\texttt{NaMaster} \cite{alonso2023namaster}, and  
\texttt{Cobaya} \cite{Torrado_2021,torrado2020cobaya}.

\bibliographystyle{spphys}       
\bibliography{main}   

\end{document}